\documentclass{aa}  
\usepackage{graphicx}       
\usepackage{natbib}
\usepackage{longtable}
\usepackage{supertabular}

\newcommand{\BCDG}{{BCDG}}

\newcommand{\FIR}{\emph{FIR}}
\newcommand{\HI}{\textsc{H\,i}}
\newcommand{\Ha} {H$\alpha$}
\newcommand{\Hb} {H$\beta$}
\newcommand{\HII}{\textsc{H\,ii\ }}

\newcommand{\IMF}{{IMF}}
\newcommand{\ISM}{{\sc ISM}}

\newcommand{\Mo} {$M_{\odot}$}
\newcommand{\Myr}{Myr}          

\newcommand{\NIR} {\emph{NIR}}

\newcommand{\SDSS}{{SDSS}}

\newcommand{\TDG} {{TDG}}

\newcommand{\UV}{\emph{UV}}

\newcommand{\WRBUMP} {WR bump}
\newcommand{\Zo} {$Z_{\odot}$}

\newcommand{\STARBURST}{{\sc Starburst~99}}
\newcommand{\PEGASE}{{\sc Pegase.2}}
\newcommand{\CLOUDY}{{\sc Cloudy}}
\newcommand{\MAPPINGS}{{\sc Mappings}}

\newcommand{\WHa}{$W($H$\alpha)$}

\newcommand{\tableline}{\hline}

\newcommand{\WHb}{$W$(H$\beta$)}

\newcommand{\Te}{$T_{\rm e}$}

\newcommand{\TeOiii}{$T_e$(\ion{O}{iii})}
\newcommand{\TeOii}{$T_e$(\ion{O}{ii})}

\newcommand{\CHb}{$c$(H$\beta$)}
\newcommand{\Wabs}{$W_{abs}$}

\newcommand{\abox}{12+log(O/H)}

\newcommand{\nodata}{...}
\begin{document}
   \title{Massive star formation in Wolf-Rayet galaxies\thanks{Based on observations made with NOT (Nordic Optical Telescope), INT (Isaac Newton 
Telescope) and WHT (William Herschel Telescope) operated on the island of La Palma jointly by Denmark, Finland, Iceland, Norway and Sweden (NOT) or 
the Isaac Newton Group (INT, WHT) in the Spanish Observatorio del Roque de Los Muchachos of the Instituto de Astrof\'\i sica de Canarias. 
Based on observations made at the Centro Astron\'omico Hispano Alem\'an (CAHA) at Calar Alto, 
operated by the Max-Planck Institut f\"ur Astronomie and the Instituto de Astrof\'{\i}sica de Andaluc\'{\i}a (CSIC).
}}

  \subtitle{IV. Colours, chemical-composition analysis and metallicity-luminosity relations}

   \author{\'Angel R. L\'opez-S\'anchez
          \inst{1,2}
		  \and
		  C\'esar Esteban\inst{2,3}
          }

   \offprints{\'Angel R. L\'opez-S\'anchez, \email{Angel.Lopez-Sanchez@csiro.au}}

\institute{CSIRO Astronomy \& Space Science / Australia Telescope National Facility, PO\,BOX\,76, Epping, NSW\,1710, Australia \and Instituto de 
Astrof{\'\i}sica de Canarias, C/ V\'{\i}a L\'actea S/N, E-38200, La Laguna, Tenerife, Spain \and Departamento de Astrof\'{\i}sica de la Universidad 
de La Laguna, E-38071, La Laguna, Tenerife, Spain}


   \date{Received Jan 29, 2010; Accepted Mar 18, 2010}

 
  \abstract
   {}
{We have performed a comprehensive multiwavelength analysis of a sample of 20 starburst galaxies that show  a substantial
population of very young massive stars, most of them classified as Wolf-Rayet (WR) galaxies. In this paper, the forth of the series, we present the 
global analysis of the derived photometric and chemical properties.} 
  {We compare optical/\NIR\ colours and the physical properties (reddening coefficient, equivalent widths of the emission and underlying absorption 
lines, ionization degree, electron density, and electron temperature) and chemical properties (oxygen abundances and N/O, S/O, Ne/O, Ar/O, and Fe/O 
ratios) with previous observations and galaxy evolution models. 
We compile 41 independent star-forming regions --with oxygen abundances between \mbox{\abox= 7.58} and 8.75--, of which 31 have a direct estimate 
of the electron temperature of the ionized gas.}
{According to their absolute $B$-magnitude, many of them are not dwarf galaxies, but they should be during their quiescent phase. 
We found that both \CHb\ and \Wabs\ increase with increasing metallicity. 
The differences in the N/O ratio is explained assuming differences in the star formation histories. We detected a high N/O ratio in objects showing 
strong WR features (HCG~31~AC, UM~420, IRAS~0828+2816, III~Zw~107, ESO~566-8 and NGC~5253). The ejecta of the WR stars may be the origin of the N 
enrichment in these galaxies. We compared the abundances provided by the direct method with those obtained through empirical calibrations, finding that 
(i) the Pilyugin method 
is the best suited empirical calibration for these star-forming galaxies, (ii) the relations
provided by Pettini \& Pagel (2004) give acceptable results for objects with \abox$>$8.0, and (iii) the results provided by empirical calibrations 
based on photoionization models 
are systematically 0.2 -- 0.3 dex higher than the values derived from the direct method. 
The O and N abundances and the N/O ratios are clearly related to the optical/\NIR\ luminosity; the dispersion of the data is a consequence of the 
differences in the star-formation histories. The $L$--$Z$ relations tend to be tighter when using \NIR\ luminosities, which facilitates  
distinguishing tidal dwarf galaxies candidates and pre-existing dwarf objects. Galaxies with redder colours tend to have higher oxygen and nitrogen 
abundances.}
 {Our detailed analysis is fundamental to understand the nature of galaxies that show strong starbursts, as well as to know their star formation 
history and the relationships with the environment. This study is complementary --but usually more powerful-- to the less detailed analysis of large 
galaxy samples that are very common nowadays.}
 

\titlerunning{Massive star formation in Wolf-Rayet galaxies IV: Colours and metallicities}

\authorrunning{L\'opez-S\'anchez \& Esteban}

   \keywords{galaxies: starburst --- galaxies: interactions --- galaxies: dwarf --- galaxies: abundances --- galaxies: kinematics and dynamics--- 
stars: Wolf-Rayet}
   \maketitle
%

\section{Introduction}


The knowledge of the chemical composition of galaxies, in particular of dwarf galaxies, is vital for understanding their evolution, star formation 
history, stellar nucleosynthesis, the importance of gas inflow and outflow, and the enrichment of the intergalactic medium. Indeed, metallicity is a 
key ingredient for modelling galaxy properties, because it determines \UV, optical and \NIR\ \mbox{colours} at a given age (i.e., Leitherer et al. 
1999), nucleosynthetic yields (e.g., Woosley \& Weaver 1995), the dust-to-gas ratio (e.g., Hirashita et al 2001), the shape of the interstellar 
extinction curve (e.g., Piovan et al. 2006), or even the properties of the Wolf-Rayet stars \citep{Crowther07}.

The most robust method to derive the metallicity in star-forming and starburst galaxies is via the estimate of metal abundances and abundance 
ratios, in particular through the determination of the gas-phase oxygen abundance and the nitrogen-to-oxygen ratio. The relationships between current 
metallicity and other galaxy parameters, such as colours, luminosity, neutral gas content, star-formation rate, extinction or total mass, constrain
galaxy-evolution models and give clues about the current stage of a galaxy. 
For example, is still debated whether massive star formation results in the instantaneous enrichment of the interstellar medium of a dwarf galaxy, or if 
the bulk of the newly synthesized heavy elements must cool before becoming part of the interstellar medium (ISM) that eventually will form the next 
generation of stars. Accurate oxygen abundance measurements of several \HII regions within a dwarf galaxy will increase the understanding of its 
chemical enrichment and mixing of enriched material. The analysis of the kinematics of the ionized gas will also help to understand the dynamic stage 
of galaxies and reveal recent interaction features. Furthermore, detailed analyses of starburst galaxies in the nearby Universe are fundamental to 
interpret the observations of high-z star forming galaxies, such as Lyman Break Galaxies \citep{EP03}, as well as quantify the importance of 
interactions in the triggering of the star-formation bursts, which seem to be very common at higher redshifts (i.e., Kauffmann \& White 1993; Springer 
et al. 2005). 


The comparison of the metallicity (which reflects the gas reprocessed by stars and any exchange of gas between the galaxy and its environment) with 
the stellar mass (which reflects the amount of gas locked up into stars) provides key clues about galaxy formation and evolution.  These analyses have 
shown a clear correlation between mass and metallicity. In practice, luminosity has been used as substitute of mass because of the 
difficulty of deriving reliable galaxy masses, yielding to the so-called metallicity-luminosity relation (i.e., Rubin et al. 1984; Richer \& McCall 
1995; Salzer et al. 2005), although in recent years mass-metallicity relations are also explored (i.e., Tremonti et al. 2004; Kewley \& Elisson, 2008), 
and are studied even at high redshifts (i.e., Kobulnicky et al. 1999; Pettini et al. 2001; Kobulnicky \& Kewley 2004; Erb et al. 2006; Liang et al. 
2006). The evolution of such relationships are now predicted by semi-analytic models of galaxy formation within the $\Lambda$-cold dark matter 
framework that include chemical hydrodynamic simulations (De Lucia et al. 2004; Tissera et al. 2005; De Rossi et al. 2006; Dav\'e \& Oppenheimer 
2007). Ironically, today the main problem is not to estimate the mass of a galaxy but its real metallicity, so that different methods involving 
direct estimates of the oxygen abundance, empirical calibrations using bright emission-line ratios or theoretical methods based on photoionization 
models yield very different values (i.e., Yin et al. 2007; Kewley \& Elisson, 2008).

Hence precise photometric and spectroscopic data, including a detailed analysis of each particular galaxy that allows conclusions about its nature, are 
crucial to address these issues. We performed such a detailed photometric and spectroscopic study in a sample of strong star-forming galaxies, 
many of them previously classified as dwarf galaxies. The majority of these objects are  
Wolf-Rayet (WR) galaxies,
a very inhomogeneous class of star-forming objects which share at least
an ongoing or recent star formation event that has produced stars sufficiently massive
to evolve into the WR stage \citep{SCP99}. 
However, WR features in the spectra
of a galaxy provides useful information about the star-formation processes in the system. 
As the first WR stars typically appear around 2 -- 3 Myr after the starburst is initiated and
disappear within some 5 Myr \citep{MeynetMaeder05}, 
their detection gives indications about both the youth and strength of
the burst, offering the opportunity to study an approximately coeval
sample of very young starbursts \citep{SV98}. 



The main aim of our study of the formation of 
massive stars in starburst galaxies and the role that the interactions with or between dwarf galaxies and/or low surface brightness objects have in 
its triggering mechanism. In Paper~I \citep{LSE08} we described the motivation of this work, compiled the list of the 20 analysed 
WR galaxies (Table~1 of Paper~I), the majority of them showing several sub-regions or objects within or surrounding them, and presented the results 
of the optical/\NIR\ broad-band and \Ha\ photometry. In Paper~II \citep{LSE09} we presented the results of the analysis of the intermediate 
resolution long-slit spectroscopy of 16 WR galaxies of our sample -- the results for the other four galaxies were published separately. 
In many cases, two or more slit positions were used 
to analyse the most interesting zones, knots or morphological structures belonging to each galaxy or even surrounding objects. 
Paper~III \citep{LSE10} presented the analysis of the O and WR stellar populations within these galaxies. 
In this paper, the forth of the series, we globally compile and analyse the optical/\NIR\ photometric data (Sect.~2) and study the physical (Sect.~3) 
and chemical (Sect.~4) properties of the ionized gas within our galaxy sample. 
Thirty-one up to 41 regions have a direct estimate of the electron 
temperature of the ionized gas, and hence the element abundances were derived with the direct method.
Section~4 includes the analysis of the N/O ratio with the oxygen abundance, a discussion of the nitrogen enrichment in WR galaxies, a study of the 
$\alpha$-elements to oxygen ratio with the oxygen abundance, and the comparison of the results provided by the most common empirical calibrations 
with those derived following the direct method (the Appendix compiles all metallicity calibrations used in this work). Section~5 analyses the 
metallicity-luminosity relations obtained with our data. Section~6 discusses the relations between the metallicity and the optical/$\NIR$ colours. 
Finally, we list our main conclusions in Sect.~7.  

The final paper of the series (Paper~V) will compile the properties derived with data from other wavelengths (UV, FIR, radio, and X-ray) 
and complete a global analysis of all available multiwavelength data of our WR galaxy sample. We have produced the most comprehensive data 
set of these galaxies so far, involving multiwavelength results and analysed according to the same procedures.

\section{Global analysis of magnitudes and colours}

\begin{table*}[t!]
\centering
  \caption{\footnotesize{Compilation of the broad-band photometric data$^a$ for the individual galaxies analysed in this work.}}
  \label{colores}
  \tiny
  \begin{tabular}{l@{\hspace{0pt}} c@{\hspace{4pt}} c@{\hspace{8pt}} c@{\hspace{10pt}} 
  c@{\hspace{10pt}} c@{\hspace{10pt}}c@{\hspace{10pt}}c@{\hspace{10pt}} c@{\hspace{10pt}}c@{\hspace{8pt}} c@{\hspace{7pt}}c@{\hspace{7pt}}}
    \noalign{\smallskip}
    \tableline
		\noalign{\smallskip}
	Galaxy & $E(B-V)$ & $M_B^{NC}$ & $M_B$ & 
$U-B$& $B-V$ & $V-R$ & $V-J$ & $J-H$ & $H-K_s$ &
B. Age& UC Age \\ 
	\noalign{\smallskip}
      &   ($a$)  &  ($b$)  & ($c$)  & ($d$) & ($d$) & ($d$) & ($d$) & ($d$) & ($d$) & ($e$) & ($f$) \\
	
\noalign{\smallskip}    
  \tableline
\noalign{\smallskip}
	
HCG 31 AC	&       0.06&	-19.18&	-19.43&	-0.60$\pm$0.06&  0.03$\pm$0.08&	0.12$\pm$0.08&	0.20$\pm$0.10&	0.13$\pm$0.10&	0.15$\pm$0.12&	5.0&100\\
HCG 31 B    &	    0.18&	-17.96&	-18.71&	-0.38$\pm$0.08&  0.17$\pm$0.06&	0.06$\pm$0.06&	0.14$\pm$0.10&	0.13$\pm$0.10&	0.12$\pm$0.10&	7.0&100\\
HCG 31 E	&       0.06&	-15.51&	-15.76&	-0.65$\pm$0.10& -0.03$\pm$0.10&	0.20$\pm$0.09&	0.29$\pm$0.12&	0.05$\pm$0.10&	0.18$\pm$0.12&	6.0&--\\
HCG 31 F1	&       0.20&	-14.93&	-15.76&	-0.99$\pm$0.12& -0.07$\pm$0.12&-0.04$\pm$0.10& -0.17$\pm$0.14&	0.04$\pm$0.17&	0.29$\pm$0.30&	2.5&0\\
HCG 31 F2	&       0.09&	-13.97&	-14.34&	-1.01$\pm$0.12& -0.09$\pm$0.12&-0.02$\pm$0.10&	0.01$\pm$0.16&	0.08$\pm$0.30&	0.20$\pm$0.50&	2.5&0\\
HCG 31 G	&       0.06&	-18.63&	-18.88&	-0.43$\pm$0.09& -0.01$\pm$0.08&	0.14$\pm$0.08&	0.45$\pm$0.08&	0.12$\pm$0.10&	0.13$\pm$0.10&	6.0&100\\
Mkn 1087	&       0.17&	-21.45&	-22.15&	-0.41$\pm$0.08&  0.17$\pm$0.08&	0.20$\pm$0.08&	0.52$\pm$0.06&	0.20$\pm$0.06&	0.16$\pm$0.06&	6.0&100\\
Mkn 1087 N	&       0.10&	-17.65&	-18.06& \nodata       & -0.05$\pm$0.06&	0.14$\pm$0.10&	0.21$\pm$0.08&	0.18$\pm$0.08&	0.13$\pm$0.08&	7.0&--\\
Mkn 1087 \#1&	    0.07$^g$&-16.05&-16.34&	-0.75$\pm$0.15& -0.01$\pm$0.10&	0.10$\pm$0.08& \nodata       & \nodata       & \nodata       &  6.0&--\\
Mkn 1087 \#3&	    0.07$^g$&-16.91&-17.20&  0.08$\pm$0.30&  0.11$\pm$0.06&	0.26$\pm$0.06&	0.64$\pm$0.10&	0.50$\pm$0.20& \nodata       &   --&150\\
Haro 15	    &       0.11&	-20.41&	-20.87&	-0.52$\pm$0.08&  0.26$\pm$0.08&	0.32$\pm$0.08&	0.17$\pm$0.08&	0.58$\pm$0.08&	0.22$\pm$0.08&	5.0&500\\
Mkn 1199	&       0.15&	-20.06&	-20.68&	-0.44$\pm$0.06&  0.46$\pm$0.06&	0.29$\pm$0.06&	1.30$\pm$0.07&	0.55$\pm$0.08&	0.34$\pm$0.08&	8.0&500\\
Mkn 1199 NE	&       0.11&	-17.11&	-17.57&  0.16$\pm$0.08&  0.51$\pm$0.08&	0.34$\pm$0.08&	1.29$\pm$0.08&	0.62$\pm$0.10&	0.20$\pm$0.10& 12.0&500\\

Mkn 5	    &       0.20&	-14.74&	-15.57&	-0.41$\pm$0.06&  0.44$\pm$0.06&	0.30$\pm$0.06&	0.81:      	 &  0.52$\pm$0.03&	0.38$\pm$0.04&	5.0&500\\
IRAS 08208+2816&    0.17&	-20.59&	-21.29&	-0.49$\pm$0.06&  0.22$\pm$0.06&	0.35$\pm$0.08&	1.03$\pm$0.08&	0.54$\pm$0.08&	0.22$\pm$0.10&	5.5&500\\
IRAS 08339+6517&    0.16&	-20.91&	-21.57&	-0.51$\pm$0.08&  0.01$\pm$0.08&	0.26$\pm$0.08&	1.36$\pm$0.06&	0.64$\pm$0.05&	0.23$\pm$0.06&	4.5&150\\
IRAS 08339+6517 C&   0.13&	-17.67&	-18.21&	-0.16$\pm$0.10&  0.20$\pm$0.08&	0.26$\pm$0.08&	1.45$\pm$0.12&	0.21$\pm$0.25&	0.68$\pm$0.28&	5.5&250\\
POX~4	        &   0.06&	-18.54&	-18.79&	-0.68$\pm$0.03&  0.29$\pm$0.02&	0.32$\pm$0.04&	0.42$\pm$0.08&	0.28$\pm$0.08&	0.15$\pm$0.10&	3.5&250\\
POX~4 Comp	    &   0.12&	-14.86&	-15.36&	-0.02$\pm$0.06&  0.25$\pm$0.02&	0.30$\pm$0.04&	0.87$\pm$0.10&	0.7:	     &  0.3:       &  5.0&300\\

UM 420	        &   0.06&	-19.30&	-19.55&	-0.80$\pm$0.06&  0.31$\pm$0.06&	0.13$\pm$0.06&	0.77$\pm$0.12&	0.41$\pm$0.12&	0.12$\pm$0.16&	4.5&200\\
SBS 0926+606 A	&   0.08&	-16.96&	-17.29&	-0.75$\pm$0.06&  0.01$\pm$0.06&	0.14$\pm$0.06&	0.54$\pm$0.06&	0.21$\pm$0.06&	0.15$\pm$0.08&	4.8&200\\
SBS 0926+606 B	&   0.12&	-16.87&	-17.37&	-0.51$\pm$0.08&  0.08$\pm$0.06&	0.20$\pm$0.06&	0.83$\pm$0.06&	0.29$\pm$0.06&	0.18$\pm$0.08&	6.7&100\\
SBS 0948+532	&   0.24&	-17.44&	-18.43&	-1.20$\pm$0.06& -0.12$\pm$0.06&	0.16$\pm$0.06& \nodata       & \nodata       & \nodata       &  4.6&100\\
SBS 1054+365	&   0.02&	-13.98&	-14.06&	-0.34$\pm$0.06&  0.33$\pm$0.06& \nodata      &   0.92$\pm$0.08&	0.38$\pm$0.12&	0.16$\pm$0.15&	4.9&500\\
SBS 1211+540	&   0.08&	-12.94&	-13.27&	-0.61$\pm$0.06&  0.04$\pm$0.06&	0.21$\pm$0.06& \nodata       & \nodata       & \nodata       &  4.7&500\\

SBS 1319+579	&   0.02&	-18.45&	-18.53&	-0.39$\pm$0.06&  0.34$\pm$0.06&	0.19$\pm$0.06&	1.03$\pm$0.08&	0.39$\pm$0.12&	0.16$\pm$0.20&	3.7&300\\
SBS 1415+437	&   0.13&	-14.09&	-14.52&	-0.47$\pm$0.06&  0.21$\pm$0.06&	0.27$\pm$0.06&	0.98$\pm$0.08&	0.35$\pm$0.10&	0.15:	    &   3.6&250\\
III Zw 107	    &   0.21&	-19.27&	-20.14&	-0.42$\pm$0.06&  0.14$\pm$0.06&	0.22$\pm$0.06&  0.62$\pm$0.12&  0.47$\pm$0.20&  0.35$\pm$0.20 & 5.6&500\\

Tol 9	        &   0.31&	-17.98&	-19.26&	-0.34$\pm$0.06&  0.24$\pm$0.06&	0.22$\pm$0.06&	0.83$\pm$0.08&	0.68$\pm$0.10&	0.27$\pm$0.12&	5.8&500\\
Tol 1457-262 Obj1&   0.16&	-19.07&	-19.73&	-0.56$\pm$0.06&  0.23$\pm$0.06&	0.26$\pm$0.06&	0.60$\pm$0.10&	0.51$\pm$0.12&	0.22$\pm$0.12&	4.6&500\\
Tol 1457-262 Obj2&	0.16$^g$&-18.31&-18.97&	-0.42$\pm$0.06&  0.34$\pm$0.06&	0.36$\pm$0.06&	0.90$\pm$0.10&	0.58$\pm$0.12&	0.27$\pm$0.14&	5.2&500\\
Tol 1457-262 \#15&0.16$^g$&	-15.82&	-16.48&	-0.25$\pm$0.10&  0.39$\pm$0.08&	0.39$\pm$0.06&	1.10$\pm$0.20& \nodata       & \nodata       &  6.4&400\\
Tol 1457-262 \#16&0.16$^g$&	-14.03&	-14.69&	-0.10$\pm$0.15&  0.45$\pm$0.10&	0.40$\pm$0.06&	 \nodata     & \nodata       & \nodata       &  7.0&500\\
ESO 566-8	     &  0.34&	-19.47&	-20.88&	-0.48$\pm$0.06&  0.31$\pm$0.06&	0.19$\pm$0.06&	1.10$\pm$0.10&	0.60$\pm$0.12&	0.38$\pm$0.14&	4.2&500\\
ESO 566-7 	     &  0.16&	-18.69&	-19.35&	-0.21$\pm$0.08&  0.49$\pm$0.06&	0.31$\pm$0.06&	1.20$\pm$0.10&	0.71$\pm$0.16&	0.34$\pm$0.16&	4.2&500\\
NGC 5253$^h$     &  0.17&   -17.23& -17.92& -0.41$\pm$0.02&  0.27$\pm$0.02& 0.21$\pm$0.02&  0.81$\pm$0.03&  0.53$\pm$0.04&  0.19$\pm$0.05&  3.5&300\\
	\noalign{\smallskip}    
  \tableline
  \end{tabular}
  \begin{flushleft}
  
   $^a$ Colour excess, $E(B-V)$, derived from our estimates of the reddening coefficient and assuming $R_V$=3.1, $E(B-V)$=0.692\CHb.\\
  $^b$ Absolute $B$-magnitude, not corrected for extinction.\\
  $^c$ Extinction-corrected absolute $B$-magnitude, assuming $M_B$ = $M_B^{NC} - A_B$ = 
  $M_B^{NC} - 3.1\times 1.337 E(B-V)$ = 
  $M_B^{NC} - 2.868$ \CHb.\\
  $^d$ All colours have been corrected for both extinction and emission of the gas, see \citet{LSE08}.\\
  $^e$ Age of the most recent star-forming burst (derived using the \Ha\ equivalent width), in Myr.\\
  $^f$ Minimum age of the underlying stellar population (derived via the analysis of the low-luminosity component of the galaxy), in Myr.\\
  $^g$  $E(B-V)$ was estimated only considering the extinction of the Milky Way.\\
  $^h$  Optical and \NIR\ magnitudes extracted from the NED and corrected for extinction using an average value of \CHb=0.24, see \citet{LSEGRPR07}.
  \end{flushleft}
\end{table*}

Our optical and \NIR\ broad-band photometric results for the galaxy sample were presented in Paper~I. These data allowed us the analysis of the 
optical and \NIR\ magnitudes and the colours of the galaxies and surrounding dwarf objects. Table~\ref{colores} compiles the optical/\NIR\ results 
for the individual galaxies, not considering regions within them or nearby diffuse objects. This table shows the colour excess, $E(B-V)$ (derived with 
the Balmer decrement in our optical spectra, see Paper~II), the absolute $B$-magnitude (both corrected, $M_B$, and uncorrected, $M_B^{NC}$, for 
extinction), all the optical/\NIR\ colours, and the age of the most recent star-formation burst (the young population, derived from our \Ha\ images) 
and the minimum age of the old stellar population (usually estimated from the low luminosity component or regions without nebular emission using our 
optical/\NIR\ broad-band images).

Our first result from Table~\ref{colores} is that the actual number of dwarf galaxies, defined as $M_B\geq-18$, is not as high as we had 
expected considering the selection criteria of our WR galaxy sample. There are two reasons for this: (i) on the one hand, the determination of the 
magnitudes was performed in a more accurate way. As our images are deeper than those previously obtained, the integrated magnitude of a diffuse 
object is \emph{lower} than that estimated before. (ii) On the other hand, we corrected all our data for extinction, but not only considering 
the effect of the dust in the Milky Way as it is usually done, but taking into account the internal extinction derived from our spectroscopic data. 
That is why only six galaxies (Mkn~5, SBS~0926+606, SBS~1054+365, SBS~1211+540, SBS~1415+437 and NGC~5253) are strictly classified as dwarf galaxies 
following the above definition. POX~4, SBS~0948+532 and SBS~1319+579 could be also considered dwarf galaxies because $M_B\ge-19$. Table~\ref{colores} 
also lists some tidal dwarf galaxy (TDG) candidates (HCG~31~E, F1 and F2; Mkn~1087 \#1 and \#3, POX~4~Comp) and nearby external objects (Mkn~1087~N, 
Mkn~1199~NE, 
IRAS~08309+6517~C, Tol~1457-262 \#15 and \#16) surrounding a main galaxy.
However, as we remarked in Paper~I and in the analysis of the HCG~31 members \citep{LSER04a}, we must
keep in mind that the $B$-magnitude of a starburst is increased
by several magnitudes during the first 10 Myr with respect to
its brightness in the quiescent phase, so we should expect that some of the objects with $-18\geq M_B\geq-19.5$ are indeed defined as dwarf objects 
during their quiescent phase.

\begin{figure}[t!]
\includegraphics[angle=270,width=\linewidth]{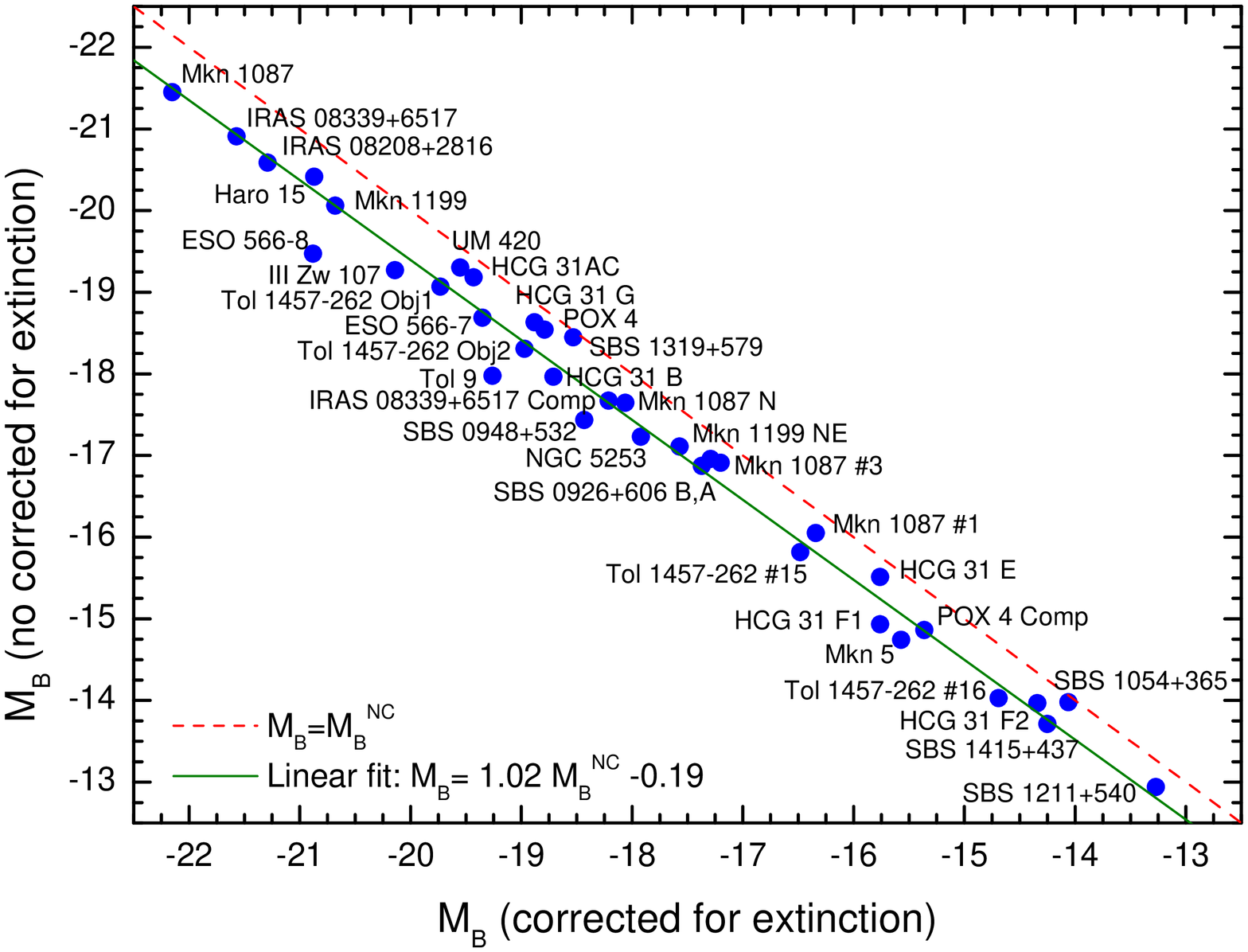} 
\protect\caption[ ]{\footnotesize{Comparison between the reddening-corrected absolute $B$-magnitudes ($M_B$) and the uncorrected ones ($M_B^{NC}$) 
for all individual galaxies analysed in this work. A linear fit to the data is shown.}}
\label{colourglobal}
\end{figure}


In order to quantify the effect of the correction for extinction, we plot the uncorrected absolute $B$-magnitude versus
the absolute $B$-magnitude corrected for extinction in Fig.~\ref{colourglobal}. We did not consider the correction for the emission of the gas in the absolute magnitude because 
(i) it is small in the $B$-filter, less than 0.10 magnitudes and, more important, (ii) we are considering the magnitude of the galaxy as a whole, 
taking into account both the star-forming bursts and regions dominated by older stellar populations that \emph{do not} possess any nebular emission.  
From Fig.~\ref{colourglobal}, we see that the magnitudes corrected for extinction are on average around 0.60 magnitudes lower than when this 
effect is not considered. As all data lie in a narrow band, we performed a linear fit, finding the following relation between both magnitudes:
\begin{eqnarray}
M_B=1.02M_B^{nc}-0.19.
\end{eqnarray}
For $M_B=-$18, the magnitude difference is $\Delta M_B\sim$0.56. Transforming this value to luminosity, the consideration of the correction for 
extinction means that one has to multiply the observed $B$-luminosity of a galaxy by a factor between 1.6 (for $M_B\sim-$16) and 1.8 (for $M_B\sim-$22). We 
note that there is a slight dependence on the extinction with the absolute magnitude of the galaxy, that is, the correction for extinction is higher at 
lower absolute magnitudes. This suggests a higher absorption of the light in brighter systems (more amount of dust). We will get the same result 
when we analyse the relation between the reddening coefficient and the warm dust mass (Paper~V). 

\begin{figure*}[t!]
\centering
\begin{tabular}{cc}
\includegraphics[angle=270,width=0.45\linewidth]{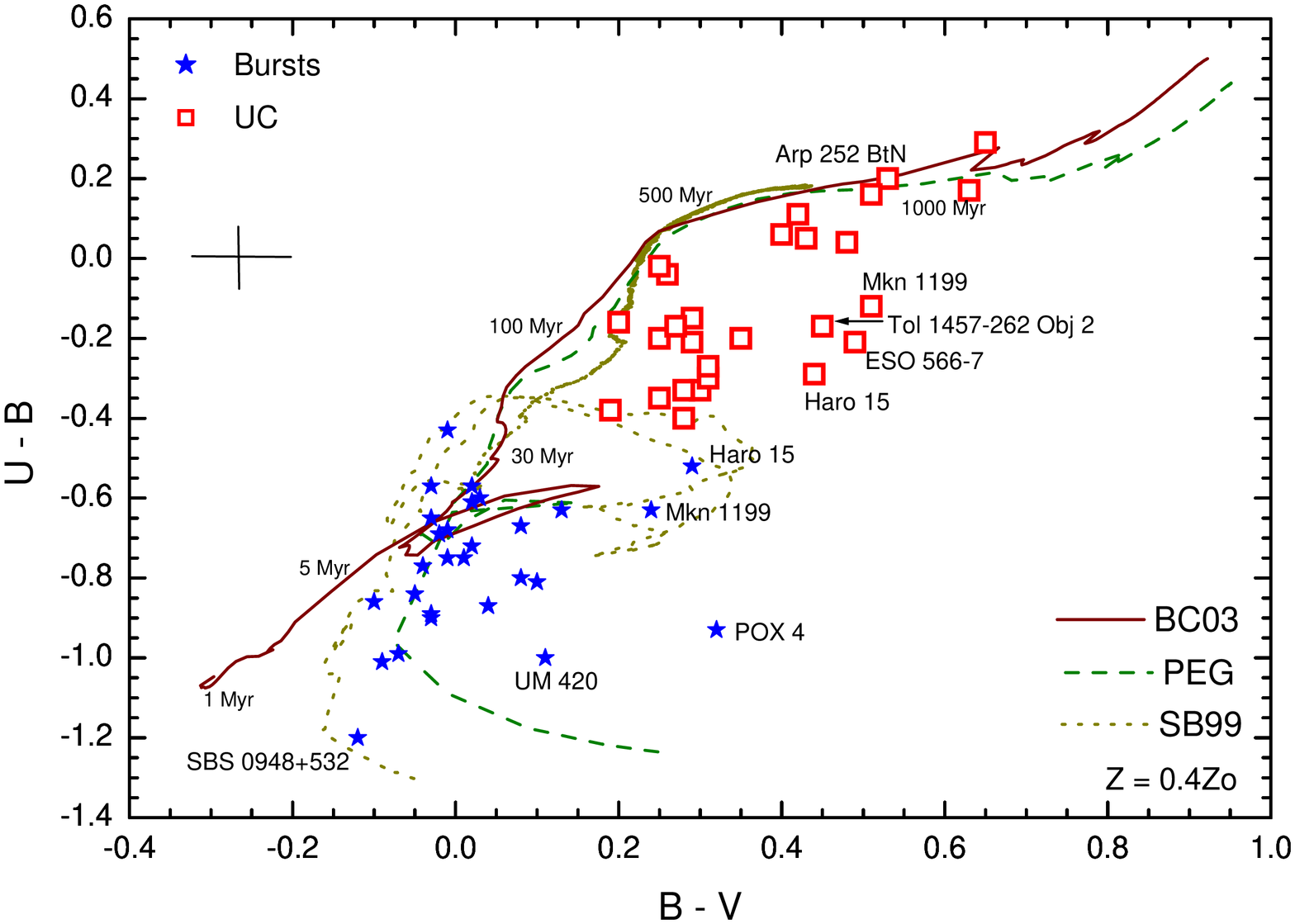}  &  
\includegraphics[angle=270,width=0.45\linewidth]{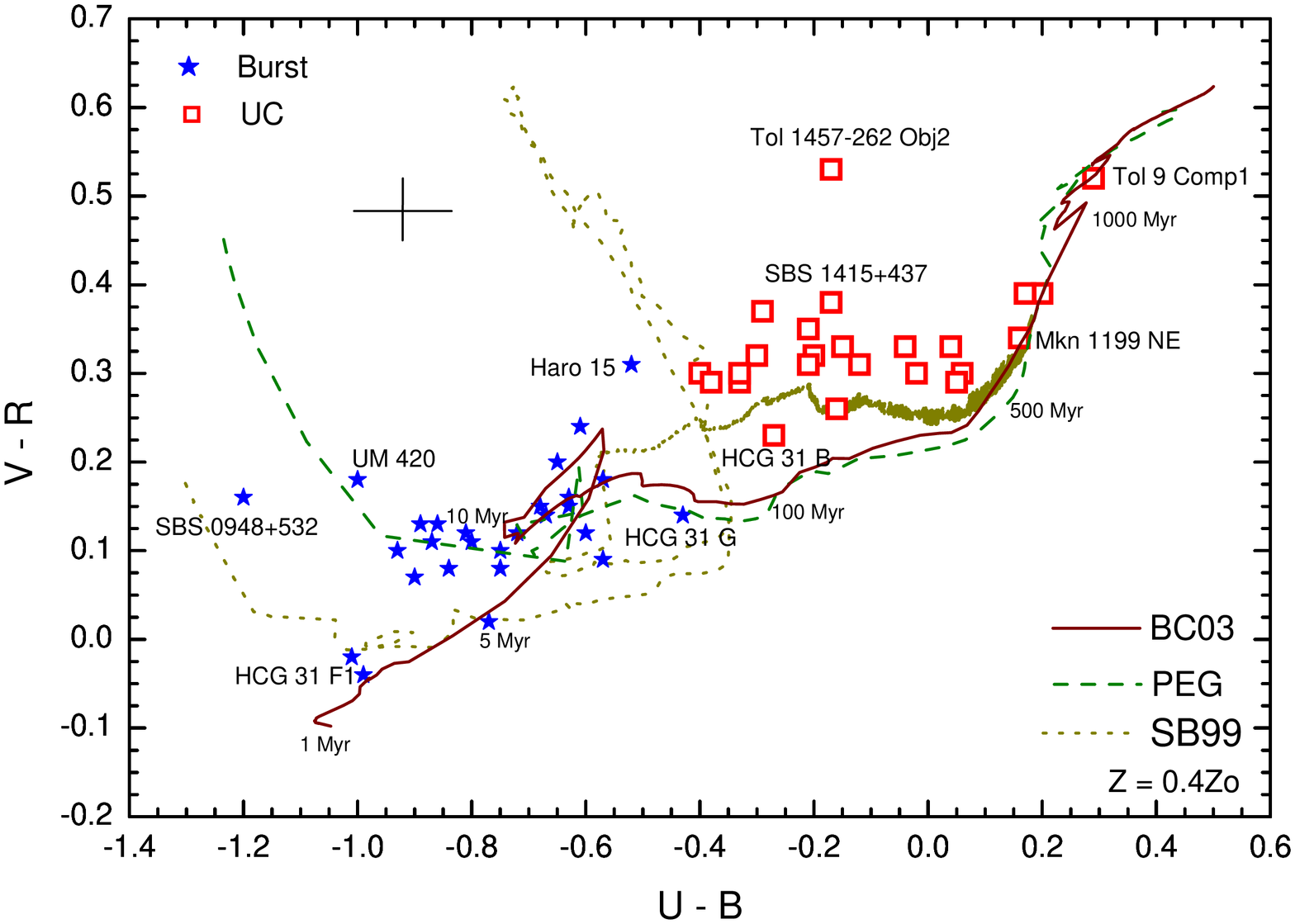}  \\  
\includegraphics[angle=270,width=0.45\linewidth]{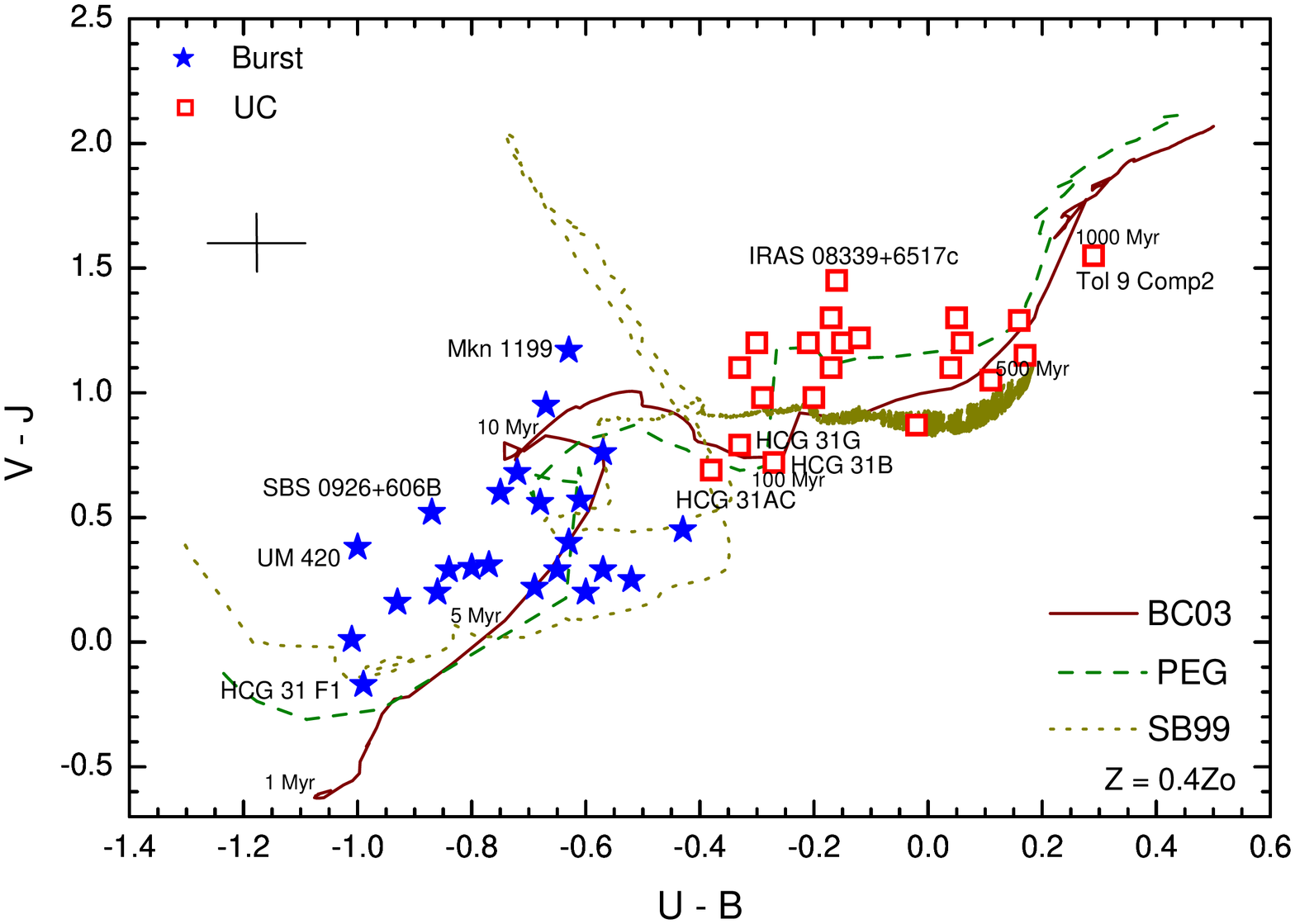}  &  
\includegraphics[angle=270,width=0.45\linewidth]{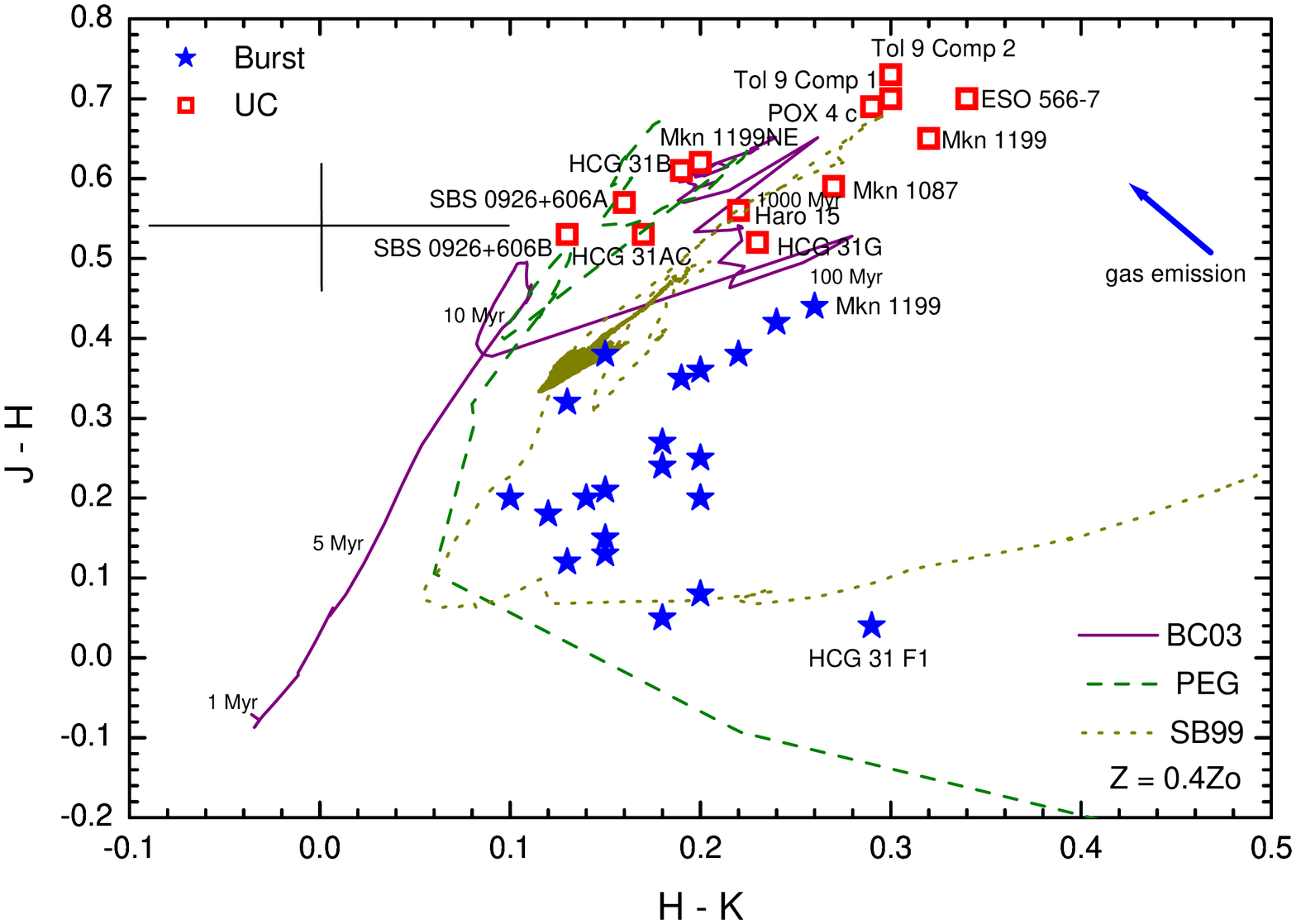}  \\  
\end{tabular}
\caption{\footnotesize{Colour-colour diagrams comparing the predictions given
by evolutionary synthesis models [continous line: BC03, \citet{BC03} models; 
discontinuous line: PEGASE.2, \citet{PEGASE97} models; 
dotted line: STARBURST~99, \citet{L99}
assuming $Z=0.4$\Zo] with the colours (corrected
for both reddening and contribution of the emission lines) of our galaxy sample 
when the burst (blue stars) and
underlying component (UC, red squares) of each system are considered
independently. The cross indicates the typical errors in our data.
Some age labels were included for the BC03 models.}}
\label{colorestodos}
\end{figure*}

As we explained in Paper~I, we compared our optical/NIR colours
(corrected for extinction and emission of the ionized gas)
with the predictions given by three different population synthesis
models, STARBURST99 \citep{L99}, PEGASE.2 \citep{PEGASE97}
and \citet{BC03}, to estimate the age of the dominant stellar population of
the galaxies, the star-forming regions, and the underlying stellar
component. We assumed an instantaneous
burst with a Salpeter \IMF\, a total mass of $10^6$ \Mo, and a
metallicity of $Z/$\Zo = 0.2, 0.4 and 1 (chosen as a function of the
oxygen abundance of the galaxy derived from our spectroscopic
data, see Paper~II) for all models. 

We found a relatively
good correspondence (see Figs.~37, 38 and 39, top, in Paper I) 
between the optical/\NIR\
data and the models, especially for compact and dwarf objects
such as HCG 31 F1 or SBS 0948+532, the ages being consistent
with a recent star-formation event ($\leq$100 Myr). We remark here
\begin{enumerate}
\item the quality of the observational data and the data reduction process, which was performed in detail and in an homogeneous way for all 
galaxies,
\item the we corrected the data for extinction and reddening, considering the \CHb\ value derived from the spectroscopic data obtained 
for each region (Paper~II). As we have seen, this correction is important and very often it is not performed in the analysis of the colours of 
extragalactic objects, which only consider the extinction of the Milky Way in the direction to the analysed galaxy,
\item and the correction of the colours for the gas emission using our spectroscopic data. This effect is not important in some galaxies, but it seems 
fundamental when analysing compact objects with strong nebular emission, such as \BCDG s or regions within a galaxy possessing an strong starburst.  
\end{enumerate}
Some discordances between the colours and the predictions of the theoretical models ($\sim$0.2 mag or even higher)
are always found in galaxies hosting a considerable population
of old stars (Mkn~1199, Mkn~5, Tol~1457-262 \#15 and \#16, ESO~566-7), because their luminosities 
barely contribute to the $U$ magnitude. Hence, the young stellar population usually dominates the $U-B$ colour, but the rest of the colours ($B-V$, 
$V-R$, $V-J$, $H-K_s$) posses an important contribution of the old stellar population.
If the bursts and the 
underlying stellar population are analysed independently, the agreement
between colours and the predictions given by the models is closer that
when considering the galaxies as a whole. The last columns in Table~\ref{colores} compile the ages estimated for the most recent star-forming event and 
the underlying population component (if possible) for all individual galaxies derived from optical/\NIR\ colours. 
Because the theoretical models are optimized to study the youngest stellar populations within the galaxies, 
in some cases we considered a lower limit of 500 Myr for the age of the underlying component (UC).

Figure~\ref{colorestodos}
shows several colour-colour diagrams comparing the predictions given
by evolutionary synthesis models with the colours (corrected
for both reddening and contribution of the emission lines) of our galaxy sample 
when the burst (blue stars) and
underlying component (UC, red squares) of each system are considered
independently. The correspondence has improved now. Indeed, all inferred 
ages of the most recent star-formation
burst are lower than 25~Myr, while the data corresponding to
the underlying component suggest ages higher than~100 Myr. Therefore, 
a proper estimate of the stellar population age
for this type of galaxy using broad-band filters is only obtained
when bursts and underlying components are independently considered.

There are still some 
discrepancies that can be explained by a lack of good separation between regions
with and without star-formation activity. The comparison of the $V-R$ colour vs. the
$U-B$ colour in data of the UC also suggest that in some galaxies the old stellar population
colours are not explained by just one single-age population, but at least two of them are needed 
(i.e., for SBS~1415+437, the UC colour may be explained by a mix of two stellar populations with 
ages of $\sim$150 Myr and $\geq$500 Myr, see Sect.~3.15.1 in Paper~I). 
However, the best method to analyse the colours and luminosities
of the host component in starburst systems (specially, in BCGs) is performing a careful 2D analysis 
of their structural parameters (i.e. Amor\'{\i}n et al. 2007; 2009). Some of the galaxies analysed by these 
authors were also studied here. Their results of the colours of the UC (the host) agree well within
the errors with those estimated here, for example, for Mkn~5 they compute $(B-V)_{UC}$=0.40$\pm$0.28 and
$(V-R)_{UC}$=0.28$\pm$0.16, while in this work we derived $(B-V)_{UC}$=0.45$\pm$0.08 and
$(V-R)_{UC}$=0.30$\pm$0.08 for the same object.

\begin{figure}[t!]
\includegraphics[angle=270,width=\linewidth]{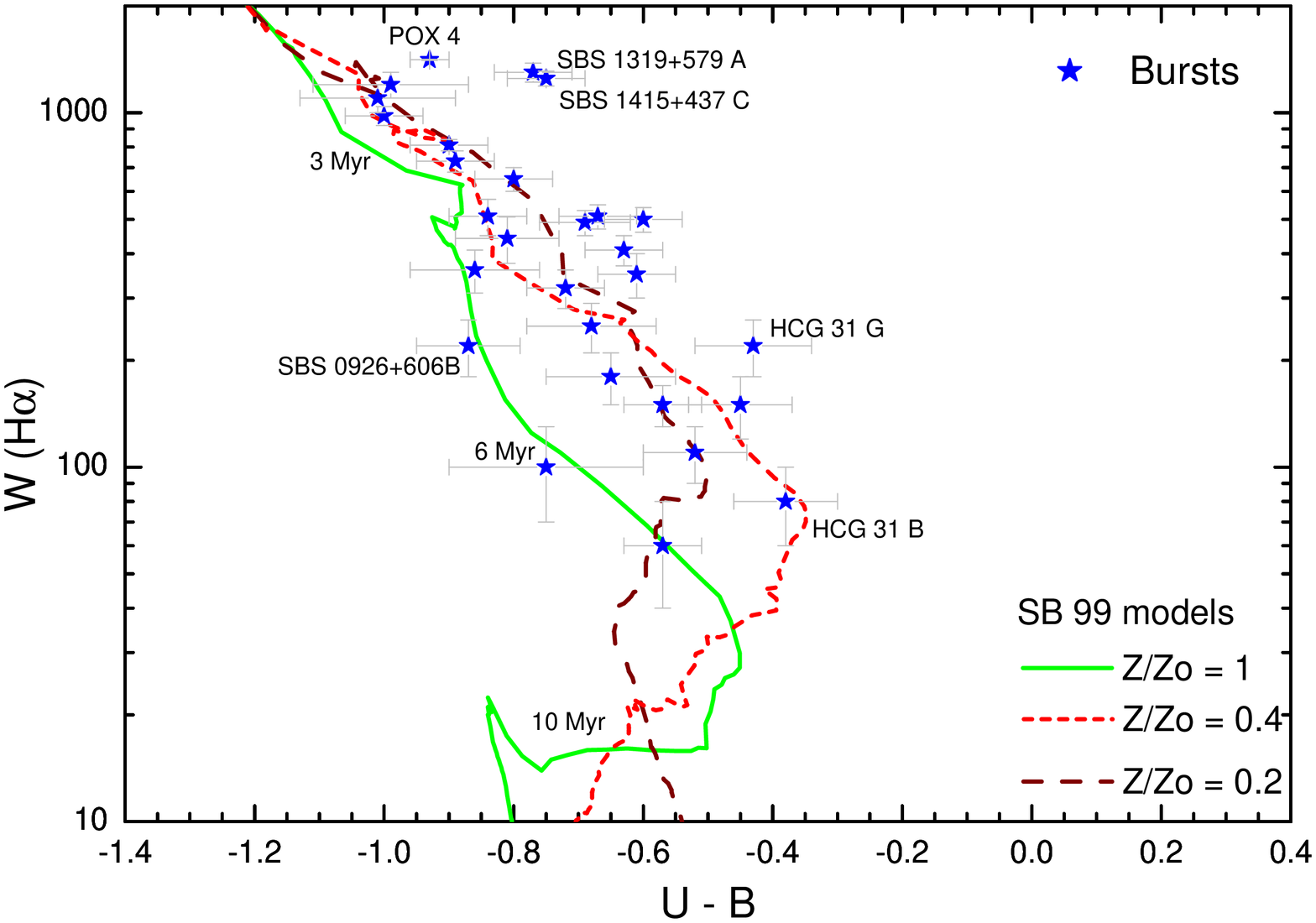} 
\protect\caption[ ]{\footnotesize{$W$(\Ha) vs. $U-B$ diagram comparing the predictions of the theoretical models of \STARBURST\ \citep{L99} for three 
different metallicities with the data corresponding to the star-forming bursts analysed in our sample galaxy.}}
\label{whaub}
\end{figure}

Figure~\ref{whaub} plots the \Ha\ equivalent width (obtained from our narrow-band images) as a function of the $U-B$ colour (obtained from our 
broad-band images) for the bursts within the analysed galaxies. We remark that the \WHa\ derived from the \Ha\ images agree quite well with those 
obtained from the optical spectroscopy (see Paper~II). This figure compares the observational data with some \STARBURST\ models \citep{L99} at 
different metallicities. As we see, the agreement is quite good for almost all objects. This also indicates both the quality of our data and the 
success of the theoretical models to reproduce the young star-forming populations. 

\begin{table*}[t!]
\centering
  \caption{\footnotesize{Physical properties of the ionized gas for the galaxies analysed in this work.}}
  \label{cfisicas}
  \tiny
  \begin{tabular}{l@{\hspace{8pt}}  c@{\hspace{8pt}}c@{\hspace{8pt}}c@{\hspace{8pt}} c@{\hspace{8pt}}c@{\hspace{8pt}}c@{\hspace{8pt}} 
c@{\hspace{8pt}} c@{\hspace{8pt}} }
  \tableline
   \noalign{\smallskip}
 Galaxy        &   \Te     & \Te\ High  & \Te\ Low  &     $n_e$  & \CHb & $W_{abs}$   &  $-W$(\Hb) \\ 
 			   & $(a)$    &  [K] & [K]  & [cm$^{-3}$] & & [\AA]  & [\AA] \\ 
\tableline
\noalign{\smallskip}
HCG 31 AC      & D  &	 9400$\pm$600	 &10800$\pm$300  &	210$\pm$70	 & 0.09$\pm$0.03 & 	2.0         & 	91.1$\pm$2.1 \\
HCG 31 B       & D  &	11500$\pm$700	 &12000$\pm$400  &	$<$100	 	 & 0.28$\pm$0.08 & 	2.0	        &   12.9$\pm$0.5 \\
HCG 31 E       & D  &	11100$\pm$1000   &11800$\pm$600  &	$<$100	 	 & 0.11$\pm$0.05 & 	2.0         & 	21.1$\pm$1.1 \\
HCG 31 F1      & D  &	12600$\pm$1100   &12600$\pm$700  &	$<$100	 	 & 0.32$\pm$0.06 & 	2.0         & 	218$\pm$13   \\
HCG 31 F2      & D  &	12300$\pm$1300   &12400$\pm$800  &	$<$100	 	 & 0.14$\pm$0.05 & 	2.0         & 	256$\pm$30 \\
HCG 31 G       & D  &	11600$\pm$700	 &12000$\pm$400  &	$<$100	 	 & 0.09$\pm$0.05 & 	2.0         & 	37.0$\pm$1.6 \\
Mkn 1087       & EC &	 7100$\pm$1000   & 8000$\pm$1000 &	220$\pm$50	 & 0.17$\pm$0.02 & 	1.7$\pm$0.2 & 	22.3$\pm$0.9 \\
Mkn 1087 N     & EC &	10900$\pm$1000   &10600$\pm$1000 &	115$\pm$50	 & 0.17$\pm$0.02 & 	0.2$\pm$0.1 & 	25.0$\pm$1.7 \\
Haro 15 C      & EC &	 9500$\pm$800	 & 9600$\pm$600  &	$<$100	 	 & 0.11$\pm$0.03 & 	2.4$\pm$0.4 & 	16.4$\pm$1.1 \\
Haro 15 A      & D  &   12850$\pm$700	 &12000$\pm$500  &	$<$100	 	 & 0.33$\pm$0.03 & 	1.3$\pm$0.3 & 	75.7$\pm$4.2 \\
Mkn 1199       & D  &	 5400$\pm$700	 & 6800$\pm$600	 &  300$\pm$100	 & 0.30$\pm$0.03 & 	1.8$\pm$0.4 & 	21.4$\pm$1.3 \\
Mkn 1199 NE    & EC &	 8450$\pm$800	 & 8900$\pm$600  &	$<$100	 	 & 0.16$\pm$0.03 & 	0.6$\pm$0.3 & 	20.2$\pm$2.3 \\
Mkn 5	       & D  &	12450$\pm$650	 &11700$\pm$450  &	$<$100	 	 & 0.17$\pm$0.02 & 	0.8$\pm$0.2 & 	75$\pm$5 \\
IRAS 08208+2816& D  &	10100$\pm$700	 &10100$\pm$500  &	$<$100	 	 & 0.11$\pm$0.02 & 	3.2$\pm$0.1 & 	80$\pm$5 \\
IRAS 08339+6517& EC &	 8700$\pm$1000   & 9100$\pm$1000 &	   100	 	 & 0.22$\pm$0.02 & 	1.8$\pm$0.2 & 	19.0$\pm$0.8 \\
IRAS 08339+6517c&EC &	 9050$\pm$1000   & 9300$\pm$1000 &	   100	 	 & 0.18$\pm$0.03 & 	1.5$\pm$0.2 & 	 7.5$\pm$0.2  \\
POX 4	       & D  &	14000$\pm$500	 &12800$\pm$400  &	250$\pm$80	 & 0.08$\pm$0.01 & 	2.0$\pm$0.1 & 	200$\pm$9 \\
POX 4 Comp     & EC &	12400$\pm$1000	 &11700$\pm$700  &	$<$100	 	 & 0.06$\pm$0.03 & 	1.4$\pm$0.3 & 	12$\pm$4 \\
UM 420	       & D  &	13200$\pm$600	 &12200$\pm$500  &	140$\pm$80	 & 0.09$\pm$0.01 & 	2.0$\pm$0.1 & 	169$\pm$10 \\
SBS 0926+606 A & D  &	13600$\pm$700	 &12500$\pm$500  &	$<$100	 	 & 0.12$\pm$0.03 & 	0.7$\pm$0.1 & 	125$\pm$6 \\
SBS 0926+606 B & EC &	11500$\pm$1000   &11000$\pm$800  &	$<$100	 	 & 0.18$\pm$0.04 & 	1.0$\pm$0.3 & 	18$\pm$3 \\
SBS 0948+532   & D  &	13100$\pm$600	 &12200$\pm$400  &	250$\pm$80	 & 0.35$\pm$0.03 & 	0.3$\pm$0.1 & 	213$\pm$11 \\
SBS 1054+365   & D  &	13700$\pm$900	 &12600$\pm$700  &	$<$100	 	 & 0.02$\pm$0.02 & 	0.8$\pm$0.1 & 	89$\pm$7 \\
SBS 1054+365 b & EC &	11800$\pm$1100   &11300$\pm$900  &	300$\pm$200	 & 0.03$\pm$0.03 & 	0.3$\pm$0.1 & 	8$\pm$3 \\
SBS 1211+540   & D  &	17100$\pm$600	 &15000$\pm$400  &	320$\pm$50	 & 0.12$\pm$0.01 & 	1.3$\pm$0.1 & 	135$\pm$10 \\
SBS 1319+579 A & D  &	13400$\pm$500	 &12400$\pm$400  &	$<$100	 	 & 0.03$\pm$0.01 & 	0.0$\pm$0.1	&   285$\pm$14 \\
SBS 1319+579 B & D  &	11900$\pm$800	 &11300$\pm$600  &	$<$100	 	 & 0.11$\pm$0.03 & 	0.4$\pm$0.1 & 	42$\pm$4 \\
SBS 1319+579 C & D  &	11500$\pm$600	 &11050$\pm$400  &	$<$100	 	 & 0.02$\pm$0.02 & 	0.2$\pm$0.1 & 	94$\pm$6 \\
SBS 1415+579 C & D  &	16400$\pm$600	 &14500$\pm$400  &	$<$100	 	 & 0.01$\pm$0.02 & 	0.8$\pm$0.1 & 	222$\pm$11 \\
SBS 1415+579 A & D  &	15500$\pm$700	 &13850$\pm$500  &	$<$100	 	 & 0.15$\pm$0.03 & 	1.0$\pm$0.2 & 	130$\pm$8 \\
III Zw 107 A   & D  &	10900$\pm$900	 &10500$\pm$800	 &  200$\pm$60	 & 0.68$\pm$0.04 & 	2.0$\pm$0.3 & 	44$\pm$2 \\
Tol 9	       & D  &	 7600$\pm$1000   & 8300$\pm$700  &	180$\pm$60	 & 0.50$\pm$0.05 & 	7.5$\pm$0.8 & 	33$\pm$2 \\
Tol 1457-262 A & D  &	14000$\pm$700	 &12500$\pm$600	 &  200$\pm$80	 & 0.57$\pm$0.03 & 	1.4$\pm$0.2 & 	101$\pm$6 \\
Tol 1457-262 B & D  &	15200$\pm$900    &14200$\pm$700	 &	$<$100	 	 & 0.00$\pm$0.05 & 	0.0$\pm$0.1 & 	82$\pm$7 \\
Tol 1457-262 C & D  &	13400$\pm$1100   &12400$\pm$1000 &	200$\pm$100	 & 0.15$\pm$0.02 & 	0.7$\pm$0.1 & 	92$\pm$9 \\
ESO 566-8	   & D  &	 8700$\pm$900	 & 9100$\pm$800  &	300$\pm$100	 & 0.49$\pm$0.03 & 	1.3$\pm$0.1 & 	95$\pm$7 \\
ESO 566-7	   & EC &	 7900$\pm$1000   & 8500$\pm$900  &	100$\pm$50	 & 0.23$\pm$0.05 & 	2.7$\pm$0.2 & 	13$\pm$2 \\
NGC 5253 A	   & D  &	12100$\pm$260	 &11170$\pm$520  &	580$\pm$110	 & 0.23$\pm$0.02 & 	1.3$\pm$0.1 & 	234$\pm$5 \\
NGC 5253 B     & D  &	12030$\pm$260	 &11250$\pm$490  &	610$\pm$100	 & 0.38$\pm$0.03 & 	1.7$\pm$0.1 & 	254$\pm$5 \\
NGC 5253 C	   & D  &	10810$\pm$230	 &10530$\pm$470  &	370$\pm$80	 & 0.25$\pm$0.03 & 	0.8$\pm$0.1 & 	94$\pm$3 \\
NGC 5253 D	   & D  &	11160$\pm$510	 &10350$\pm$650	 &  230$\pm$70	 & 0.10$\pm$0.02 & 	0.6$\pm$0.1 & 	39$\pm$2 \\	
\noalign{\smallskip}
\tableline
  \end{tabular}
  \begin{flushleft}
  $^a$  In this column we indicate if \Te\ was computed using the direct method (D) or via empirical calibrations (EC).
  \end{flushleft}
  \end{table*}

However, if we compare the \WHa\ and the $U-B$ colour considering the total extension of each galaxy (and not only the burst component), this 
agreement is less good. In this case, data with a fixed \WHa\ have a redder $U-B$ colour than that predicted by the models. This is explained 
because the $U-B$ colour is slightly contaminated with the light of older stellar populations or regions with no nebular emission. Hence, as we 
emphasized before, it is important to distinguish between the pure starburst regions and the underlying component to get a good estimate of the 
properties of these galaxies and, in particular, the strong star-forming regions.



The age of the last starburst event experienced by each galaxy and the minimum age of its old stellar populations are compiled in the last two 
columns of Table~\ref{colores}. Except for a few objects (HCG~31 members E, F1 and F2 and Mkn~1087 members N and \#1) for which it was not possible 
to estimate the colours of the UC, all analysed galaxies show an older stellar population underlying the bursts. Indeed, in many cases the colours of 
the UC suggest ages older than 500 Myr. This clearly indicates that all galaxies have experienced a previous star-formation events long time before 
those they are now hosting, as concluded in many other previous results (i.e. Cair\'os et al. 2001a,b; Bergvall \& \"Ostlin, 2002; Papaderos et 
al. 2006; Amor\'{\i}n et al. 2009). However, as we previously said \citep{LSER04a}, this seems not to be true in the particular case of members F 
of HCG~31, which clearly show no evidences of underlying old stellar populations. This was recently confirmed by deep \emph{Hubble Space 
Telescope} imaging \citep{Gallagher10}, and hence these two objects are very likely experiencing their very first star-formation event.

\section{Physical properties of the ionized gas}

Table~\ref{cfisicas} compiles all the high- and low-ionization electron temperatures of the ionized gas, \Te, electron density $n_e$, reddenning 
coefficient \CHb, equivalent width of the underlying stellar absorption in the Balmer \HI\ lines \Wabs, and the \Hb\ equivalent width, for the 
galaxies analysed in this work (see Paper~II). Thirty-one up to 41 of the objects listed in Table~\ref{cfisicas} have a direct estimate of the electron 
temperature of the ionized gas. For most, this was computed using the [\ion{O}{iii}] ratio involving the nebular [\ion{O}{iii}] 
$\lambda\lambda$4959,5007 and the auroral [\ion{O}{iii}] $\lambda$4363 emission lines. In most objects, the low-ionization electron temperature was 
not computed directly but assuming the relation between \TeOiii\ and \TeOii\ provided by \citet{G92}. Half of the objects of our galaxy sample (22) 
have electron densities lower than 100 cm$^{-3}$.

\begin{figure}[t!]
\centering
\includegraphics[angle=270,width=\linewidth]{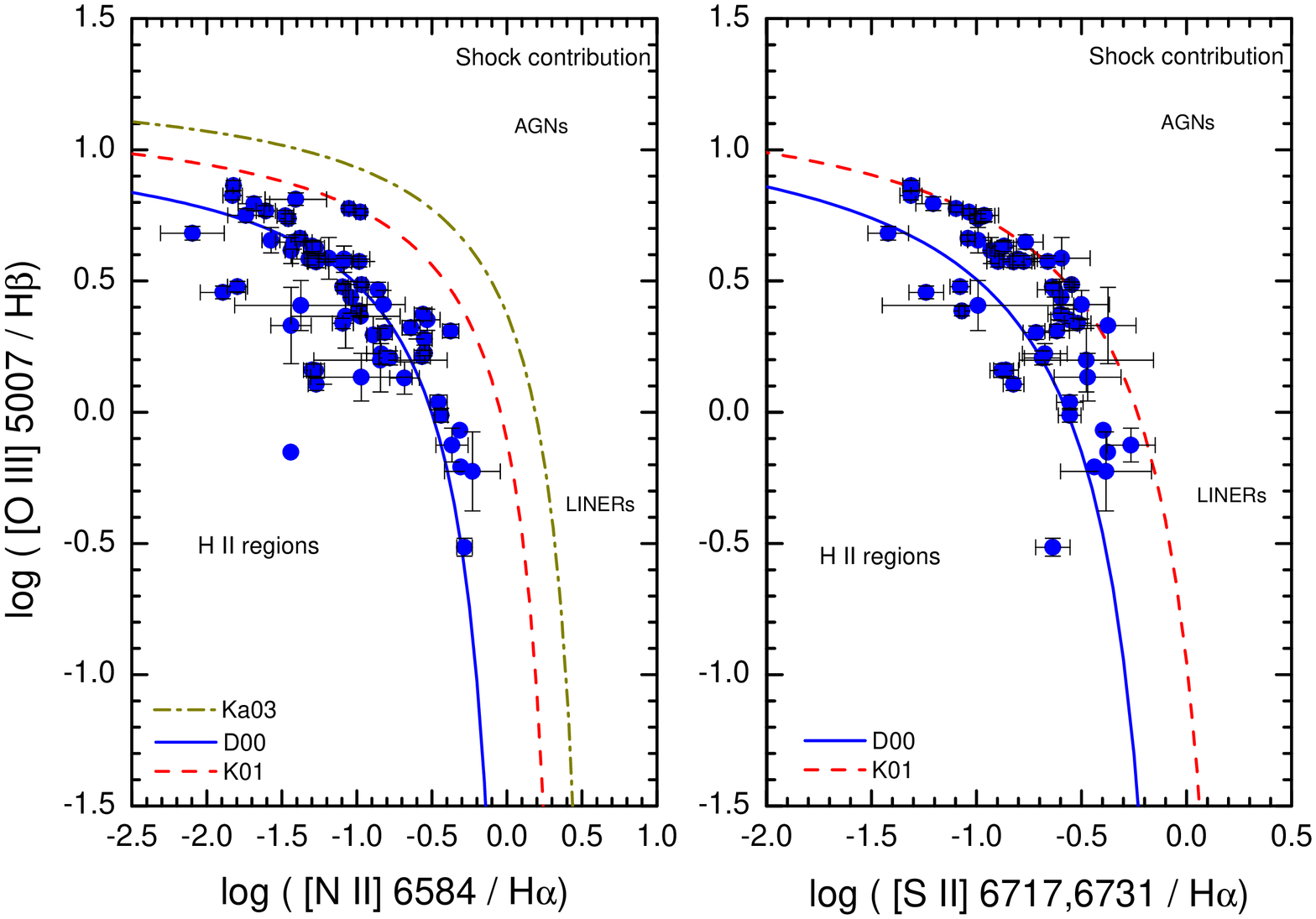} 
\caption{\footnotesize{Comparison of some observational flux ratios obtained for all available regions analysed in this work with the diagnostic 
diagrams proposed by Dopita et al. (2000), blue continuous line (D00), and Kewley et al. (2001), red discontinuous line (K01). The left panel also 
shows the empirical relation provided by Kauffmann et al. (2003) with a dotted-dashed dark yellow line (Ka03).}}
\label{diagnosticoG}
\end{figure}

We explored possible correlations among some of the different quantities compiled in Table~\ref{cfisicas} as well as the oxygen abundance (see 
Table~\ref{abtotal}) computed for the objects.
First, we checked the nature of the ionized gas of the sample galaxies. Figure~\ref{diagnosticoG} plots the typical diagnostic diagrams between bright 
emission lines and the predictions given by the photoionized models provided by \citet{Do00} for extragalactic \HII regions (that assume 
instantaneous star-formation within star-forming regions) and the \citet{KD01} models for starburst galaxies (which consider continuous star 
formation and more realistic assumptions about the physics of starburst galaxies). The dividing line given by the \citet{KD01} models represents an upper 
envelope of positions of star-forming galaxies. As we see, in all cases the data are found below the theoretical prediction given by this line. This 
indicates that photoionization is the main excitation mechanism of the gas.
We will get the same result when we compare the relation between the \FIR\ and the radio-continuum luminosities (Paper~V). 
It is interesting to notice that the observational points included in the diagnostic diagram involving the [\ion{O}{iii}]/\Hb\ and [\ion{N}{ii}]/\Ha\ ratios are located close to 
the prediction given by the \citet{Do00} models, while points included in the diagnostic diagram that considers the [\ion{O}{iii}]/\Hb\ and 
[\ion{S}{ii}]/\Ha\ ratios are found very close to the upper envelope given by \citet{KD01}.
The left panel in Figure~\ref{diagnosticoG} includes the empirical relation between the [\ion{O}{iii}]/\Hb\ and the [\ion{N}{ii}]/\Ha\ ratios provided by 
\citet{Kauffmann03} analysing a large data sample of star-forming galaxies from the Sloan Digital Sky Survey (SDSS; York et al. 2000). As can be 
seen, the comparison of the \citet{Kauffmann03} relation with our data points also indicates that our objects are experiencing a pure star-formation 
event, despite a clear offset between both datasets.

\begin{figure}[t!]
\includegraphics[angle=270,width=\linewidth]{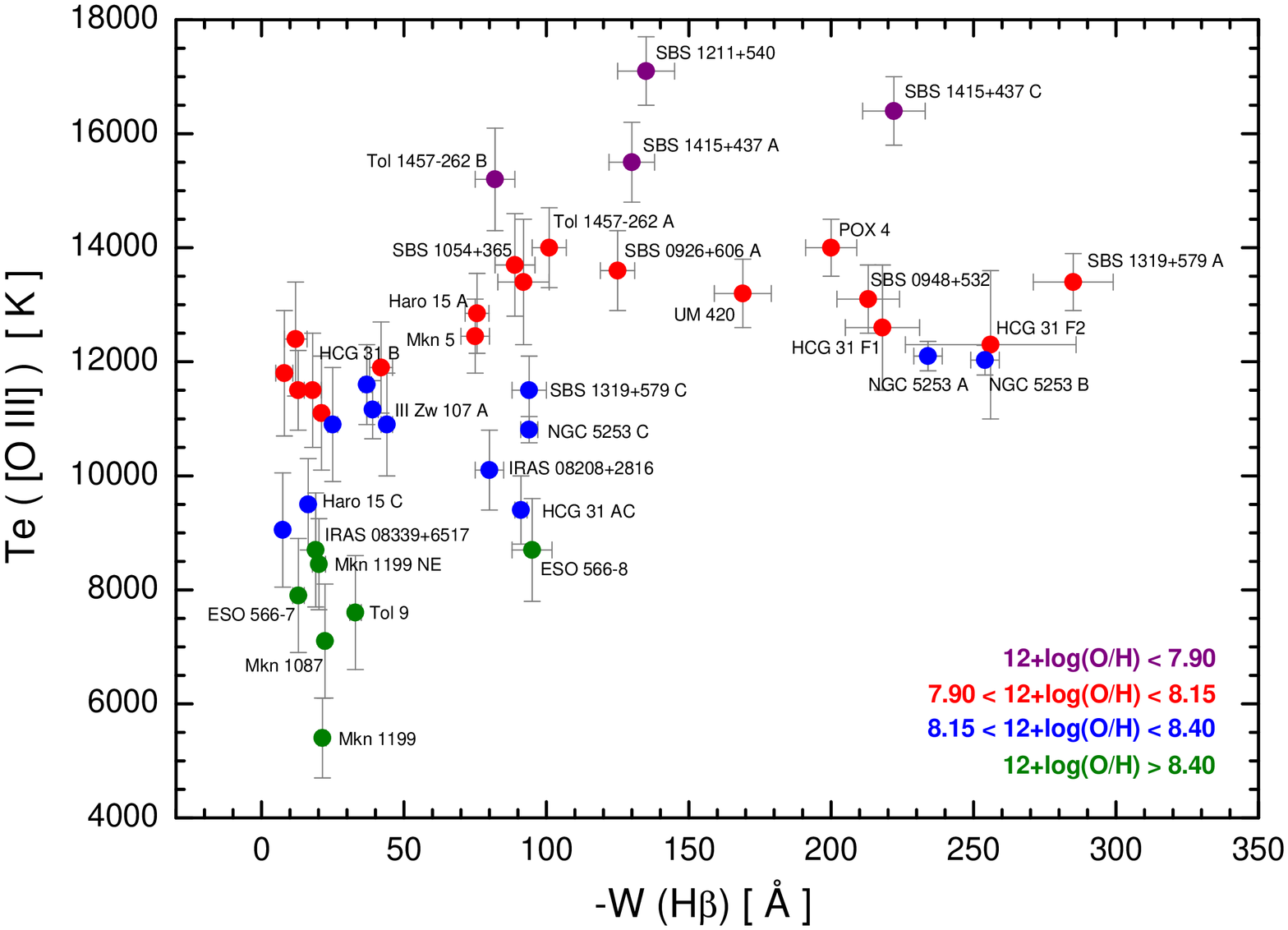} 
\protect\caption[ ]{\footnotesize{High ionization electron temperature, \TeOiii, vs. the \Hb\ equivalent width. The colour-dots indicate the 
oxygen abundance range of each object. Some objects have been labeled.}}
\label{to3whb}
\end{figure}

\WHb\ is a good indicator for the age of the most recent star-formation event. The hydrogen ionizing
flux of a star cluster gradually decreases as the most massive stars disappear with time, and hence the width of \Hb\ decreases with time (see 
Papers~I, II and III). Figure~\ref{to3whb} plots the relation between the high ionization electron temperature and the \Hb\ equivalent width. 
As we see, $|$\WHb$|$ increases with increasing \Te, but we must remember that there is a strong correlation between the electron temperature 
and the oxygen abundance, as high-metallicity \HII regions cool more efficiently than low-metallicity \HII regions. To study this effect, we used  
colours to plot four metallicity ranges in Fig.~\ref{to3whb}. These colours indicate the oxygen abundance range of each object: $<$7.90, 7.90--8.15, 
8.15--8.40 and $>$8.40, in units of \abox. 
Although now it is not so evident, it still seems that regions with larger $|$\WHb$|$ tend to have higher \Te. This indicates that younger bursts 
have a larger ionization budgets and are therefore capable to heat the ionized gas to higher electron temperatures. Another effect that we should
considered here is that galaxies with higher metallicity (and hence with lower electron temperature) usually have a higher absorption in the \Hb\ 
line than low-metallicity objects because of older stellar populations. Indeed, it is interesting to note that objects in the 
metallicity range 7.90$<$\abox$<$8.15 may have any value of \WHb. This is very probably because within this metallicity range lie both dwarf 
objects with no underlying old stellar population (i.e., HCG~31~F, SBS~0948+532) and galaxies which possess a considerable amount of old stars (i.e, HCG~31~B, Mkn~5).

\begin{figure}[t]
\includegraphics[angle=270,width=\linewidth]{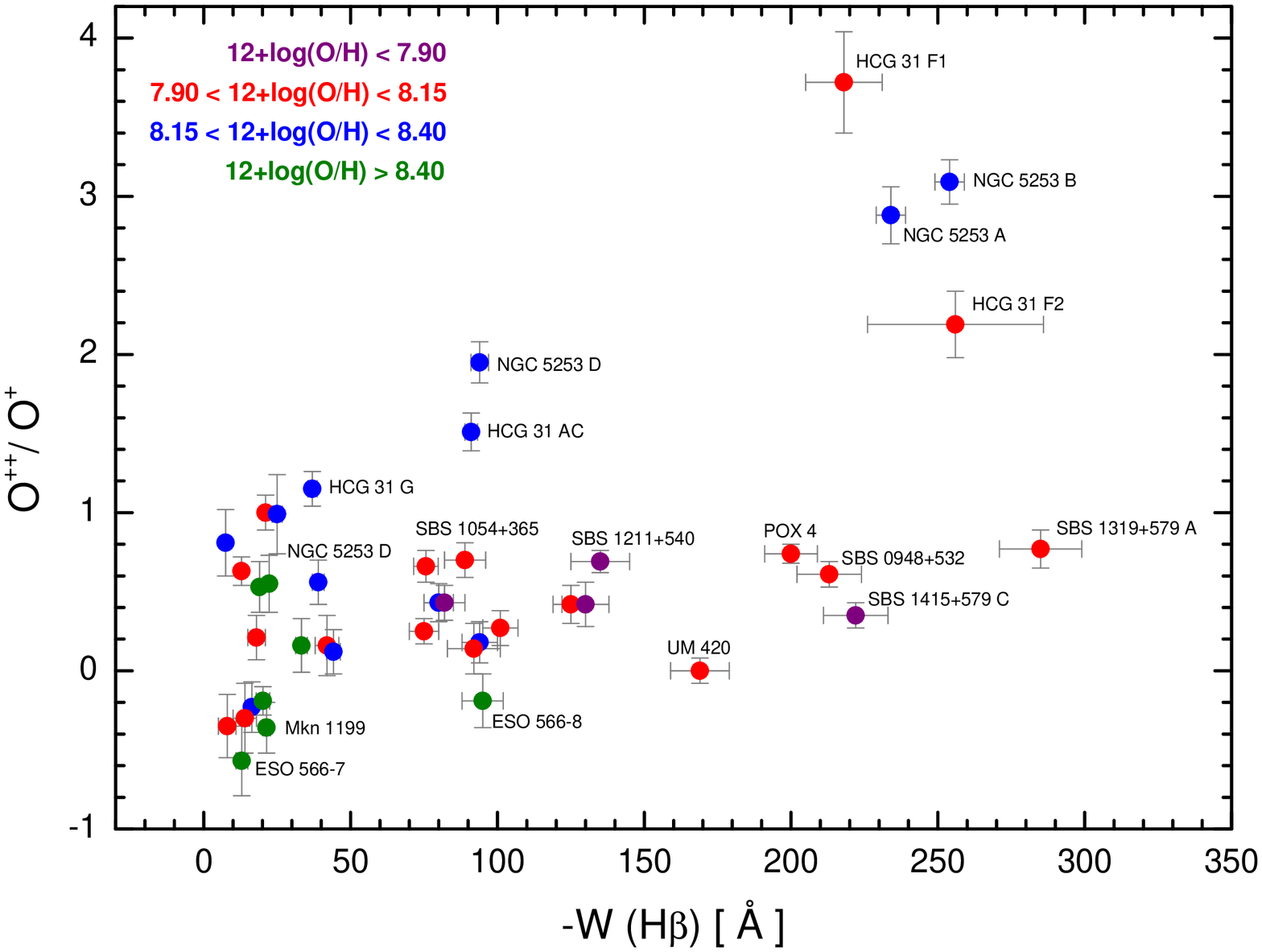}  
\protect\caption[ ]{\footnotesize{O$^{++}$/O$^+$ ratio vs. $W$(\Hb). The colour-dots indicate the oxygen abundance range of each object. Some 
objects have been labeled.}}
\label{whbo3o2}
\end{figure}
  

Figure~\ref{whbo3o2} shows that the ionization degree of the ionized gas of the starburts seems to increase with the increasing of $|$\WHb$|$. This 
is a relation similar to that found in Fig.~\ref{to3whb} and also indicates that younger bursts harbour a higher proportion of massive stars and 
therefore their associated \HII regions have larger ionization parameters. That is evident in NGC~5253~A, B and HCG~31~F, which show both the 
highest values of the $|$\WHb$|$  and the O$^{++}$/O$^+$ ratio and possess the youngest star-formation bursts (see Table~\ref{colores}). 
The increasing of the O$^{++}$/O$^+$ ratio as increasing  $|$\WHb$|$ seems to be independent of the metallicity, although galaxies with higher 
metallicity tend to show the lowest ionization degrees.

\begin{figure}[t!]
\includegraphics[angle=270,width=\linewidth]{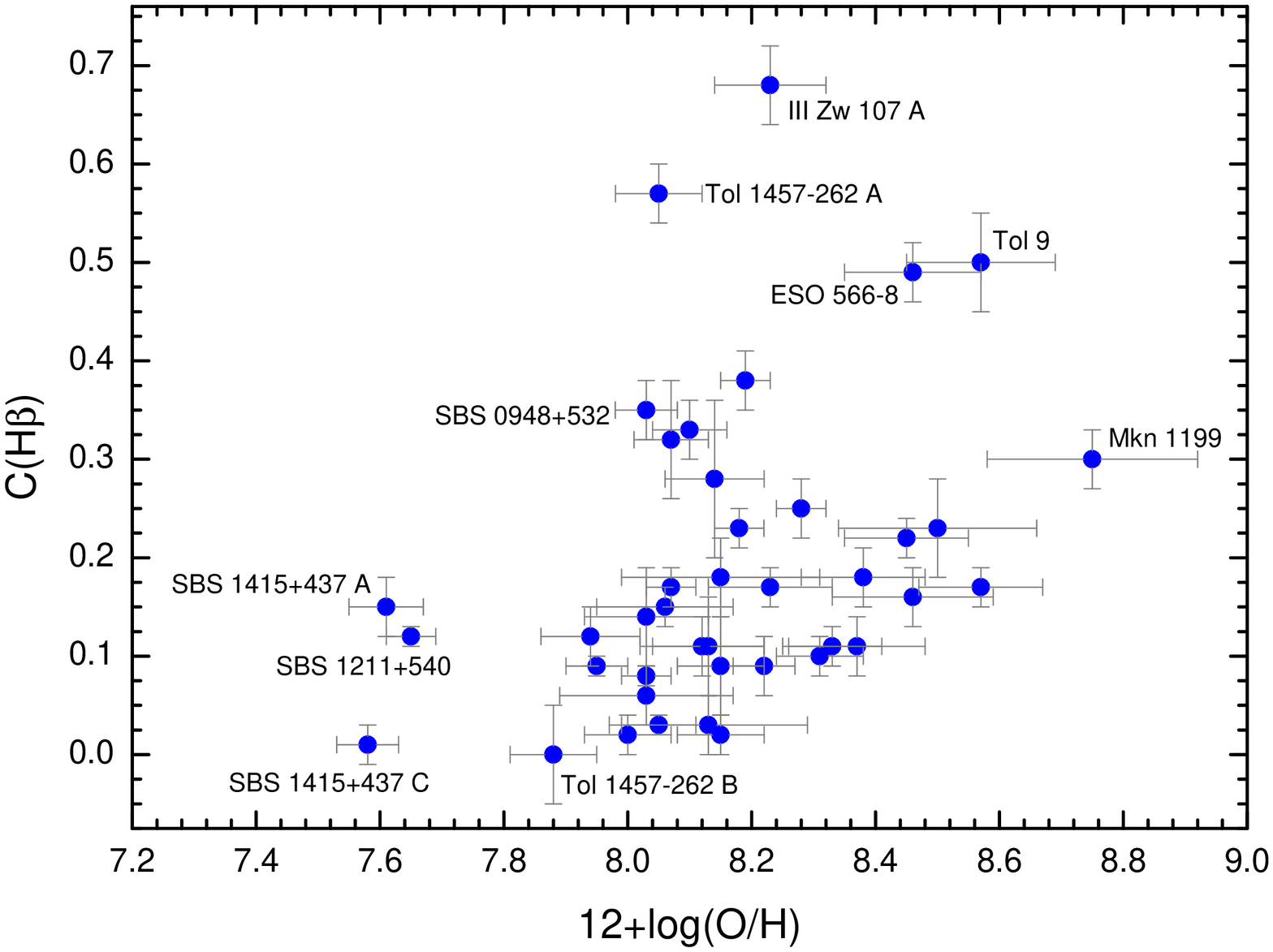}  
\protect\caption[ ]{\footnotesize{Reddening coefficient, \CHb, vs. the oxygen abundance for the regions analysed in this work. Some objects have been 
labeled.}}
\label{chbabox}
\end{figure} 

\begin{figure}[t!]
\includegraphics[angle=270,width=\linewidth]{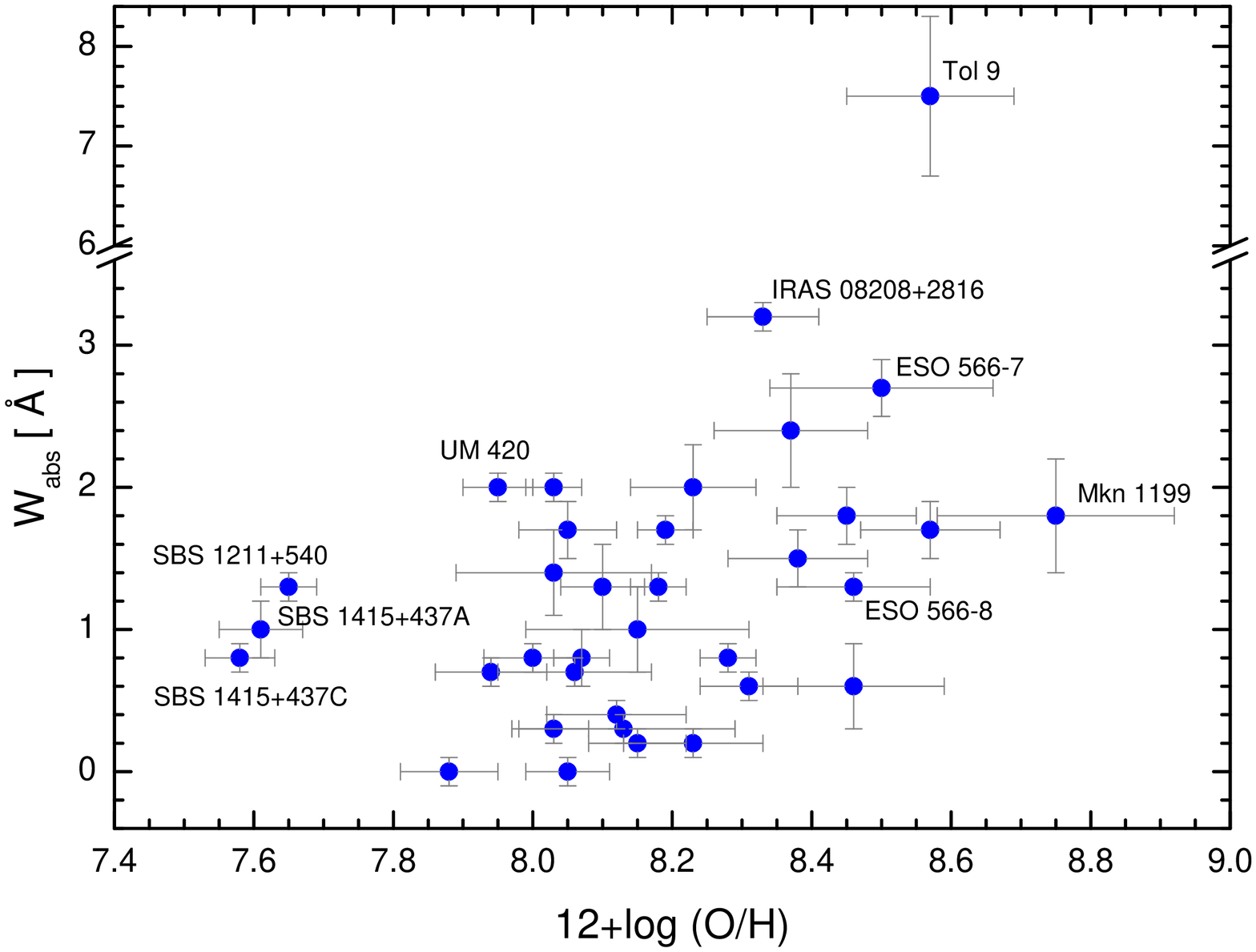}  
\protect\caption[ ]{\footnotesize{Equivalent width of the underlying stellar absorption in the Balmer \HI\ lines, \Wabs, vs. the oxygen abundance for 
the regions analysed in this work. Notice that the y-axis has been broken for clarity. 
The value of \Wabs\ for UM~420 is very probably overestimated, because this galaxy is observed through the external areas of the spiral disc of the 
foreground galaxy UCG~01809 (see Fig.~13 in Paper~I).}}
\label{wabsabox}
\end{figure}  


The dependence of the reddening coefficient as a function of other parameters is also interesting. Figure~\ref{chbabox} plots \CHb\ vs. the oxygen 
abundance. Although the dispersion of the data is rather scattered, we see a clear dependence: the reddening coefficient is higher at higher 
metallicities. We should expect this result, because galaxies with higher oxygen abundance are chemically more evolved 
and should contain a larger proportion of dust particles that absorbs the nebular emission. 

Another interesting relation is shown in Fig.~\ref{wabsabox}, which plots the equivalent width of the stellar absorption underlying the \HI\ Balmer 
lines (\Wabs) as a function of the oxygen abundance. We can see that objects with higher metallicities show larger \Wabs. More metallic galaxies 
correspond to more massive and chemically evolved systems, which means that they have consumed a larger fraction of their gas and the stellar component should be 
comparatively more important.
The data corresponding to the lowest metallicity objects analysed in this work (SBS~1415+579 and SBS~1211+540) show a value of \Wabs\ relatively high 
what is to be for them. This suggests a considerable underlying stellar population in these very low-metallicity galaxies, as we 
already discussed (see Sect.~3.15 and Sect~3.13 of Paper~I). 

\begin{figure}[t!]
\includegraphics[angle=270,width=\linewidth]{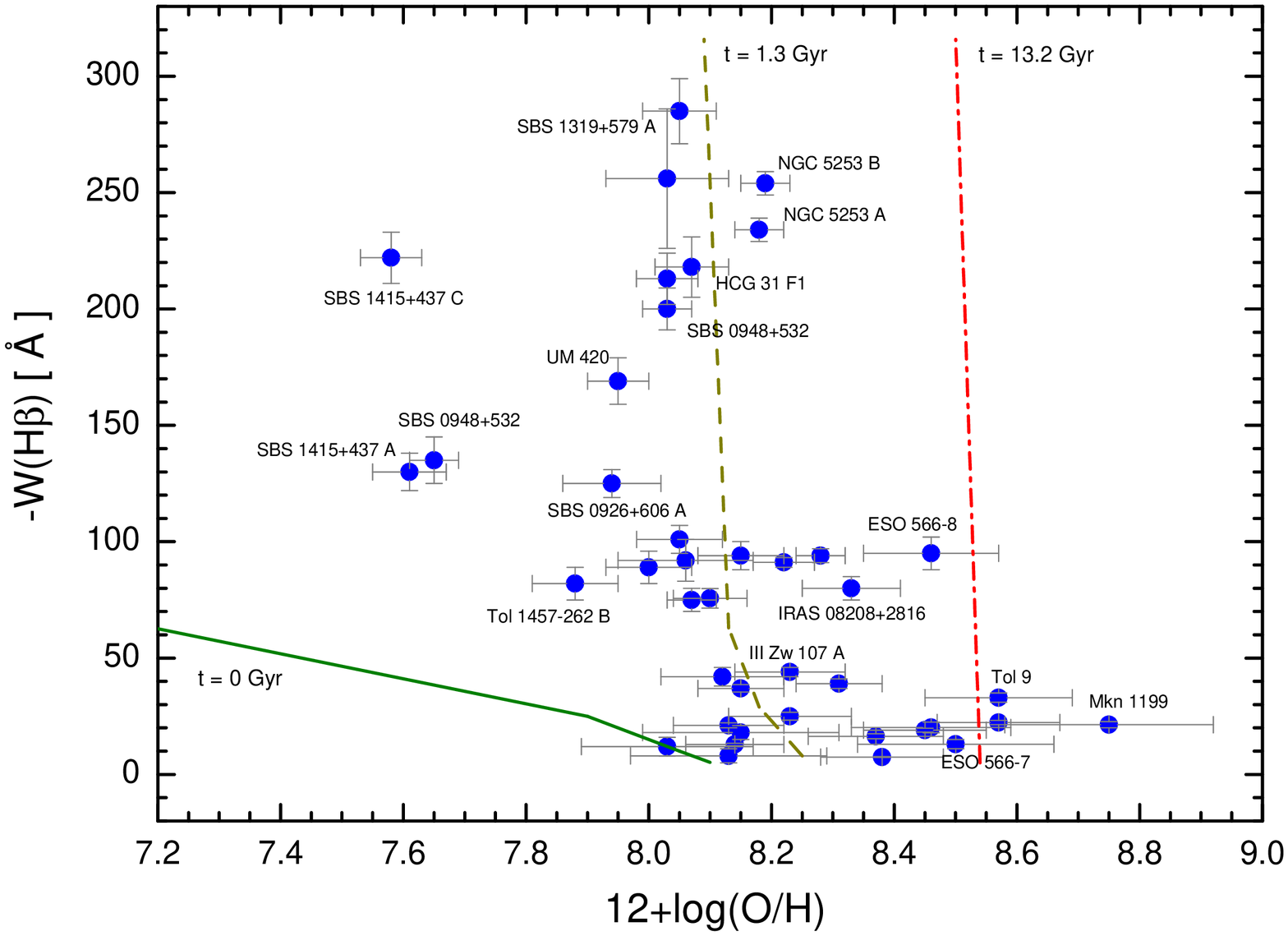}  
\protect\caption[ ]{\footnotesize{\Hb\ equivalent width vs. oxygen abundance. Some objects have been labeled. Some chemical evolution models of 
\HII galaxies provided by \citet{MartinManjon08} are also plotted. See text for details.}} 
\label{whbabox}
\end{figure}

Figure~\ref{whbabox} plots \WHb\ vs. the oxygen abundance. The very large dispersion of \WHb\ for \abox\ of about 8.0 is remarkable, but it also 
seems clear that galaxies with O/H ratios higher than that value tend to have $|$\WHb$|\leq$ 100~\AA\ and, conversely, galaxies with lower oxygen 
abundances show $|$\WHb$|\geq$ 100~\AA. This behaviour may be be related to the results in Fig.~\ref{wabsabox}, in the sense that more metallic 
objects tend to have a higher underlying stellar absorption that can lead to an underestimation of $|$\WHb$|$. 
Indeed, Fig.~\ref{whbabox} also compares our observational data with the predictions given by the chemical evolution models of \HII galaxies 
provided by \citet{MartinManjon08}. They assumed the star formation as a set of successive bursts, each galaxy experiencing 11 star-formation bursts 
along its evolution of 13.2 Gyr. Figure~\ref{whbabox} includes the results for the first ($t$=0 Gyr), second ($t$=1.3 Gyr) and last ($t$=13.2 Gyr) 
bursts for a model that considers an attenuated bursting star-formation mode and that 1/3 of the gas is always used to form stars in each time-step. 
As we see, all the strong starbursting systems are located between the positions of the first and second burst models, confirming that although the 
dominant stellar population is certainly very young, previous star-formation events in the last 500-1000 Myr are needed to explain our observational 
data points. This agrees well with the minimum ages of the underlying stellar component we derived using our photometric data (see Sect.~2 and last 
column of Table~\ref{colores}). The \citet{MartinManjon08} models also explain the large dispersion of \WHb\ for \abox\ of about 8.0, as well as the 
trend that more metal rich galaxies have lower values of $|$\WHb$|$ because of the effects of the underlying stellar populations.

\begin{figure}[t]
\includegraphics[angle=270,width=\linewidth]{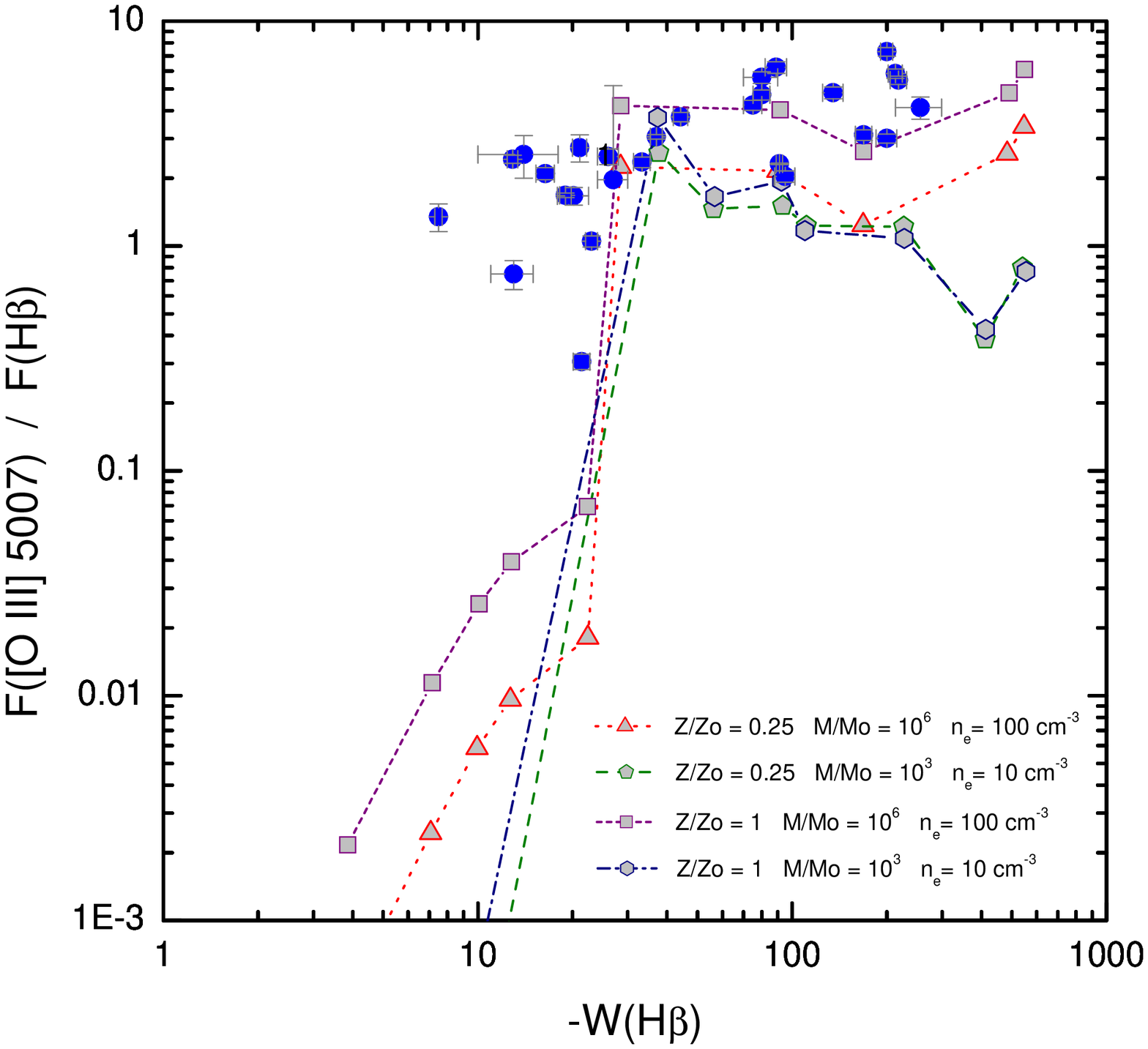}  
\protect\caption[ ]{\footnotesize{$F$([\ion{O}{iii}] $\lambda$5007)/$F$(\Hb) vs. \WHb\ and comparison with the predictions given by four different 
ionization models by \citet*{SSL01}. See text for details.}}
\label{stasinska}
\end{figure}

Another indication of the effect of the underlying evolved stellar population  is found in Fig.~\ref{stasinska}, which compares the [\ion{O}{iii}] 
$\lambda$5007 line flux with the \WHb\ of the sample galaxies with the model predictions by \citet{SSL01}. Although a general good correspondence is 
found, some of the objects are slightly displaced to the left --lower \WHb-- of the models predictions, suggesting that perhaps the measured values 
of $|$\WHb$|$ are underestimated for some of them that precisely coincide with those with a larger oxygen abundance, as we also concluded 
before.

\section{Chemical abundances of the ionized gas}

\begin{table*}[t!]
\centering
  \caption{\footnotesize{Chemical properties of the ionized gas for the galaxies analysed in this work.}}
  
  \label{abtotal}
  \tiny
  \begin{tabular}{l@{\hspace{8pt}} c@{\hspace{6pt}}       c@{\hspace{12pt}}  c@{\hspace{12pt}}  c@{\hspace{12pt}}       
c@{\hspace{12pt}}c@{\hspace{12pt}}c@{\hspace{12pt}}c@{\hspace{12pt}}}
  \tableline
   \noalign{\smallskip}
Galaxy & \Te$^a$ & 12+$\log$(O/H) & $\log\frac{\rm O^{++}}{\rm O^+}$ & $\log$(N/O)& $\log$(S/O) & $\log$(Ne/O) & $\log$(Ar/O) &  $\log$(Fe/O) \\ 
\noalign{\smallskip}
\tableline
\noalign{\smallskip}

HCG 31 AC	    & D  &8.22$\pm$0.05 &	1.51$\pm$0.12 &	-1.12$\pm$0.08 & \nodata       &-0.93$\pm$0.12 &\nodata        &-2.12$\pm$0.21 \\
HCG 31 B	    & D  &8.14$\pm$0.08 &	0.63$\pm$0.09 &	-1.39$\pm$0.10 &-1.67$\pm$0.14 &-0.42$\pm$0.13 &\nodata        &-1.87$\pm$0.32 \\
HCG 31 E	    & D  &8.13$\pm$0.09 &   1.00$\pm$0.11 &	-1.26$\pm$0.12 &-1.58$\pm$0.15 &-0.42$\pm$0.14 &\nodata        &-1.77$\pm$0.32 \\
HCG 31 F1	    & D  &8.07$\pm$0.06 &	3.72$\pm$0.32 &	-1.27$\pm$0.11 &-1.69$\pm$0.15 &-0.80$\pm$0.17 &\nodata        &-1.9:	 \\
HCG 31 F2	    & D  &8.03$\pm$0.10 &	2.19$\pm$0.21 &	-1.43$\pm$0.16 &-1.67$\pm$0.18 &-0.76$\pm$0.20 &\nodata        & \nodata   \\
HCG 31 G	    & D  &8.15$\pm$0.07 &	1.15$\pm$0.11 &	-1.31$\pm$0.10 &-1.67$\pm$0.22 &-0.56$\pm$0.14 &\nodata        &-2.0:	 \\
Mkn 1087	    & EC &8.57$\pm$0.10 &	0.55$\pm$0.18 &	-0.81$\pm$0.12 &-1.78$\pm$0.16 &-0.45$\pm$0.17 &\nodata        & \nodata   \\
Mkn 1087 N	    & EC &8.23$\pm$0.10 &	0.99$\pm$0.25 &	-1.46$\pm$0.15 & \nodata       &-0.52$\pm$0.19 &\nodata        & \nodata   \\
Haro 15 C	    & EC &8.37$\pm$0.10 &  -0.23$\pm$0.16 & -1.03$\pm$0.15 &-1.71$\pm$0.18 &-0.65$\pm$0.18 &\nodata        &-2.2:	 \\
Haro 15 A	    & D  &8.10$\pm$0.06 &	0.66$\pm$0.10 &	-1.35$\pm$0.11 &-1.89$\pm$0.15 &-0.68$\pm$0.12 &\nodata        &-1.6:    \\
Mkn 1199	    & D  &8.75$\pm$0.12 &  -0.36$\pm$0.16 & -0.62$\pm$0.10 &-1.54$\pm$0.14 &-0.58$\pm$0.17 &\nodata        &-1.86$\pm$0.26 \\
Mkn 1199 NE     & EC &8.46$\pm$0.13 &  -0.19$\pm$0.09 & -1.20$\pm$0.11 &-1.54$\pm$0.17 &-0.65$\pm$0.18 &\nodata        & \nodata   \\
Mkn 5	        & D  &8.07$\pm$0.04 &	0.25$\pm$0.08 &	-1.38$\pm$0.07 &-1.62$\pm$0.11 &-0.80$\pm$0.13 &-2.31$\pm$0.12 &-1.96$\pm$0.18 \\
IRAS 08208+2816 & D  &8.33$\pm$0.08 &	0.43$\pm$0.12 &	-0.89$\pm$0.11 &-1.64$\pm$0.16 &-0.67$\pm$0.13 &-2.51$\pm$0.15 &-1.95$\pm$0.17 \\
IRAS 08339+6517	& EC &8.45$\pm$0.10 &	0.53$\pm$0.16 &	-0.94$\pm$0.14 & \nodata	   &-0.45$\pm$0.18 &\nodata        &\nodata   \\
IRAS 08339+6517c& EC &8.38$\pm$0.10 &	0.81$\pm$0.21 &	-1.13$\pm$0.17 & \nodata	   &-0.55:	       &\nodata        &\nodata   \\
POX 4	        & D  &8.03$\pm$0.04 &	0.74$\pm$0.06 &	-1.54$\pm$0.06 &-1.80$\pm$0.10 &-0.78$\pm$0.10 &\nodata        &-2.17$\pm$0.11 \\
POX 4c   	    & EC &8.03$\pm$0.14 &  -0.30$\pm$0.22 & -1.60$\pm$0.20 & \nodata	   &-0.60:         &\nodata        &\nodata   \\
UM 420	        & D  &7.95$\pm$0.05 &	0.00$\pm$0.08 &	-1.11$\pm$0.07 &-1.66$\pm$0.13 &-0.71$\pm$0.13 &\nodata        &-2.16$\pm$0.13 \\
SBS 0926+606 A 	& D  &7.94$\pm$0.08 &	0.42$\pm$0.12 &	-1.45$\pm$0.09 &-1.60$\pm$0.13 & \nodata	   &-2.34$\pm$0.13 &-1.99$\pm$0.16 \\
SBS 0926+606 B  & EC &8.15$\pm$0.16 &	0.21$\pm$0.14 &	-1.35$\pm$0.12 & \nodata	   &  \nodata	   &\nodata        &\nodata   \\
SBS 0948+532    & D  &8.03$\pm$0.05 &	0.61$\pm$0.08 &	-1.42$\pm$0.08 &-1.69$\pm$0.14 &-0.73$\pm$0.12 &\nodata        &-1.78$\pm$0.10 \\
SBS 1054+365    & D  &8.00$\pm$0.07 &	0.70$\pm$0.11 &	-1.41$\pm$0.08 &-1.79$\pm$0.15 &-0.67$\pm$0.11 &-2.29$\pm$0.14 &\nodata    \\
SBS 1054+365 b	& EC &8.13$\pm$0.16 &  -0.35$\pm$0.20 & -1.47$\pm$0.20 & \nodata       &   \nodata     &\nodata        &\nodata   \\
SBS 1211+540    & D  &7.65$\pm$0.04 &	0.69$\pm$0.07 &	-1.62$\pm$0.10 &-1.47$\pm$0.14 &-0.75$\pm$0.10 &\nodata        &\nodata   \\
SBS 1319+579 A	& D  &8.05$\pm$0.06 &	0.77$\pm$0.12 &	-1.53$\pm$0.10 &-1.76$\pm$0.10 &	 \nodata   &-2.41$\pm$0.11 &\nodata    \\
SBS 1319+579 B	& D  &8.12$\pm$0.10 &	0.16$\pm$0.19 &	-1.49$\pm$0.12 &-1.76$\pm$0.14 &	\nodata    &   \nodata     &\nodata   \\
SBS 1319+579 C	& D  &8.15$\pm$0.07 &	0.18$\pm$0.13 &	-1.38$\pm$0.10 &-1.60$\pm$0.11 &	 \nodata   &   \nodata     &-2.3:	  \\
SBS 1415+437 C	& D  &7.58$\pm$0.05 &	0.35$\pm$0.08 &	-1.57$\pm$0.08 &-1.62$\pm$0.12 &	 \nodata   &-2.31$\pm$0.13 &-1.91$\pm$0.13\\
SBS 1415+437 A	& D  &7.61$\pm$0.06 &	0.42$\pm$0.14 &	-1.57$\pm$0.09 &-1.72$\pm$0.14 &	 \nodata   &    \nodata    &-1.9:	 \\
III Zw 107 A	& D  &8.23$\pm$0.09 &	0.12$\pm$0.14 &	-1.16$\pm$0.10 &-1.82$\pm$0.15 &-0.73$\pm$0.15 &-2.46$\pm$0.13 &-2.3:	 \\
Tol 9	        & D  &8.57$\pm$0.10 &	0.16$\pm$0.17 &	-0.81$\pm$0.11 &-1.62$\pm$0.12 &-0.72$\pm$0.14 &-2.55$\pm$0.15 &-2.1:	 \\
Tol 1457-262 A	& D  &8.05$\pm$0.07 &	0.27$\pm$0.11 &	-1.57$\pm$0.11 &-1.88$\pm$0.13 &-0.88$\pm$0.18 &-2.50$\pm$0.13 &-2.2:	 \\
Tol 1457-262 B	& D  &7.88$\pm$0.07 &	0.43$\pm$0.11 &	-1.61$\pm$0.12 &-1.72$\pm$0.18 &-0.88$\pm$0.20 &-2.44$\pm$0.18 &-1.90$\pm$0.22 \\
Tol 1457-262 C	& D  &8.06$\pm$0.11 &	0.14$\pm$0.16 &	-1.59$\pm$0.16 & \nodata	   &-0.84$\pm$0.22 &-2.45$\pm$0.20 &     \nodata \\
ESO 566-8	    & D  &8.46$\pm$0.11 &  -0.19$\pm$0.17 & -0.76$\pm$0.12 & \nodata	   &-0.56$\pm$0.19 &-2.17$\pm$0.19 &-2.5:	 \\
ESO 566-7	    & EC &8.50$\pm$0.16 &  -0.57$\pm$0.22 & -0.82$\pm$0.16 & \nodata	   &   \nodata     &-2.49$\pm$0.25 &     \nodata \\
NGC 5253 A	    & D  &8.18$\pm$0.04 &	2.88$\pm$0.18 &	-0.91$\pm$0.07 &-1.58$\pm$0.08 &-0.71$\pm$0.08 &-2.19$\pm$0.07 &-2.10$\pm$0.12 \\
NGC 5253 B	    & D  &8.19$\pm$0.04 &	3.09$\pm$0.14 &	-1.02$\pm$0.07 &-1.60$\pm$0.08 &-0.70$\pm$0.08 &-2.21$\pm$0.07 &-2.18$\pm$0.11 \\
NGC 5253 C	    & D  &8.28$\pm$0.04 &	1.95$\pm$0.13 &	-1.50$\pm$0.08 &-1.69$\pm$0.09 &-0.74$\pm$0.08 &-2.30$\pm$0.08 &-2.46$\pm$0.14 \\
NGC 5253 D	    & D  &8.31$\pm$0.07 &	0.56$\pm$0.14 &	-1.49$\pm$0.10 &-1.74$\pm$0.13 &-0.70$\pm$0.15 &-2.30$\pm$0.13 &-2.25$\pm$0.16 \\
\noalign{\smallskip}
\tableline
  \end{tabular}
  \begin{flushleft}
  $^a$  In this column we indicate if \Te\ was computed using the direct method (D) or via empirical calibrations (EC).
  \end{flushleft}
  \end{table*}

Table~\ref{abtotal} compiles the oxygen abundance, the O$^{++}$/O$^{+}$ ratio and the N/O, S/O, Ne/O, Ar/O, and Fe/O ratios (all in logarithmic 
units) for all galaxies analysed in this work. As we already said, 31 up to the 41 independent regions within the WR galaxy sample analysed here have 
a direct estimate of the electron temperature (as indicated in Table~\ref{abtotal}). Figure~\ref{compabun} (left) plots the N/O ratio vs. the 
oxygen abundance for all our data with a direct estimate of the electron temperature and its comparison with previous samples involving similar 
objects and \Te-based \citep{IT99,Izotov04}. This figure shows that the position of our data agrees with that obtained using other observations. The 
errors we estimated in our objects are in general higher than 
those reported by \citet{IT99} and \citet{Izotov04} basically because we used different criteria for estimating observational errors, which are more 
conservative as well as more realistic in our opinion. We also point out that our data are always of higher spectral and spatial resolution 
than those obtained by the aforementioned authors, and have a similar or even higher signal-to-noise ratio in many cases. This is an important point 
to be clarified because non-specialist in the spectra of ionized nebulae may interpret that lower quoted uncertainties are synonymous of better 
observational data, and this may not always be the case. 
Although the errors in the electron temperatures derived using the empirical methods are large, relative atomic abundances (such as the N/O ratio) 
are less sensitive to the choice of \Te. Therefore they are used in many occasions to compare with the results provided by previous observations or  
with the predictions given by theoretical models.

\subsection{The N/O ratio}
  
  \begin{figure*}[t]
\centering
\begin{tabular}{cc}
\includegraphics[angle=270,width=0.45\linewidth]{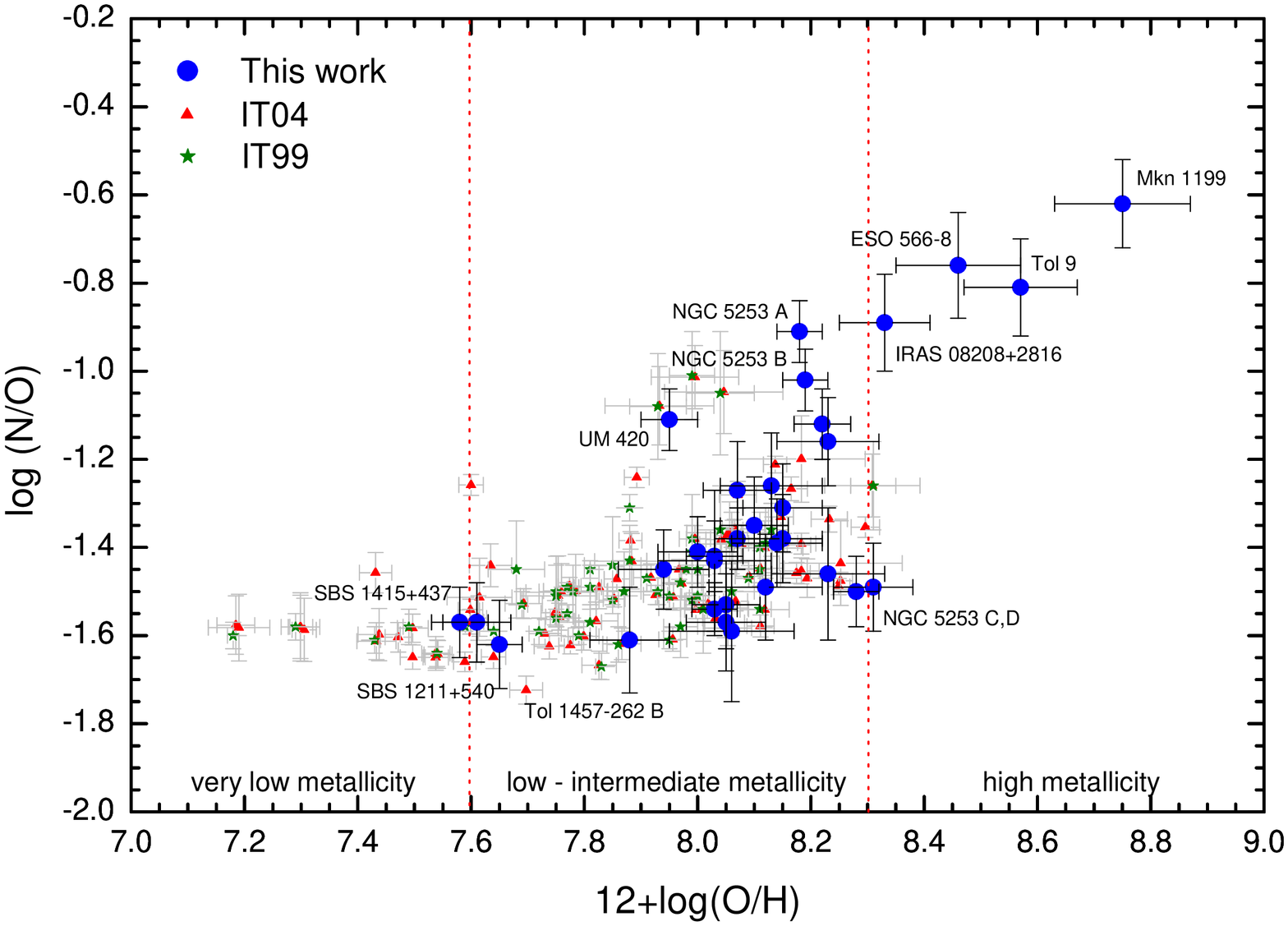} &    
\includegraphics[angle=270,width=0.45\linewidth]{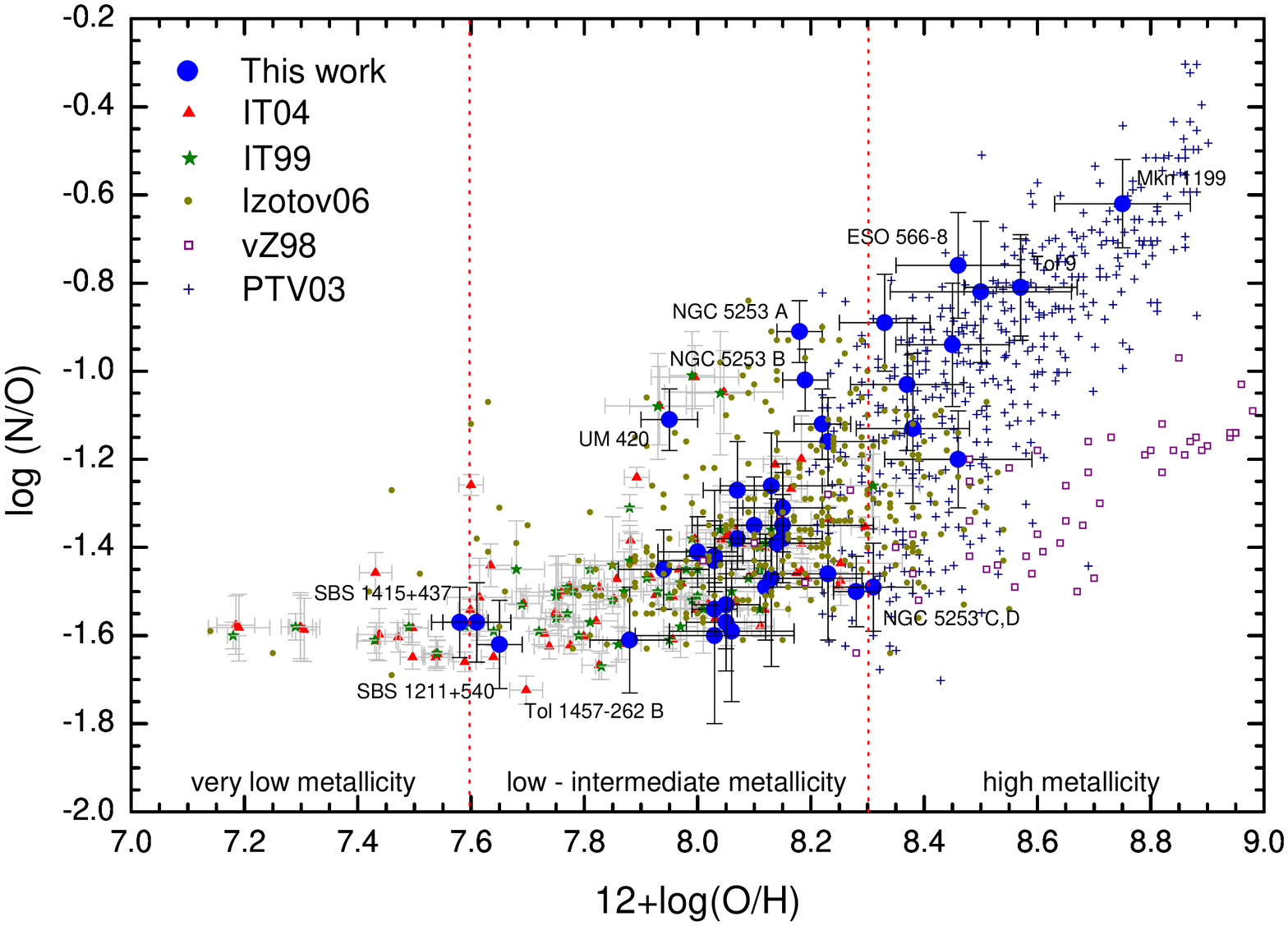} \\ 
\end{tabular}
\protect\caption[ ]{\footnotesize{(\emph{Left}) N/O ratio vs. oxygen abundance for all objects with a direct estimate of \Te. Our data are compared 
with those compiled by IT99 (Izotov \& Thuan 1999) and Izotov et al. (2004). (\emph{Right}) N/O ratio vs. oxygen abundance for all galaxies analysed 
in this work (also included objects which chemical abundances are derived using empirical calibrations). For comparison, we have included other 
galaxy samples from the literature whose O/H ratio has been determined using empirical calibrations: Izotov et al. (2006), van Zee et al. (1998), 
IT99 (Izotov \& Thuan 1999), Izotov et al. (2004) and PTV03 (Pilygin et al. 2003).}}
\label{compabun}
\end{figure*} 
  
Figure~\ref{compabun} (right) plots the N/O ratio vs. \abox\ for all objects analysed in this work; the chemical abundances were derived either from 
the direct method or via empirical calibrations. We compare our data with the two galaxy samples previously indicated \citep{IT99,Izotov04} and with 
other galaxy samples whose data have been obtained using empirical calibrations: \citet{Izotov06}, whose data were extracted from the \SDSS, and 
\citet{vZee98}, who study data from \HII regions within spiral galaxies with chemical abundances computed via the direct method or using the 
\citet{McGaugh94} empirical calibration. 
In some sense, the N/O ratio of a galaxy is an indicator of the time that
has elapsed since the bulk of star formation occurred, or of the nominal \emph{age} of the galaxy 
as suggested by \citet{EP78}. Following the position of our data points in Fig.~\ref{compabun} we see that they follow the expected trend:
\begin{enumerate}
\item The N/O ratio is rather constant for \abox$\leq$7.6. In our case, for the galaxies SBS~1211+540 and SBS~1415+437 we derive log N/O$\sim-$1.6, 
similar values as those found by \citet{IT99}. These authors explained the constant N/O ratio in very low-metallicity objects assuming that 
the nitrogen is produced only as a primary element in massive, short-life stars. 
However, other authors have claimed that this may be not completely true (i.e., Henry et al. 2000; Pilyugin et al. 2003; Moll\'a et al. 2006) because 
of the lack of a clear mechanism that produces N in massive stars besides the effect or the stellar rotation \citep{MeynetMaeder05}. Furthermore, 
these galaxies already host old stellar populations, and hence low- and intermediate-mass stars should be also releasing N to the ISM. \citet{Henry00} 
explained the constancy of the N/O ratio in metal-poor galaxies by a historically low star-formation rate, where almost all the nitrogen is produced by 
4--8 \Mo\ stars.
Additionally, \citet{Izotov06} suggested that the low dispersion of the data in this part of the diagram is probably explained by the low number of 
WR stars that are expected at very low-metallicity regimes.
\item However, there is an important dispersion of the data in the interval 7.6$\leq$\abox$\leq$8.3. This problem has been analysed by several 
authors in the past (i.e., Kobulnicky \& Skillman, 1998; Izotov \& Thuan, 1999; Pilyugin et al. 2003; Moll\'a et al. 2006). Two main scenarios have 
been proposed for explaining this dispersion:
\begin{enumerate}
\item A loss of heavy element via galactic winds. In particular, it should be a loss of $\alpha$-elements via supernova explosions. 
$\alpha$-elements, such as oxygen, are produced in massive short-lived stars \citep{EP78,ClaytonPantelaki93}. Hence, the effect of supernova 
explosions would produce a underabundance of oxygen \citep{EP95}, increasing the N/O ratio. However, the observational evidence for low-mass galaxies 
with galactic winds is still unclear (i.e. Marlowe et al. 1995; Bomans et al. 2007; Dubois \& Teyssier 2008; van Eymeren et al. 2008, 2009, 2010) and 
even the numerical simulations give very discrepant results (i.e., Mac Low \& Ferrara 1999; Springel \& Hernquist 2003; Tenorio-Tagle et al. 2006; 
Dubois \& R. Teyssier 2008).  
\item A delayed release of nitrogen and elements produced in low-mass long-lived stars 
compared to the \mbox{$\alpha$-elements.} 
The N/O ratio drops and the O/H ratio increases as supernovae release the $\alpha$-elements into the \ISM. 
Consequently, the chemical properties of these galaxies would vary very fast (few tens of \Myr) during the starburst phase \citep{KS98}. 
The delayed-release hypothesis also predicts that \BCDG s with high N/O ratios 
are experiencing their first burst of massive star formation after a relatively long quiescent
interval (oxygen has still not been completely delayed by massive stars and mixed with the surrounding ISM), 
while those objects with low N/O ratios have had little or no quiescent interval. 
However, recent chemical evolution models 
suggest that the releasing and mixing of the oxygen occurs almost instantaneously, and hence the delayed-release scenario cannot explain \BCDG s 
with high N/O ratio. If this is the case, the most plausible explanation of the high N/O ratio observed in these objects is the chemical pollution 
due to the winds of WR stars, which are indeed ejecting N to the ISM, as we will discuss below.
\end{enumerate}
\item For moderate high-metallicities objects, \abox$\geq$8.3, the N/O ratio clearly increases with increasing oxygen abundance. This trend seems to 
be a consequence of the metallicity-dependence of nitrogen production in both massive and intermediate-mass stars (e.g., Pilyugin et al. 2003), 
the N/O ratio increases at higher metallicities. Hence, nitrogen is essentially a secondary element in this metallicity regime 
\citep*{Torres-Peimbert89,Vila-Costas93,Henry00,vZH06}.
Besides the uncertainties (that are higher than those estimated in other objects because the \Te\ error is higher at higher metallicities) our data 
agree with the tendency found in other galaxy samples, as that compiled by \citet{PTV03}. 
However, notice that the galaxy sample compiled by \citet*{vZee98} does not agree with our data, as her data have a systematically lower N/O ratio. 
In some cases, differences higher than 0.5 dex in the N/O ratio are found for a particular oxygen abundance. This discrepancy may be partially 
explained by the fact that empirical calibrations from photoionized models --\citet*{vZee98} used \citet{McGaugh91} models-- seem to overestimate 
the actual oxygen abundance by at least 0.2 dex (see below), and hence the derived N/O ratio is lower than the actual value.
\end{enumerate}
In summary, Fig.~\ref{compabun} can be explained assuming the very different star-formation histories that each individual galaxy has experienced 
\citep{PTV03,vZH06}. A galaxy with a constant SFR will have a lower net N/O yield than a galaxy with declining SFR, because more oxygen has been 
released into the \ISM\ due to the ongoing star-formation activity. This observational result completely agrees with the predictions given by 
chemical evolution models that consider the effect of the star-formation history in the N/O--O/H diagram, as those presented by \citet{Molla06}.

\subsection{Nitrogen enrichment in WR galaxies}

From Fig.~\ref{compabun}, it is evident that there are some objects in the low-intermediate metallicity regime with a higher N/O ratio than expected 
for their oxygen abundance. 
An excess of nitrogen abundance has been reported in a few cases (e.g. Kobulnicky et al. 1997, Pustilnik et al. 2004). Remarkably, the common 
factor observed in all galaxies with a high N/O ratio is the detection of Wolf-Rayet features. Indeed, as we demonstrated in our analysis of NGC~5253 
\citep{LSEGRPR07}, the ejecta of WR stars may be the origin of a localized N (and probably also He) enrichment of the \ISM\ in these galaxies.

The analysis of the WR populations within our sample galaxy was performed in Paper~III. The numbers of WNL and WCE stars were computed assuming 
metallicity-dependent WR luminosities \citep{CH06}.  
We detected the blue \WRBUMP\ (the broad \ion{He}{ii} $\lambda$4686 emission line) in all objects with high N/O ratio: UM~420, NGC~5253~A,B,  
HCG~31~AC, IRAS~08208+2816, III~Zw~107 and ESO~566-8, indicating that these bursts host an important WNL star population. 

\begin{figure}[t]
\includegraphics[angle=270,width=\linewidth]{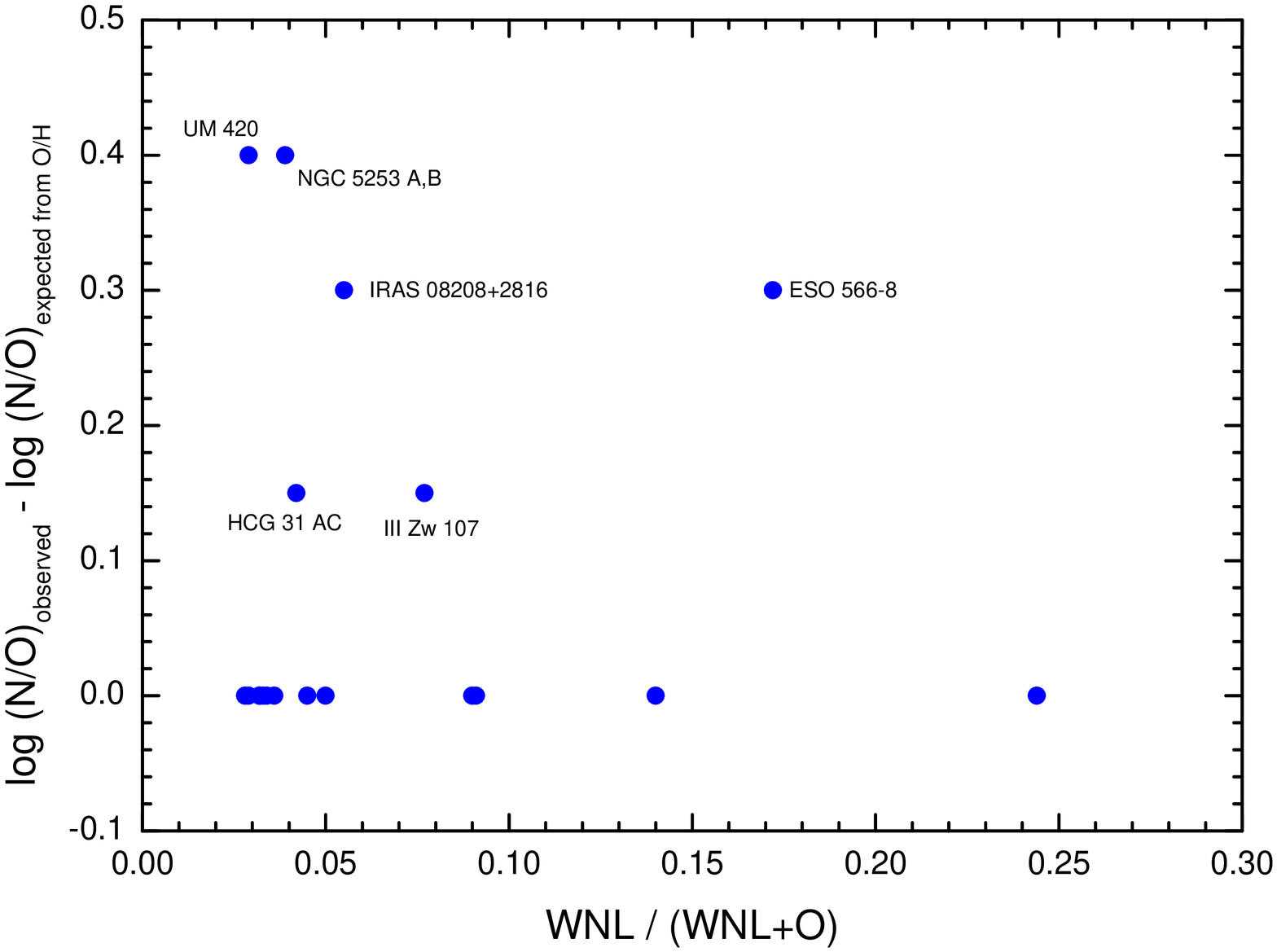}  
\protect\caption[ ]{\footnotesize{Observed nitrogen overabundance, $\Delta(N/O)=\log(N/O_{\rm observed})-\log(N/O_{\rm expected\ from\ O/H}$), vs. 
the WNL/(WNL+O) ratio for our sample galaxies (derived in Paper~III). Some important objects have been labeled.}}
\label{nowr}
\end{figure}

Figure~\ref{nowr} plots the observed nitrogen overabundance, $\Delta(N/O)=\log(N/O_{\rm observed})-\log(N/O_{\rm expected\ from\ O/H}$),  as a 
function of the derived WNL/(WNL+O) ratio. We do not see any clear trend in this diagram, merely that 
the objects with a high N/O ratio do not show a particularly high WNL/(WNL+O) ratio.  

Galaxies HCG~31~AC and III~Zw~107 seem to show a slight nitrogen excess [$\Delta(N/O)\sim0.15$ dex]. Three of the galaxies with high N/O ratio 
compiled by \citet{Pustilnik04} are plotted in Fig.~\ref{nowr}: NGC~5253 (already discussed), Mkn~1089 (HCG~31~AC) and UM~420. Our data confirm the 
nitrogen overabundance in UM~420 [$\Delta(N/O)\sim0.4$ dex], but not a significant N/O ratio in HCG~31~AC [$\Delta(N/O)\sim0.15$ dex; 
\citet{Pustilnik04} quoted $\sim0.5$ dex]. We also find a relatively high N/O ratio in ESO~566-8 [$\Delta(N/O)\sim0.3$ dex].

The very rare occurrence of objects with a large N overabundance suggests the general idea of the short-time scales for the localized
N pollution and its fast dispersal.  \citet{BKD08} used SDSS data to find that for $|$\WHb$|\geq 100$ \AA, WR galaxies show a high N/O compared to 
non-WR galaxies. Quantitatively these authors found that on average $\Delta\log$(N/O)$_{\rm [WR - nonWR]}$=0.13$\pm$0.04. They interpreted this 
result as a rapid enrichment of the \ISM\ from WR winds. \citet*{BKD08} also found that WR galaxies are in general more metal-rich at a given
$|$\WHb$|$ than similar galaxies without WR features, which likely is a reflection of WR stars being more abundant at higher metallicities 
(see Fig.~5 of Paper~III).

\begin{figure}[t]
\includegraphics[angle=270,width=\linewidth]{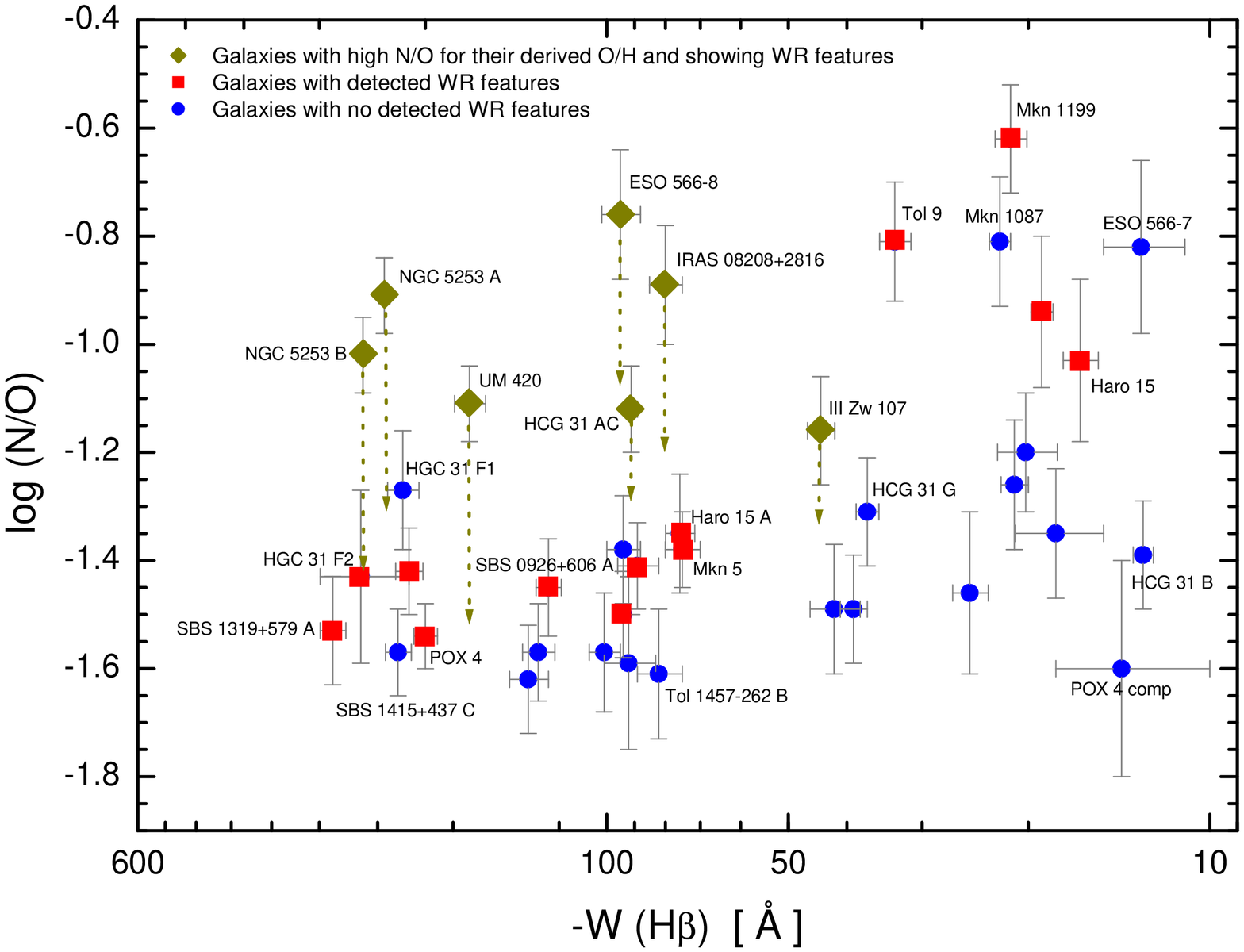}  
\protect\caption[ ]{\footnotesize{\Hb\ equivalent width vs. the N/O ratio. Blue circles indicate regions with no detected WR features. Red squares are 
galaxies in which WR features are observed. Yellow diamonds correspond to objects with a high N/O ratio for their oxygen abundance, all showing WR 
features. The  $\Delta(N/O)=\log(N/O_{\rm observed})-\log(N/O_{\rm expected\ from\ O/H}$) difference is shown in every galaxy with dotted yellow 
lines.}}
\label{nowhb}
\end{figure}

Although we do not dismiss the statistical analysis performed by \citet*{BKD08} comparing WR and non-WR galaxy data, we would like to warn about the 
use of data with low spectral resolution in order to derive an accurate nitrogen abundance in individual objects. This is commonly done via the 
analysis of the [\ion{N}{ii}] $\lambda$5683 emission line, very close to \Ha. Lack of sufficient spectral resolution will derive a blending of both 
lines and a probable over-estimation of the [\ion{N}{ii}] $\lambda$5683 flux, that would be even higher if broad low-intensity wings in the \Ha\ 
profile exist, which are actually rather common in WR galaxies (e.g. M\'endez \& Esteban 1997). On the other hand, as explained by \citet{Izotov06}, 
the bright doublet [\ion{O}{ii}] $\lambda\lambda$3726,3729 is not observed in the SDSS spectra for nearby galaxies, and hence the estimate of the 
total oxygen abundance has to be done via the [\ion{O}{ii}] $\lambda\lambda$7319,7330, which is much fainter, very dependent on the electron density 
and severely affected by sky emission features.
Therefore their associated errors are usually larger than for the  [\ion{O}{ii}] $\lambda\lambda$3726,3729 lines.
More data and a re-analysis of the chemical abundances in galaxies where WR features are detected, with a similar analysis of a sample of non-WR 
galaxies, are needed to get any definitive results. 

\begin{figure*}[t!]
\centering
\begin{tabular}{cc}
\includegraphics[angle=270,width=0.45\linewidth]{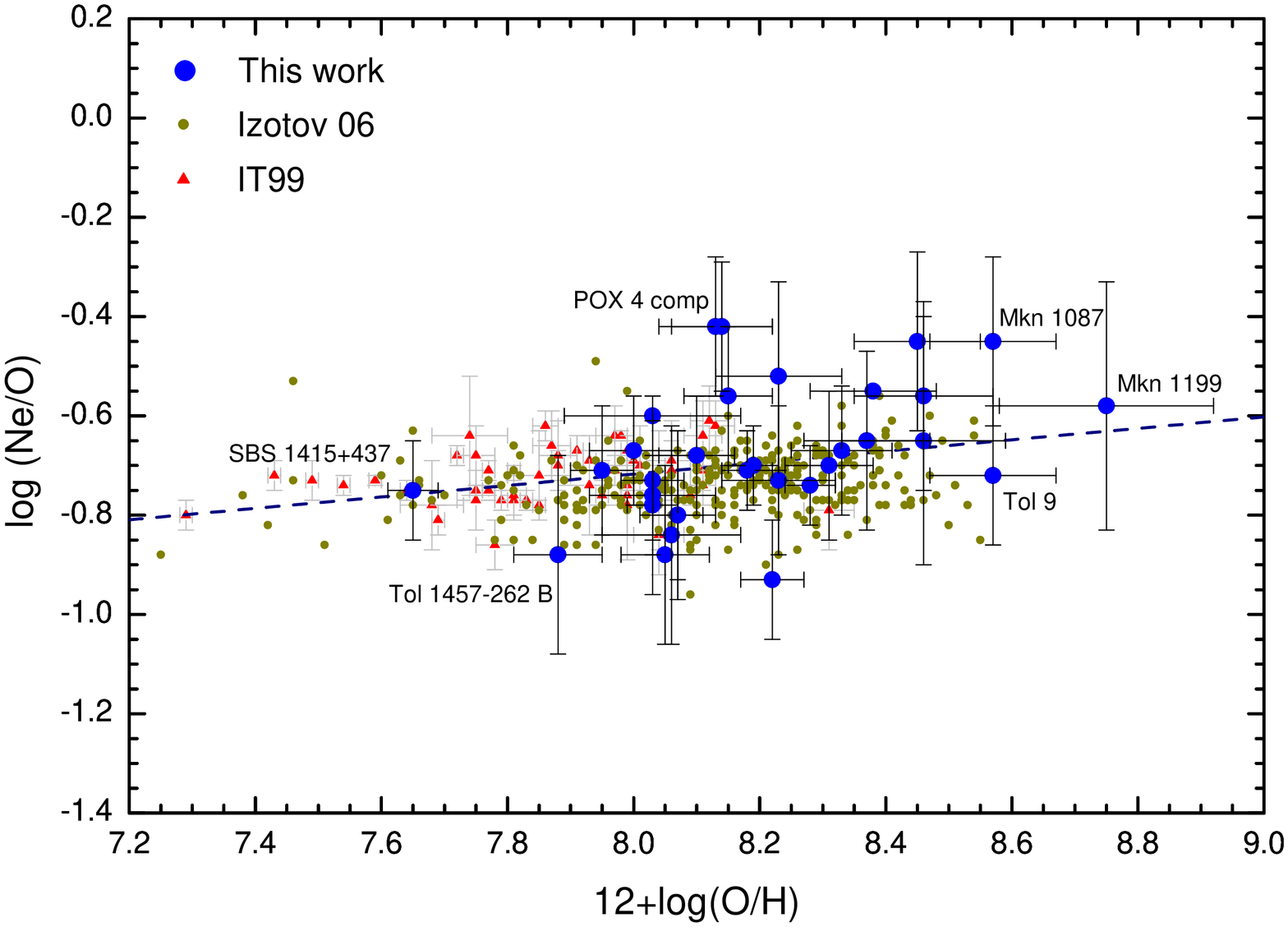} &  
\includegraphics[angle=270,width=0.45\linewidth]{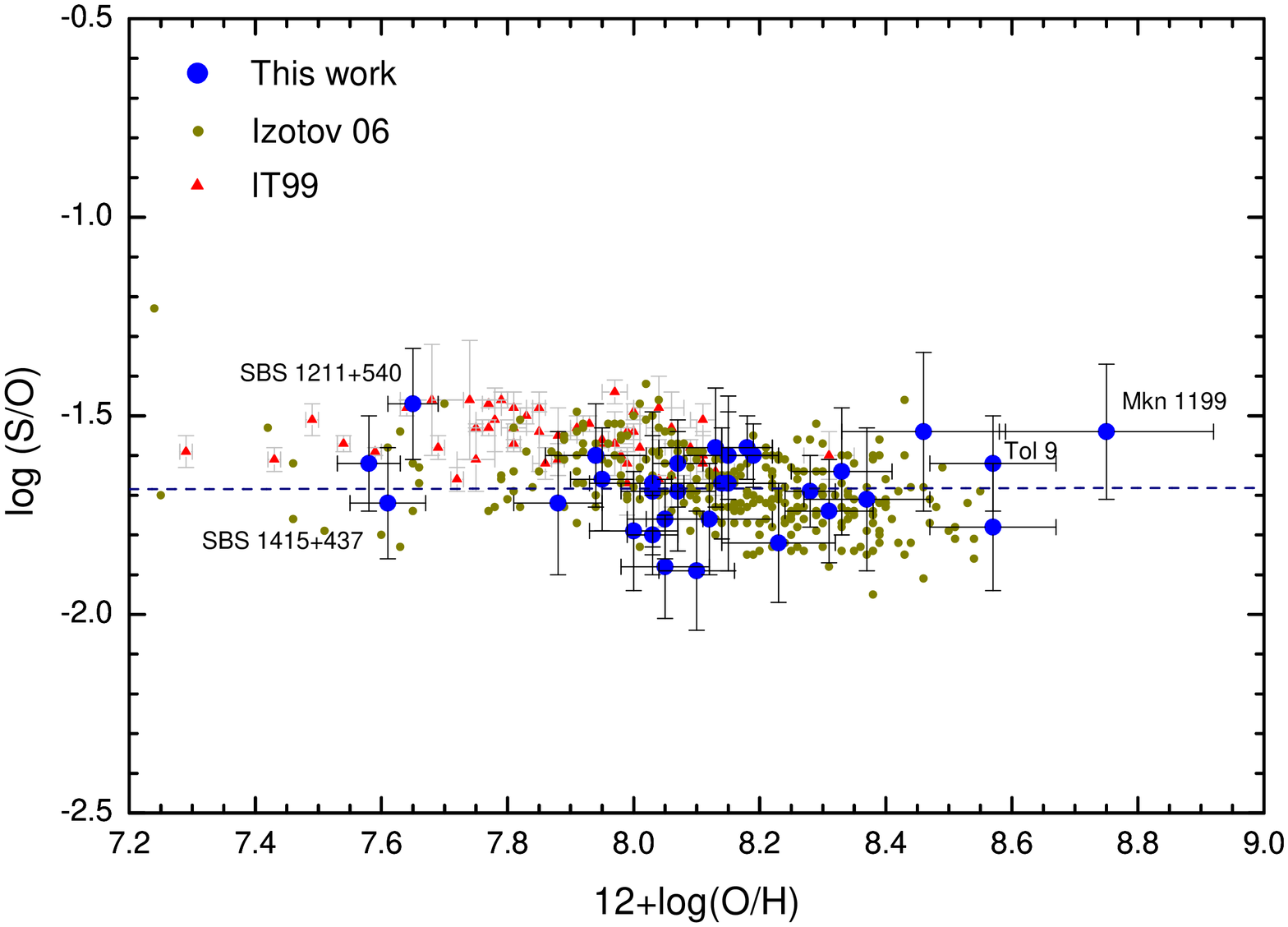} \\  
\includegraphics[angle=270,width=0.45\linewidth]{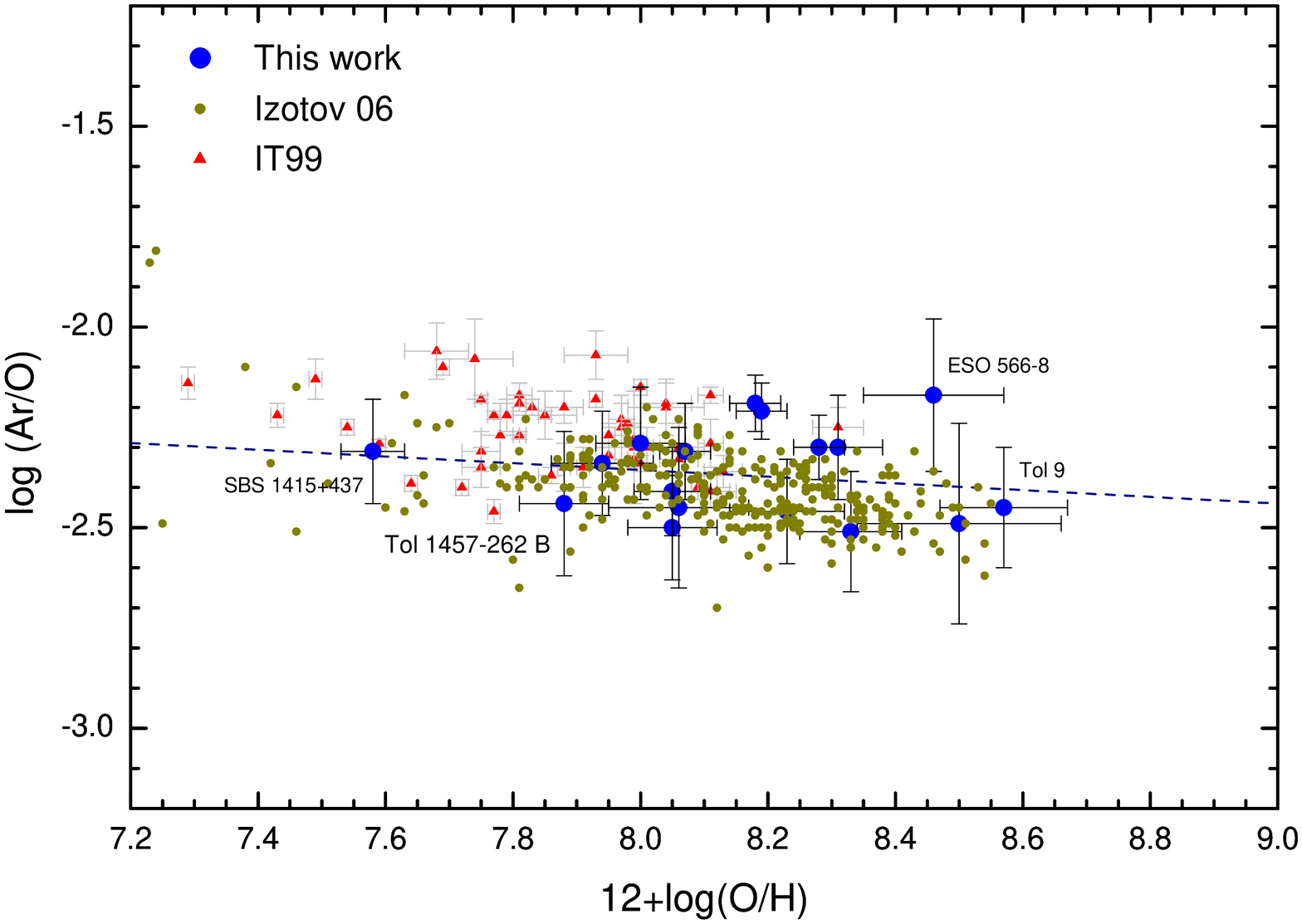} &  
\includegraphics[angle=270,width=0.45\linewidth]{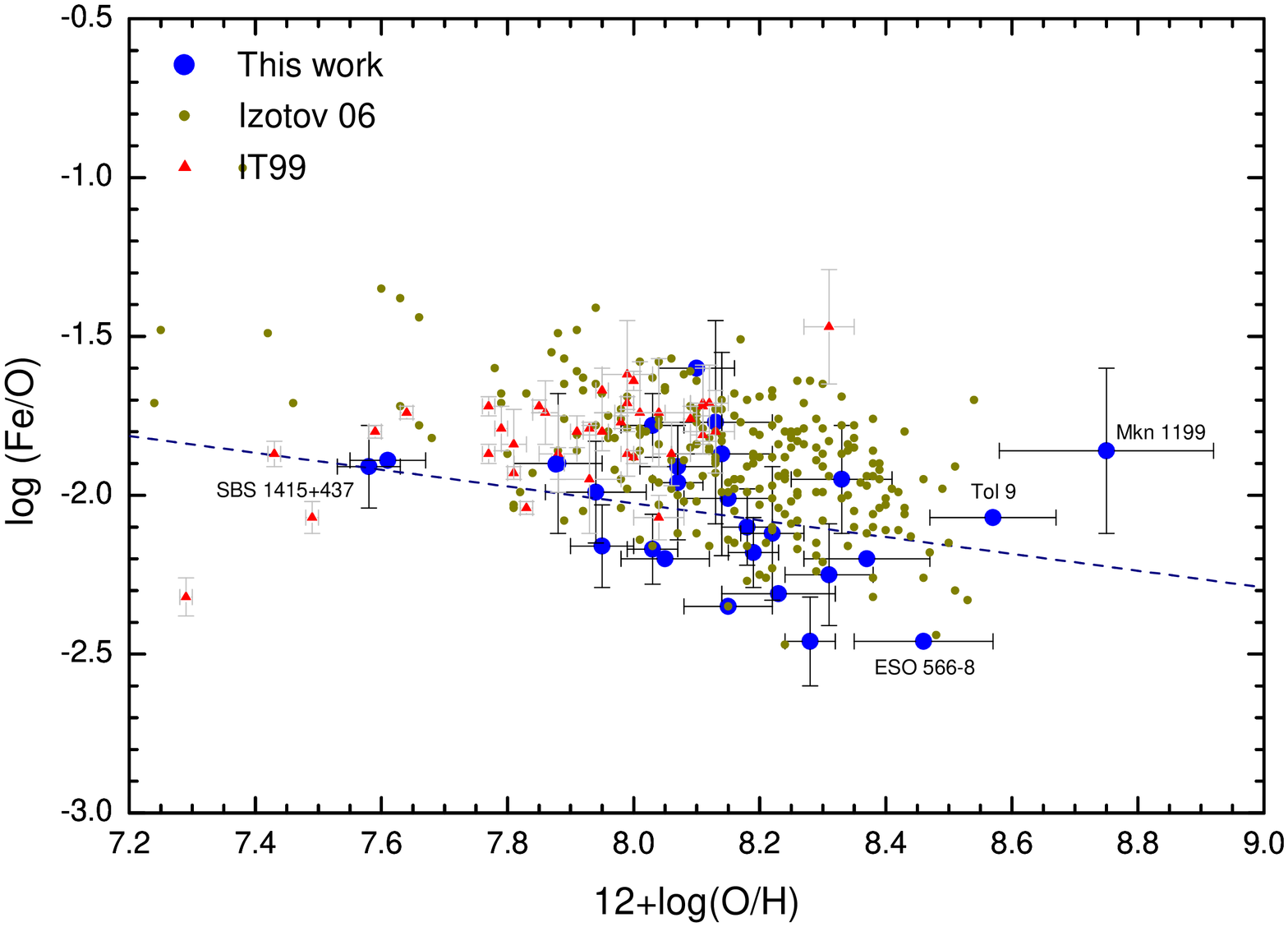} \\  
\end{tabular}
\protect\caption[ ]{\footnotesize{Ne/O, S/O, Ar/O ang Fe/O ratios vs the oxygen abundance for the galaxies analysed in this work. We compare with the 
results provided by the galaxy samples considered by \citet{IT99} and \citet{Izotov06}. The dashed dark-blue lines indicate a fit to our data.}}
\label{compabun2}
\end{figure*}

Some authors (i.e. Izotov et al. 2006) suggest that there is a dependence between the N/O ratio and the \Hb\ equivalent width: the N/O ratio should 
increase with decreasing $|$\WHb$|$.
This trend was also observed by \citet*{BKD08}, but only for objects with $|$\WHb$|\geq 100$ \AA. Figure~\ref{nowhb} plots the N/O ratio vs. \WHb\ 
for the objects analysed here. In this figure we distinguish between objects with high N/O ratio for their oxygen abundance and WR features 
(yellow diamonds), and galaxies with a normal N/O ratio with (red squares) or without (blue circles) WR features. 

From Fig.~\ref{nowhb} it is evident that non-WR galaxies only show high N/O ratios when their $|$\WHb$|<$ 50 \AA. Their N/O ratio becomes low for 
larger equivalent widths. 
\citet*{BKD08} suggested that the non-detection of high N/O ratios in objects with small equivalent widths is consistent with very young bursts where 
the WR stars have not yet had a change to enrich the surrounding ISM to a noticeable degree. 
However, this is probably a consequence of both the complex star-formation histories and the high relative importance of the old underlying stellar 
populations in these systems. 

On the other hand it is remarkable that the WR galaxies with $|$\WHb$|<$50 \AA\  show systematically high N/O ratios, but a large dispersion when 
$|$\WHb$|>$ 50 \AA. This dispersion becomes substantially smaller when we consider the N/O ratio 
expected for their O abundance for those objects with a high $\Delta$(N/O).
If the high N/O ratios in these galaxies are produced by the chemical pollution due to winds of WR stars, objects would 
move to the right in Fig.~\ref{nowhb} once the burst is finished
because of the decreasing of $|$\WHb$|$. 
If the chemical pollution is very localized, we should also expect 
that objects with a N excess would move towards lower values of the N/O ratio as the fresh released material is dispersed and mixed with the 
existing gas of the galaxy.
Hence, detailed analyses of galaxies with a localized high N/O ratio, such as we performed in NGC~5253 \citep{LSEGRPR07}, are fundamental to 
solve these unsolved questions, which will definitely be key elements to the evolution of the galaxies.

\begin{table*}[t!]
\centering
  \caption{\footnotesize{Results of the comparison between the oxygen abundance given by several empirical calibrations and the oxygen abundance derived here following the direct (\Te) method.}}
  \label{dispempi}
  \begin{tabular}{l@{\hspace{8pt}}  c@{\hspace{8pt}}c@{\hspace{8pt}}c@{\hspace{8pt}} c@{\hspace{8pt}}c@{\hspace{8pt}} c@{\hspace{8pt}}  c 
c@{\hspace{8pt}}  c@{\hspace{8pt}} c@{\hspace{8pt}} c c@{\hspace{8pt}} c@{\hspace{8pt}} } 
  \tableline
   \noalign{\smallskip}
Parameter   & \multicolumn{6}{c@{\hspace{4pt}}}{$R_{23}$} & & \multicolumn{3}{c@{\hspace{4pt}}}{$N_2$} & & 
\multicolumn{2}{c@{\hspace{4pt}}}{$O_3N_2$} \\

   \cline{2-7} \cline{9-11} \cline{13-14}
         \noalign{\smallskip}
		 
Calibration$^a$ &  P01   & PT05  &  N06  & M91  & KD02 & KK04  & & D02  &  PP04 &   N06 & & PP04  &  N06    \\ 

\tableline
\noalign{\smallskip}    
	Average$^b$ &  0.07  & 0.08 &  0.14  & 0.15 & 0.28 & 0.27  & & 0.14    & 0.12    &  0.18 & & 0.12 & 0.21    \\ 
 $\sigma^c$   &  0.05  & 0.07 &  0.12  & 0.11 & 0.18 & 0.13  & & 0.10    & 0.10    &  0.14 & & 0.10 & 0.16   \\ 
 Notes$^d$      & B/A (1)& B/A  &   (2)  & S.H. & S.H. & S.H.  & & S.H. (3)& B/A (4) &  S.L.(5)  & & B/A (6) & S.L.      \\ 
\noalign{\smallskip}
\tableline
  \end{tabular}
  \begin{flushleft}
  $^a$ The names of the calibrations are the same as in Table~\ref{abempirica2}.\\
  $^b$ Average value (in absolute values) of the difference between the abundance given by empirical calibrations 
and that obtained using the direct method. The names of the calibrations are the same as in Table~\ref{abempirica2}. \\
 $^c$ Dispersion (in absolute values) of the difference between the abundance given by empirical calibrations and that obtained using the direct method.\\
 $d$ We indicate if the empirical calibration gives results both below and above the direct value (B/A), if they are systematically higher than the direct value (S.H.) or if they are systematically lower than the direct value (S.L.). Some additional notes are: \\ 
  (1) Higher deviation in the low branch. \\ 
  (2) This calibration provides lower oxygen abundances in low-metallicity regions and higher oxygen abundances in high-metallicity regions.\\ 
  (3) Systematically higher only for \abox$>$8.2. \\ 
  (4) Higher deviation for  \abox$<$8.0. Considering \abox$>$8.0, we get average=0.08 and $\sigma$=0.06. \\ 
  (5) Higher deviation at lower oxygen abundances. \\
  (6) Higher deviation for \abox$<$8.0. Assuming \abox$>$8.0, average=0.09 and $\sigma$=0.06. 
  \end{flushleft}
\end{table*}

\subsection{The $\alpha$-elements to oxygen ratio}

Figure~\ref{compabun2} plots the Ne/O, S/O, Ar/O, and Fe/O ratios as a function of \abox\ for all objects analysed in this work. We compare our 
observational results with those found in samples with similar characteristics \citep{IT99,Izotov06}. We remark again that although the estimates 
of our errors are higher than those provided by other samples, they are not a consequence of the quality of our data but derive from the different 
formalism we used to estimate the uncertainties.
In any case, our results completely agree with those previously reported in the literature.  
The relative abundance ratios of the $\alpha$-elements
(neon, sulfur, and argon) to oxygen are approximately constant,
as expected because all four elements (O, Ne, S and Ar) are mainly produced by massive stars. 
For our sample, the mean $\log$(Ne/O), $\log$(S/O), and $\log$(Ar/O) ratios are $-0.70\pm0.13$, $-1.68\pm0.10$  and $-2.37\pm0.12$, respectively. 
These values are comparable to the values reported for starbursting dwarf galaxies
(e.g., Izotov \& Thuan, 1999; Izotov et al. 2006) and for other dwarf irregular galaxies (e.g., van Zee et al. 1998, van Zee \& Haynes 2006).
For example, \citet{IT99} found mean values of $-0.72\pm0.06$, $-1.56\pm0.06$,  and $-2.26\pm0.09$ for the $\log$(Ne/O), $\log$(S/O), and 
$\log$(Ar/O) ratios,  respectively. 

Despite their higher uncertainties, we notice that the Ne/O ratio seems to increase slightly with increasing oxygen abundance. This effect was 
reported by \citet{Izotov06}, who interpreted it as due to a moderate depletion of oxygen onto grains in the most metal-rich galaxies. \citet{Verma03}  
observed an underabundance of sulphur in relatively high-metallicity starburst galaxies. These authors interpreted this effect as a consequence of 
the depletion of sulphur onto dust grains. However, we do not see any underabundance of sulphur; quite the opposite, the sulphur abundance is well 
correlated with both neon and argon abundances.

\begin{figure*}[t!]
\centering
\begin{tabular}{cc}
\includegraphics[angle=270,width=0.4\linewidth]{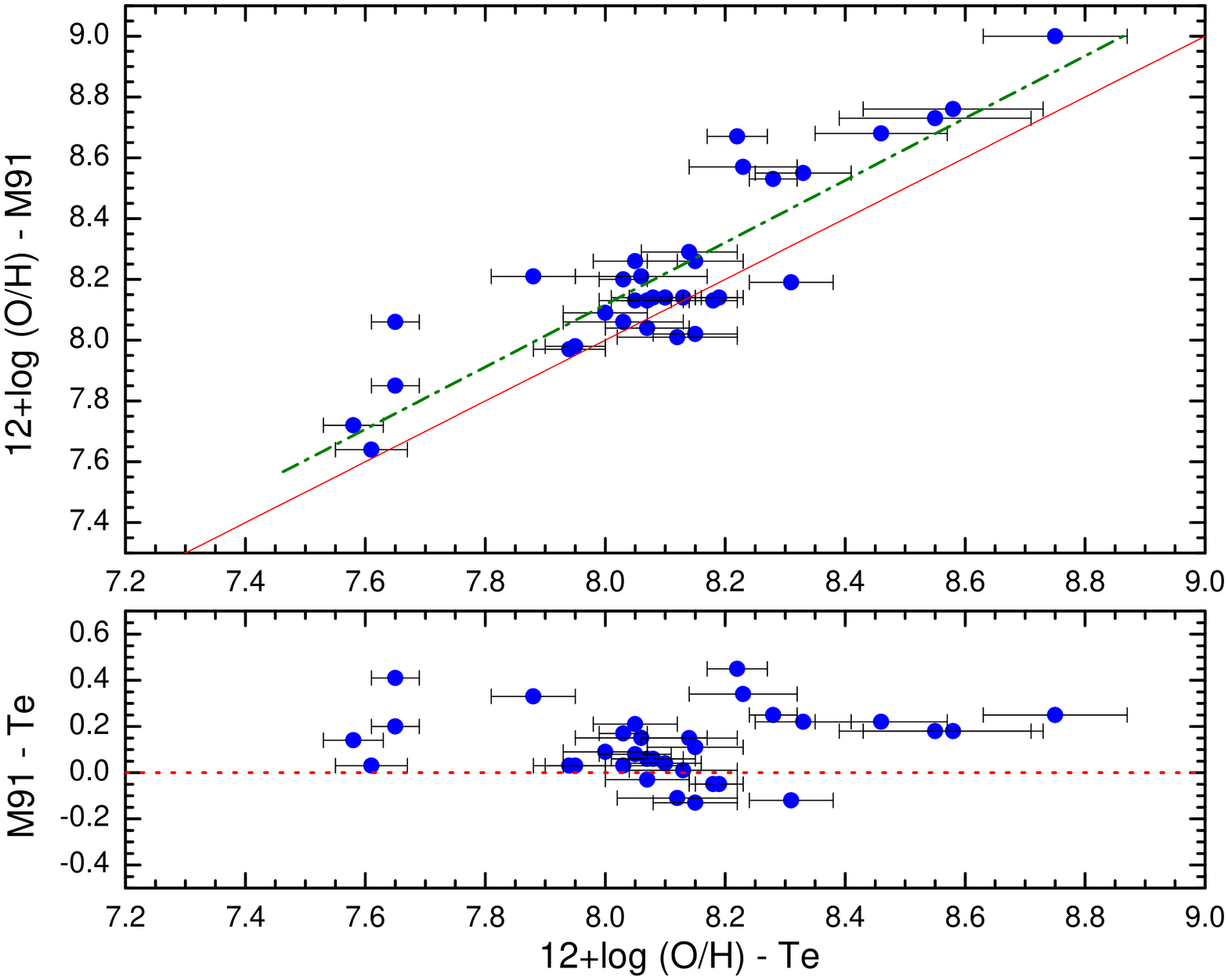} &  
\includegraphics[angle=270,width=0.4\linewidth]{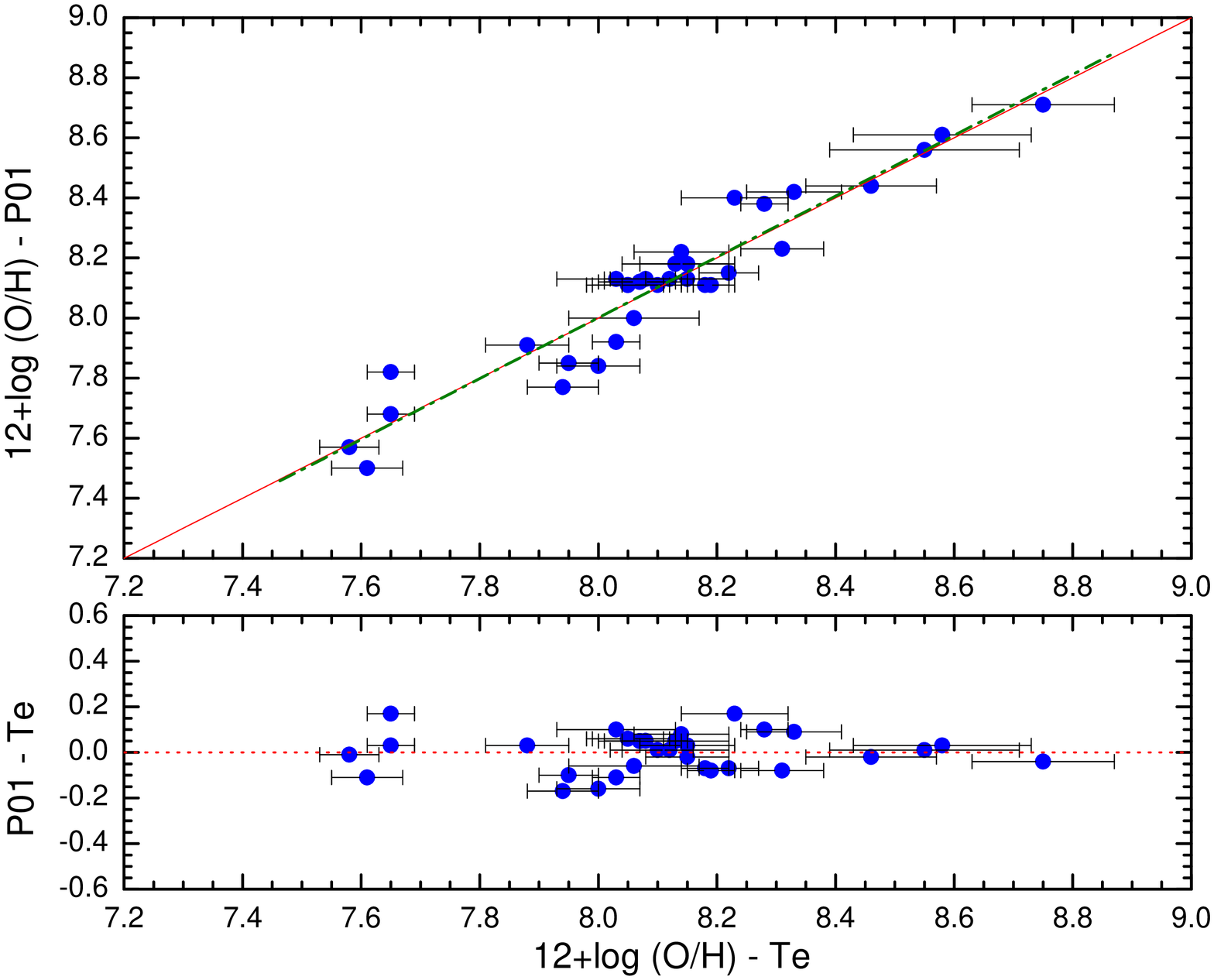} \\  
\includegraphics[angle=270,width=0.4\linewidth]{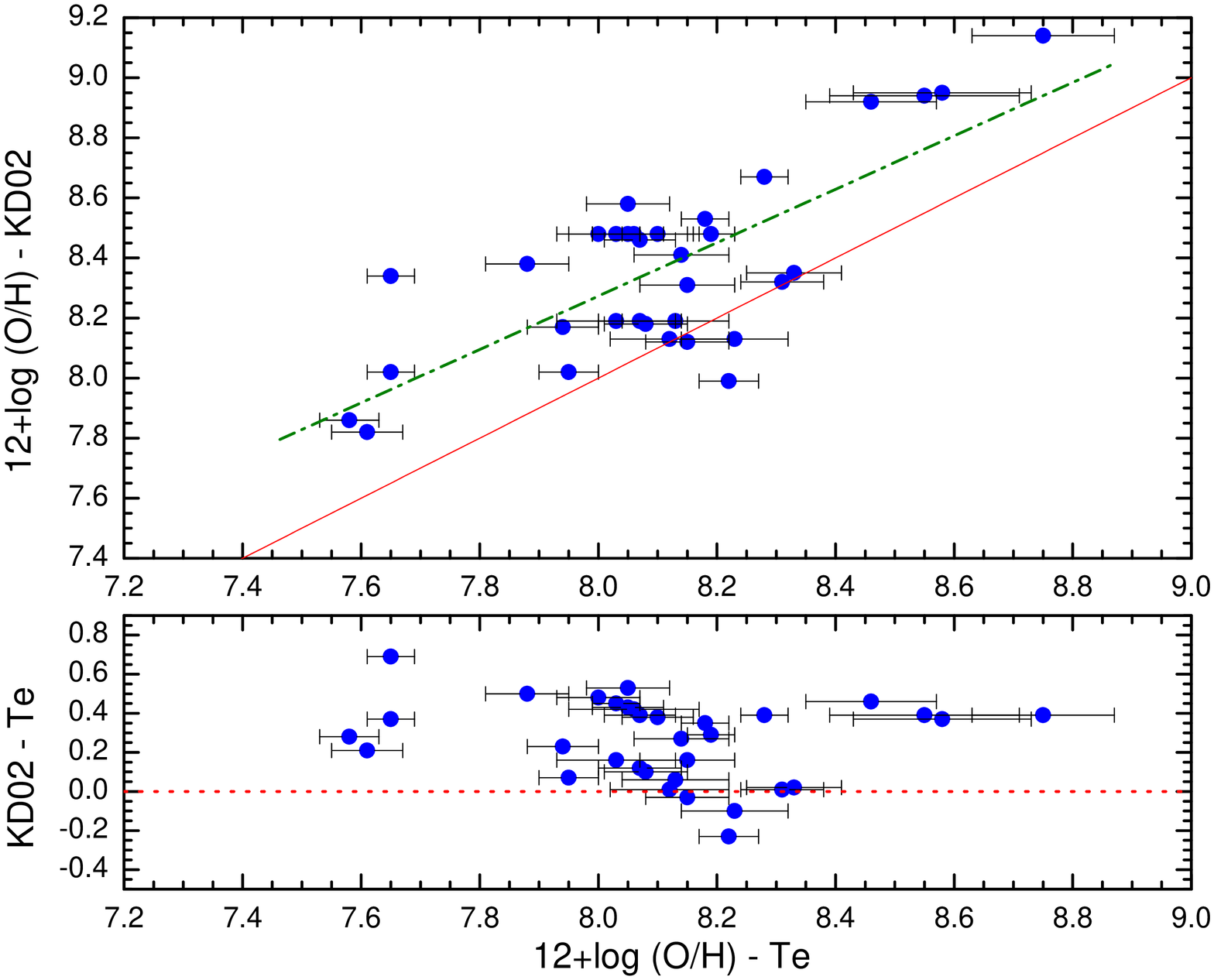} &  
\includegraphics[angle=270,width=0.4\linewidth]{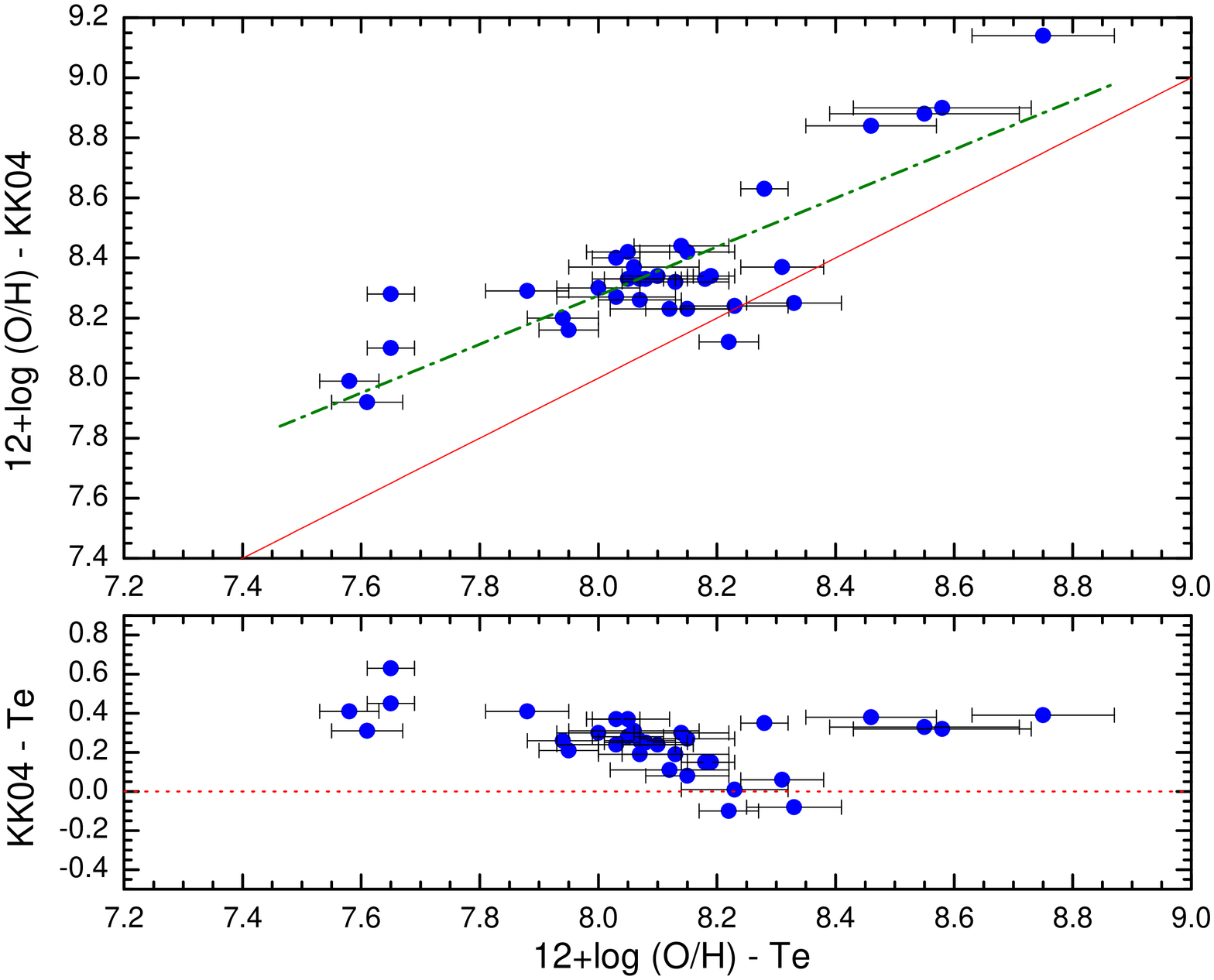} \\  
\includegraphics[angle=270,width=0.4\linewidth]{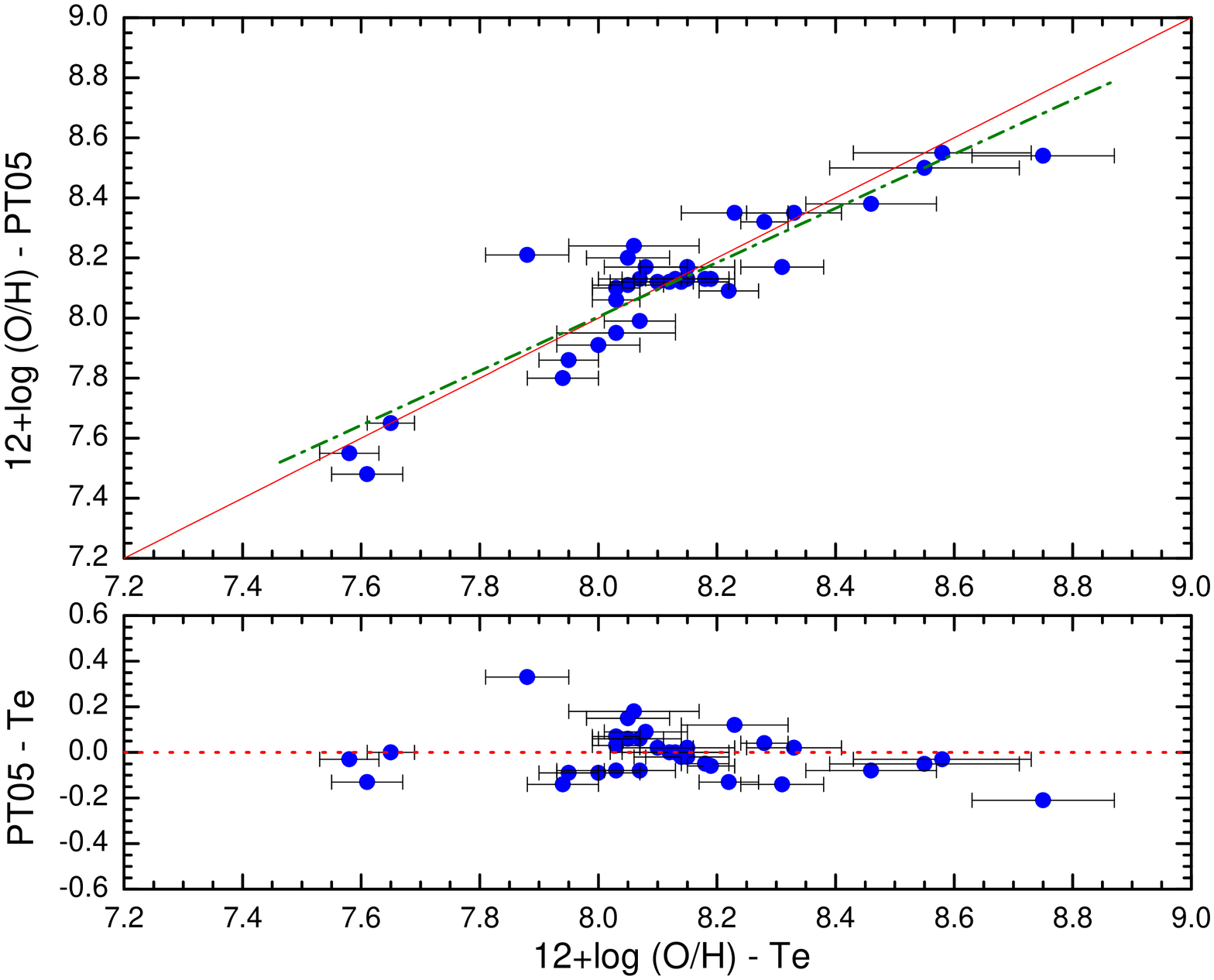} &  
\includegraphics[angle=270,width=0.4\linewidth]{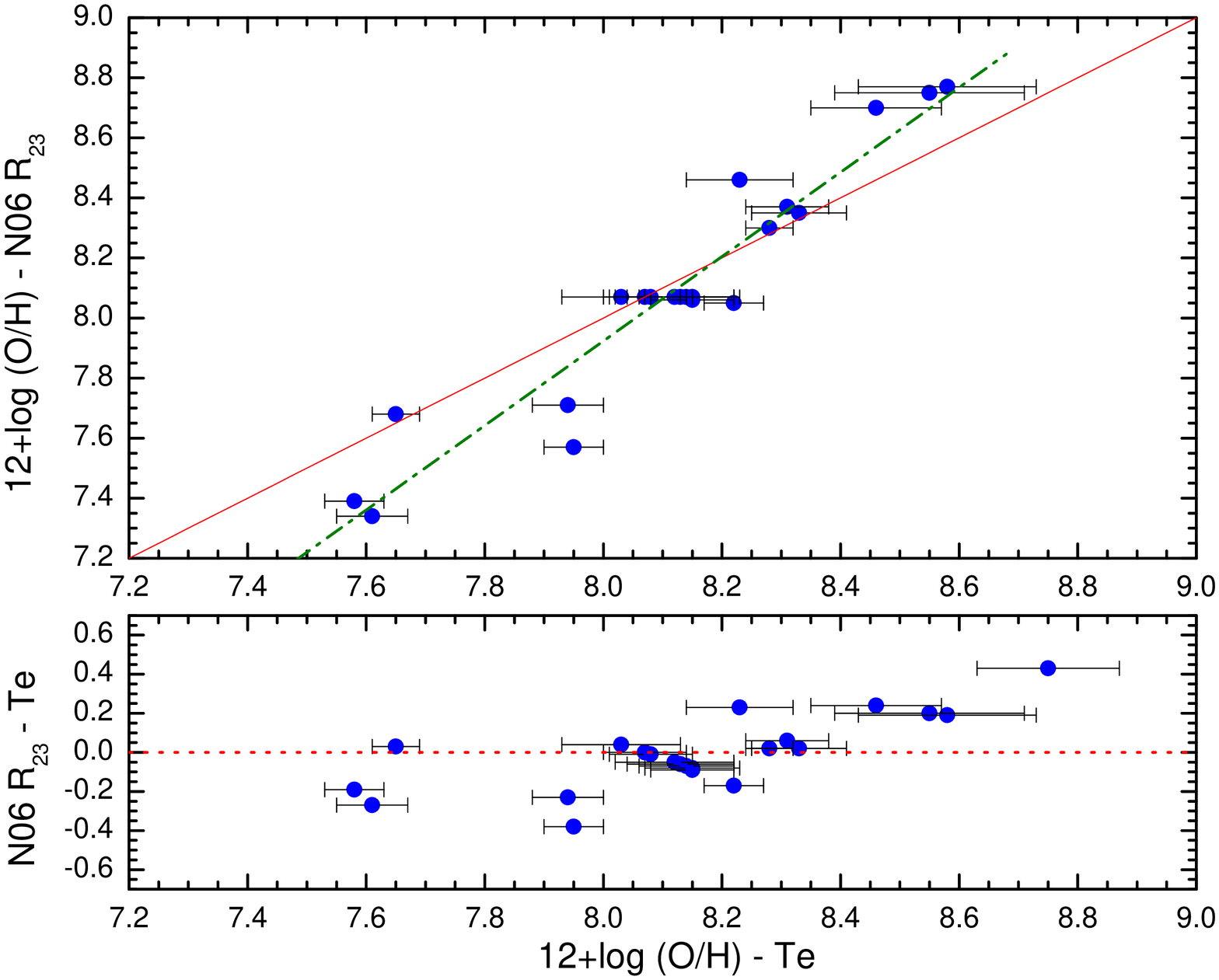} \\  
\end{tabular}
\protect\caption[ ]{\footnotesize{Comparison between the oxygen abundances derived using the direct method (\Te, always plotted in the $x-axis$) with 
those estimated using six different empirical calibrations that consider the $R_{23}$ parameter: M91: \citet{McGaugh91}; P01: \citet{P01a,P01b}; 
KD02: \citet{KD02}; KK04: \citet{KK04}; PT05: \citet{PT05}; N06: \citet*{Nagao06}. The bottom panel of each diagram indicates the difference between 
empirical and direct data.}}
\label{figemp}
\end{figure*}

On the other hand, it seems that the Fe/O ratio slightly decreases with increasing oxygen abundance. This effect was previously observed by 
\citet{Izotov06}, who suggested that it may be a consequence of the iron depletion in dust grains, which is more important for galaxies with higher 
metallicities.  

Finally we remark how important is to use high-quality data and a proper estimate of all the physical parameters (including \Te\ via the direct 
method) in an homogeneous sample of objects to get reasonable conclusions about these topics (i.e., this work; H\"agele et al. 2008).

\subsection{Comparison with empirical calibrations}

  
We used the data of the 31 regions for which we have a direct estimate of \Te\ and, hence, a direct estimate of the oxygen abundance, to 
check the reliability of several empirical calibrations.
A recent review of 10 metallicity calibrations, including theoretical and empirical methods, was presented by \citet{KE08}.
Appendix~A gives an overview of the most common empirical calibrations and defines the typical parameters that are used to estimate the oxygen 
abundance following these relations. These parameters are ratios between bright emission lines, the most commonly used are $R_{23}$, $P$, $y$, $N_2$, 
and $O_3N_2$ (see definitions in Appendix~A).  
Table~\ref{abempirica1} lists the values of all these parameters derived for each region with a direct estimate of the oxygen abundance (see 
Paper~II for details). Table~\ref{abempirica1} also includes the value derived for the $q$ parameter (in units of cm s$^{-1}$) obtained from the 
optimal calibration provided by \citet{KD02}. The results for the oxygen abundances derived for each object and empirical calibration are listed in 
Table~\ref{abempirica2}. This table also indicates the branch (high or low metallicity) considered in each region when using the $R_{23}$ parameter 
(see Appendix~A), although, as is clearly specified in the table, for some objects with 8.00$\leq$\abox$\leq$8.3 we assumed the average value
found for the lower and upper branches.

Looking at the data compiled in Table~\ref{abempirica2} the huge range of oxygen abundance found for the same object using different 
calibrations is evident. As \citet{KE08} concluded, it is critical to use the same metallicity calibration when comparing properties from different data sets or 
investigate luminosity-metallicity or mass-metallicity relations. 
Furthermore, abundances derived with such strong-line methods may be significantly biased if the objects under study have different structural 
properties (hardness of the ionizing radiation field, morphology of the nebulae) than those used to calibrate the methods \citep{Stasinska09}.

\begin{figure*}[t!]
\centering
\begin{tabular}{cc}
\includegraphics[angle=270,width=0.4\linewidth]{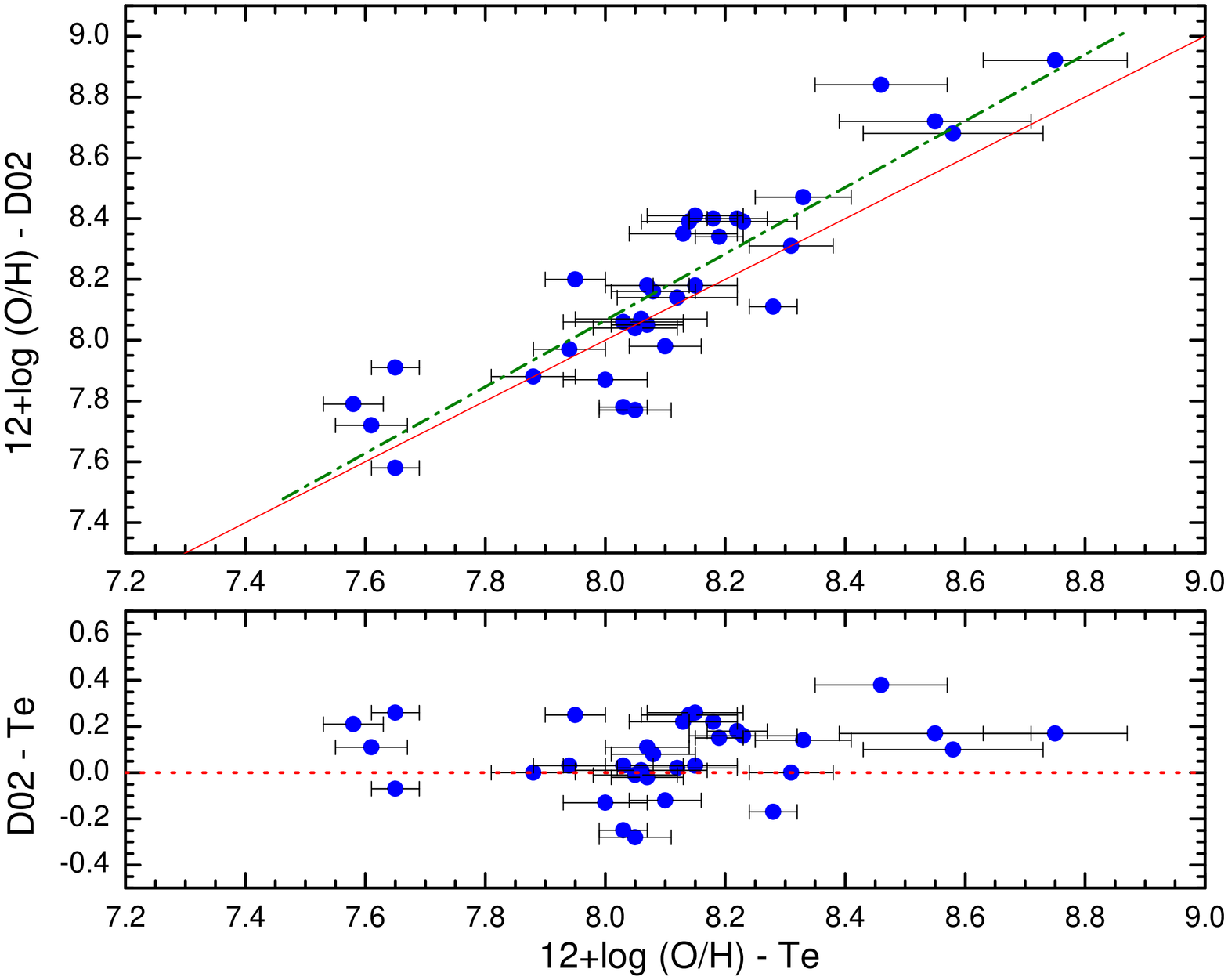} &  
\includegraphics[angle=270,width=0.4\linewidth]{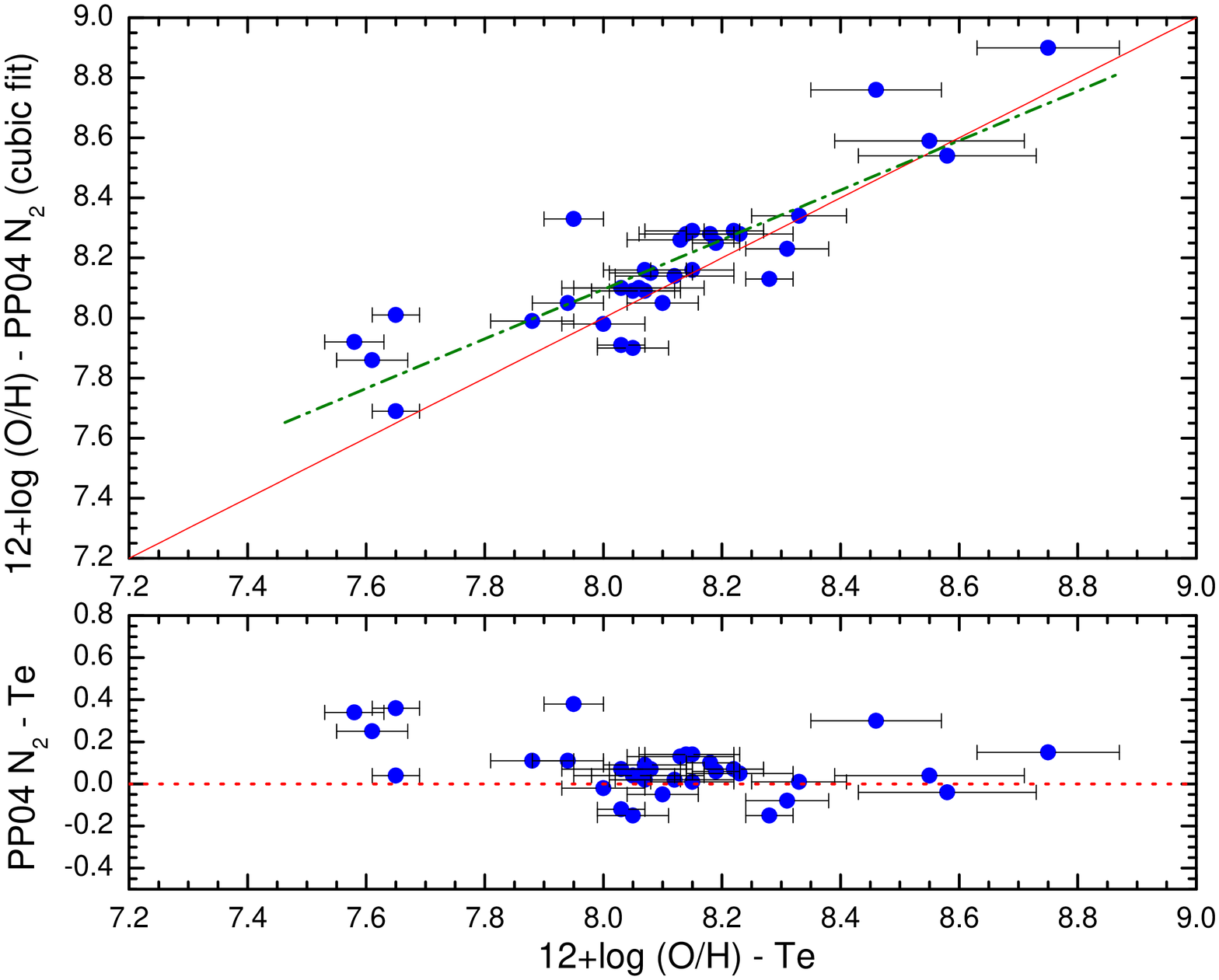} \\  
\includegraphics[angle=270,width=0.4\linewidth]{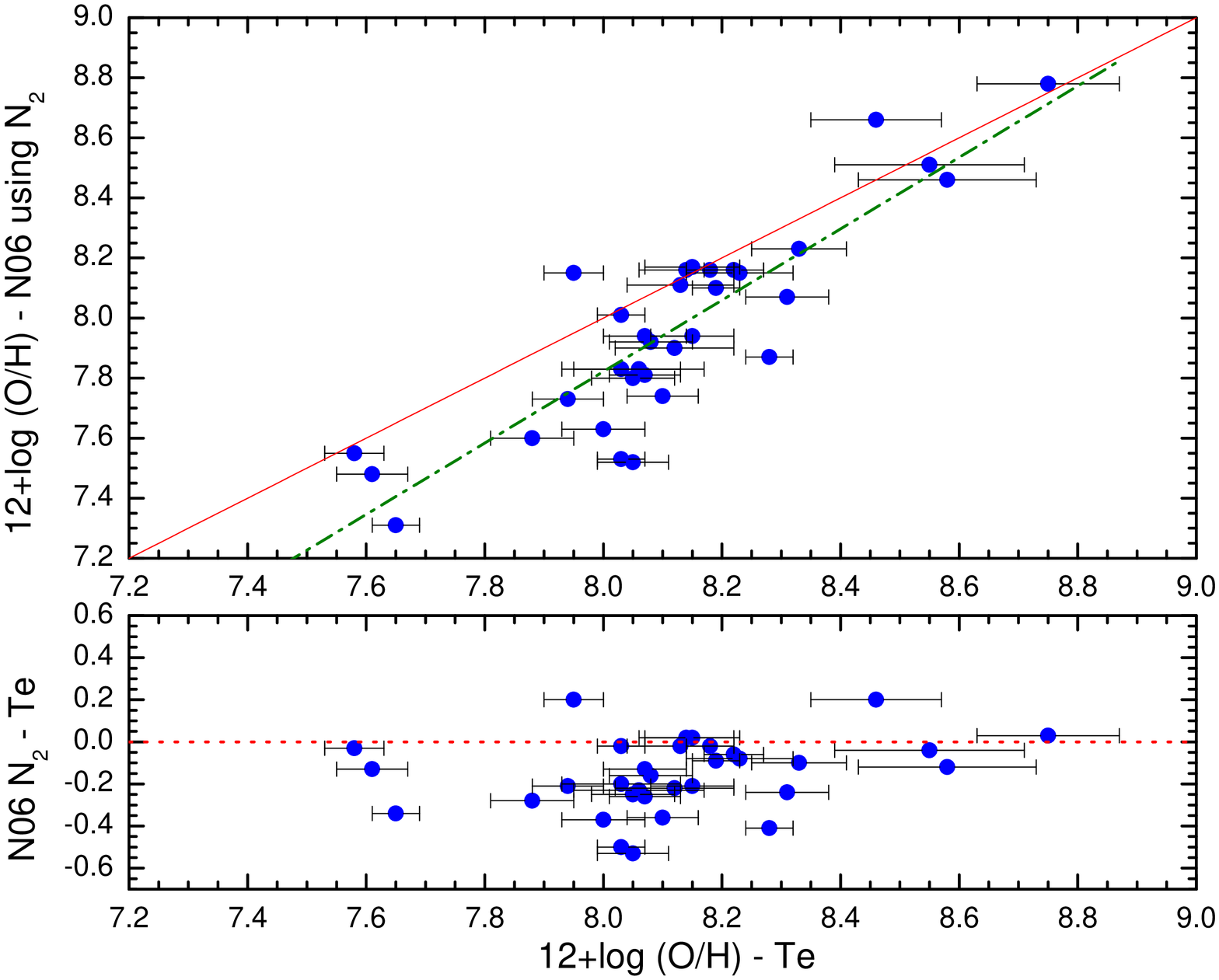} &  
\includegraphics[angle=270,width=0.4\linewidth]{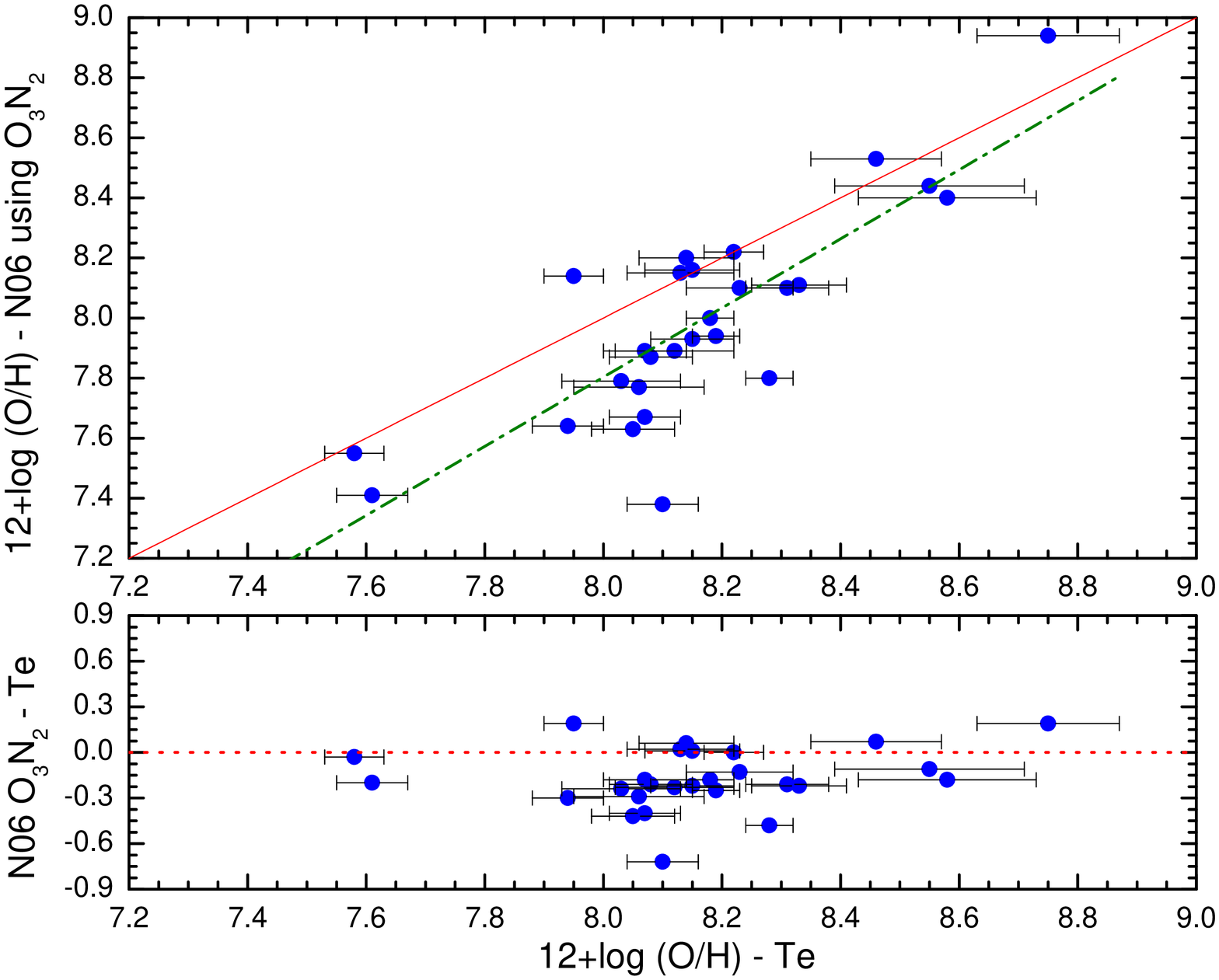} \\  
\end{tabular}
\protect\caption[ ]{\footnotesize{Comparison between the oxygen abundances derived using the direct method (\Te, in the $x-axis$) with those computed 
using the calibrations using the $N_2$ parameter --D02: \citet{D02}; PP04b: \citet{PP04} considering their cubic fit to $N_2$; N06: 
\citet*{Nagao06}-- and the O$_3$N$_2$ parameter following the cubic fit provided by \citet*{Nagao06}. The bottom panel of each diagram indicates the 
difference between empirical and direct data.}}
\label{figemp2}
\end{figure*}

Figures~\ref{figemp} and \ref{figemp2} plots the ten most common calibrations and their comparison with the oxygen abundance obtained using the 
direct method.
We performed a simple statistic analysis of the results to quantify the goodness of these empirical calibrations. Table~\ref{dispempi} compiles 
the average value and the dispersion (in absolute values) of the difference between the abundance given by empirical calibration and that obtained 
using the direct method. We check that the empirical calibration that provides the best results is that proposed by \citet{P01a,P01b}, which gives 
oxygen abundances very close to the direct values (the differences are lower than 0.1 dex in the majority of the objects), and furthermore it 
possesses a low dispersion. 
We note however that the largest divergences found using this calibration are in the low-metallicity regime. The update of this calibration 
presented by \citet{PT05} seems to partially solve this problem, the abundances provided by this calibration also agree very well with those derived 
following the direct method. We therefore conclude that the \citet{PT05} calibration is nowadays the best suitable method to derive the oxygen 
abundance of star-forming galaxies when auroral lines are not observed.    

On the other hand, the results given by the empirical calibrations provided by \citet{McGaugh91}, \citet{KD02} and \citet{KK04}, that are based on 
photoionization models, are systematically higher than the values derived from the direct method. This effect is even more marked in the last two 
calibrations, which usually are between 0.2 and 0.3 dex higher than the expected values. These empirical calibrations also have a higher dispersion 
than that estimated for \citet{P01a,P01b} or \citet{PT05} calibrations. \citet{Yin+07} also found high discrepancies when comparing the theoretical 
metallicities using the theoretical models of \citet{Tremonti04} with the \Te-based metallicites obtained from \citet{P01a,P01b} and \citet{PT05}.

One of the possible explanations for the different metallicities obtained between the direct method and those derived from the empirical calibrations 
based on photoionization models are temperature fluctuations in the ionized gas. Temperature gradients or fluctuations indeed cause 
the true metallicities based on the \Te-method to be underestimated (i.e. Peimbert 1967; Stasinska 2002,2005; Peimbert et al. 2007). 
Temperature fluctuations can also explain our results for NGC~5253 \citep{LSEGRPR07}: the ionic abundances of O$^{++}$/H$^+$ and 
C$^{++}$/H$^+$ derived from recombination lines are systematically 0.2 -- 0.3 dex higher than those determined from the direct method --based on the 
intensity ratios of collisionally excited lines. This abundance discrepancy has been also found in Galactic \citep{GRE07} and other extragalactic 
\citep{Esteban09} \HII regions and interestingly this discrepancy is in all cases of the same order as the differences between abundances 
derived from the direct methods and empirical calibrations based on photoionization models.

The conclusion that temperature fluctuations do exist in the ionized gas of starburst galaxies is very important for the analysis of the chemical 
evolution of galaxies and the Universe. Indeed, if that is correct, the majority of the abundance determinations in extragalactic objects following the 
direct method, including those provided in this work, have been underestimated by at least 0.2 to 0.3 dex. Deeper observations of a large 
sample of star-forming galaxies --that allow us to detect the faint recombination lines, such as those provided by \citet{Esteban09}-- and more 
theoretical work --including a better understanding of the photoionization models, such as the analysis provided by \citet{KE08}-- are needed to confirm 
this puzzling result.

On the other hand, we have checked the validity of the recent relation provided by \citet*{Nagao06}, which merely considers a cubic fit between the 
$R_{23}$ parameter and the oxygen abundance. This calibration was obtained combining data from several large galaxy samples, the majority from the 
SDSS, which includes all kinds of star-forming objects. As it is clearly seen in Table~\ref{abempirica2} and in Fig.~\ref{figemp2}, the 
\citet*{Nagao06} relation is not suitable to derive a proper estimate of the oxygen abundance for the majority of the objects in our galaxy sample. 
In general, this calibration provides lower oxygen abundances in low-metallicity regions and higher oxygen abundances in high-metallicity regions. 
Objects located in the metallicity range 8.00$\leq$\abox$\leq$8.15 have systematically 12+log(O/H)$_{\rm N06}\sim$8.07 because we have to use an 
average value between the low and the high branches. Furthermore, many of the regions do not have a formal solution to the \citet*{Nagao06} equation, 
such as the maximum value for $R_{23}$ is 8.39 at \abox=8.07. We consider that the use of an ionization parameter --$P$ as introduced by 
\citet{P01a,P01b} or $q$ as followed by \citet{KD02}-- is fundamental to obtain a real estimate of the oxygen abundance in star-forming galaxies, 
especially in objects showing strong starbursts. In the same sense, the direct method and not the formulae provided by \citet{Izotov06} (which assumes 
a low-density approximation in order not to have to solve the statistical equilibrium equations of the O$^{+2}$ ion) provides a good approximation to 
the actual oxygen abundance when the auroral line [\ion{O}{iii}] $\lambda$4653 is observed.

Empirical calibrations considering a linear fit to the $N_2$ ratio \citep{D02,PP04} give results that are systematically $\sim$0.15 dex higher that 
the oxygen abundances derived from the direct method. The difference is higher at higher metallicities. 
We do not consider that this trend is a consequence of comparing different objects: both \citet*{D02} and \citet{PP04} calibrations are 
obtained using a sample of star-forming galaxies similar to those analysed in this work, many of which are WR galaxies. \citet*{D02} compared the 
$N_2$ ratio with the ionization parameter together with the results of photoionization models and concluded that most of the observed trend of $N_2$ 
with the oxygen abundance is caused by metallicity changes.
The cubic fit to $N_2$ performed by \citet{PP04} better reproduces the oxygen abundance, especially in the intermediate- and high-metallicity regime  
(\abox$>$8.0), where it has an average error of $\sim$0.08 dex. However, the cubic fit to $N_2$ provided by \citet*{Nagao06} gives systematically 
lower values for the oxygen abundance than those derived using the direct method, having an average error of $\sim$0.18 dex. 

The empirical calibration between the oxygen abundance and the $O_3N_2$ parameter proposed by \citet{PP04} gives acceptable results for objects with 
\abox$>$8.0, with the average error $\sim$0.1 dex. However, the new relation provided by \citet*{Nagao06} involving the $O_3N_2$  parameter gives 
systematically lower values for the oxygen abundance. As we commented before, we consider that the \citet*{Nagao06} calibrations are not suitable for 
studying galaxies with strong star-formation bursts. Their procedures must be taken with caution, galaxies with different ionization parameters, 
different chemical evolution histories, and different star formation histories should have different relations between the bright emission lines and 
the oxygen abundance. This issue is even more important when estimating the metallicities of intermediate- and high-redshift galaxies, because the 
majority of their properties are highly unknown.

\section{Metallicity-luminosity relations}

\begin{figure*}[t]
\centering
\includegraphics[angle=270,width=\linewidth]{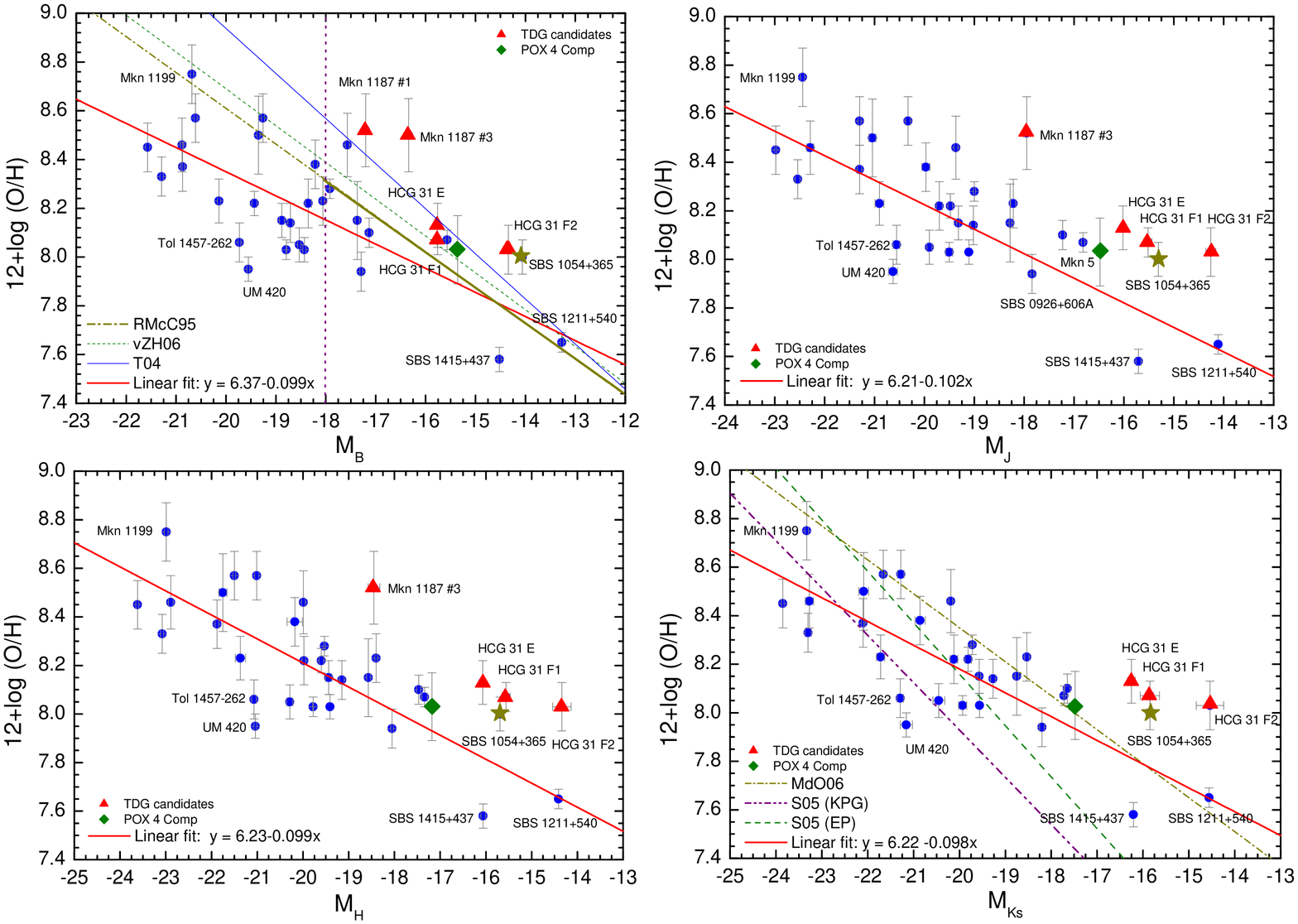}  
\protect\caption[ ]{\footnotesize{Metallicity-luminosity diagrams for the objects analysed in this work. The oxygen abundance is expressed in units 
of \abox; the luminosity is expressed as the absolute magnitude in $B$, $J$, $H$ and $K_S$ filters. Red triangles represent the \TDG\ candidates 
found in HCG~31 \citep{LSER04a} and Mkn~1087 \citep{LSER04b}. A green diamond corresponds to the dwarf object surrounding POX~4, while a yellow star 
indicates the galaxy SBS~1054+365. Linear fits to our data are shown with a continuous red line in the four panels. The $M_B$--O/H diagram includes 
the relation derived by \citet{RM95} (continuous dark yellow line) extrapolated to high luminosities (dashed-dotted dark yellow line), the relation 
provided by \citet{vZH06} (dashed green line), and the relation derived by \citet{Tremonti04} using SDSS data (dotted blue line). The $M_{K_S}$--O/H 
diagram includes the two relations found by \citet{Salzer05}, one considering the \citet{EP84} calibration (EP, dashed green line) and another 
assuming the relationships provided by \citet{Kennicutt03} (KPG, dashed-dotted purple line), and the relation found by \citet{MO06} (dashed-dotted 
yellow line).}}
\label{LMg}
\end{figure*}

\begin{figure*}[t]
\begin{tabular}{cc}
\includegraphics[angle=270,width=0.45\linewidth]{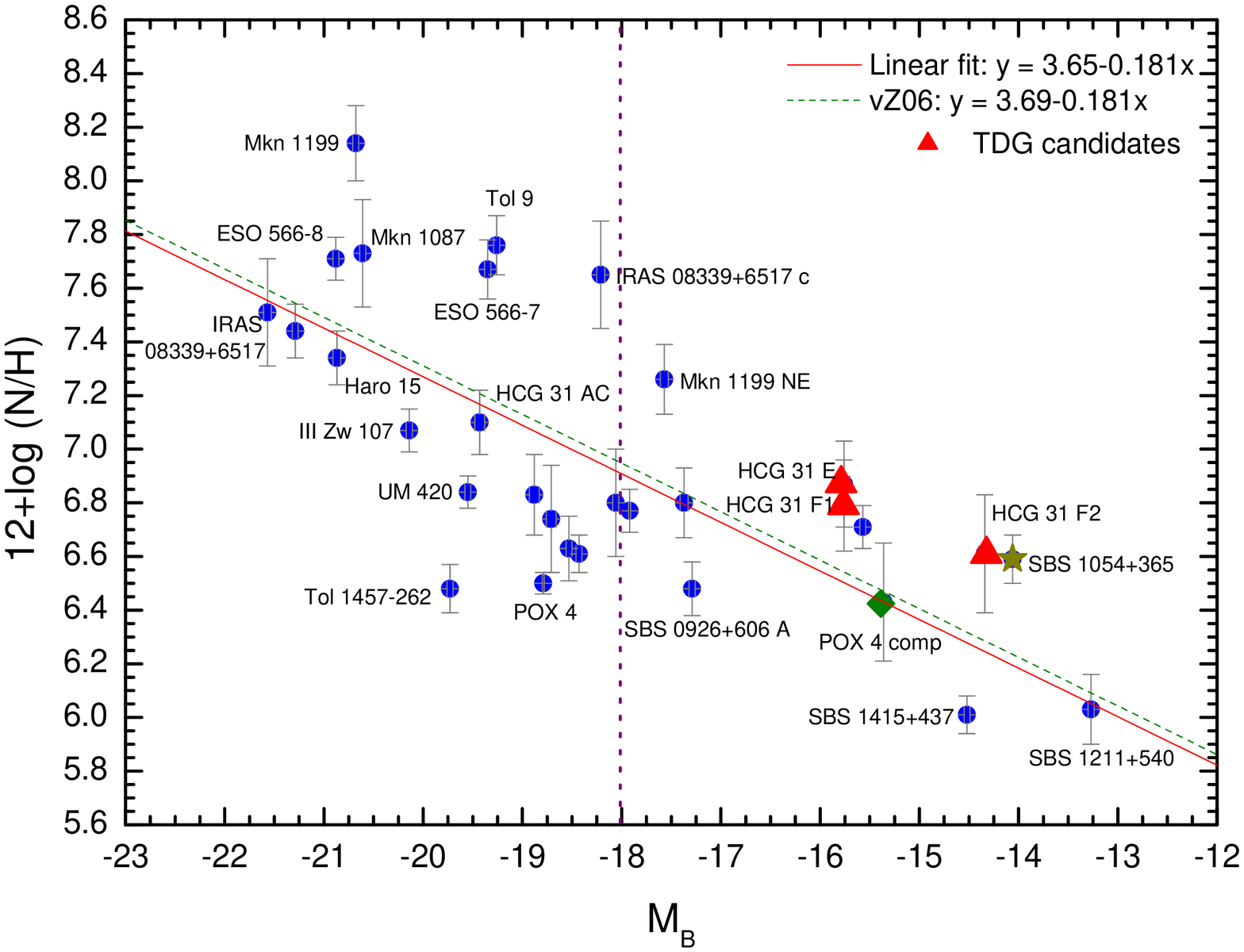} &  
\includegraphics[angle=270,width=0.45\linewidth]{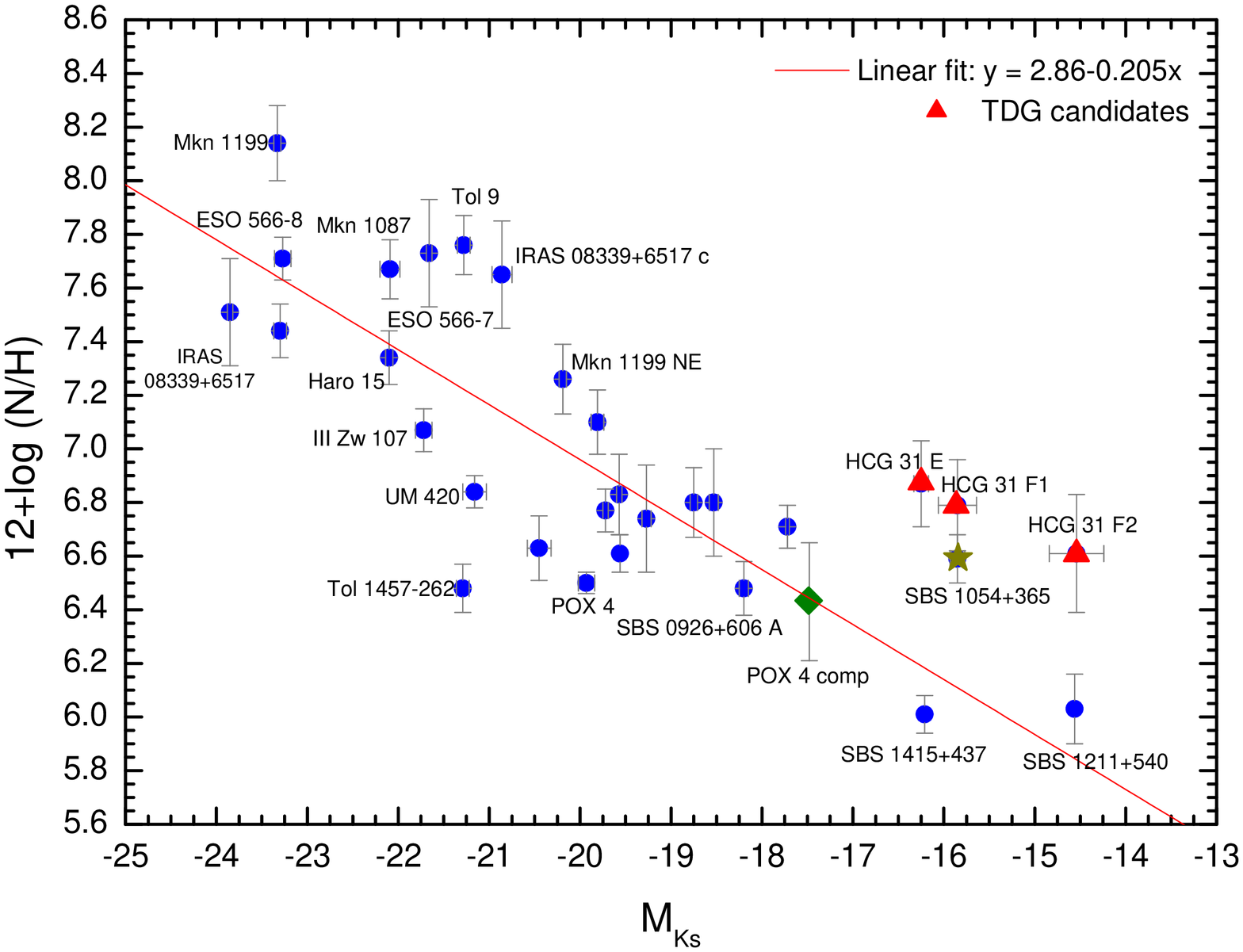} \\  
\end{tabular}
\protect\caption[ ]{\footnotesize{Nitrogen abundance vs. luminosity for the galaxies analysed in this work. The nitrogen abundance is expressed in 
units of 12+log(N/H); the luminosity is expressed as the absolute magnitude in $B$ (left) and $K_S$ (right) filters. Red triangles represent the 
\TDG\ candidates found in HCG~31 \citep{LSER04a} and Mkn~1087 \citep{LSER04b}. A green diamond corresponds to the dwarf object surrounding POX~4, 
while a yellow star indicates the galaxy SBS~1054+365. The relation derived \citet{vZH06} is plotted in the $M_B$-N/H relation with a green dashed 
line. The fits to our data are shown with a continuous red line.}}
\label{luminh}
\end{figure*}

The metallicity of normal disc galaxies is strongly correlated with the galaxy mass. The first mass-metallicity relation was found for irregular and blue 
compact galaxies \citep{Lequeux79,KinmanDavidson81}. However, luminosity is often used instead of mass because obtaining reliable mass estimates is 
difficult. \citet{RFW84} provided the first evidence that metallicity is correlated with luminosity in disk galaxies. Further work using larger 
samples of nearby disk galaxies confirmed this result \citep{Bothum84,WyseSilk85,Skillman89,Vila-Costas92,ZKH94,Garnett02}. Despite the huge 
observational effort, the origin of the luminosity-metallicity is still not well understood. The two basic ideas are (i) it represents an 
evolutionary sequence -- more luminous galaxies have processed a larger fraction of their raw materials 
\citep{McGaughdeBlok97,BelldeJong00,Boselli01}-- or (ii) it is related to a mass-retention sequence --more massive galaxies retain a larger 
fraction of their processed material \citep{Garnett02,Tremonti04,Salzer05}.  
Furthermore, other factors may play a key role in the variation of the metal content of a galaxy, remarking the quick metal enrichment that strong 
star-formation events in dwarf galaxies, such as \BCDG s, may experience. In these objects, the freshly processed material may be expelled into the 
intergalactic medium via galactic winds or be mixed with the reservoirs of non-synthesized gas, in both cases decreasing the global metallicity of the 
galaxy. 
   
In addition,
luminosity-metallicity relations are very useful to discern between pre-existing dwarf galaxies and tidal dwarf galaxy (\TDG) candidates 
\citep{DM98,Duc00} because these objects should have a metallicity similar to that observed in their parent spiral galaxies \citep{WDA03} and not a 
low-metallicity as it is found in dwarf objects. 

We studied the metallicity-luminosity relation using the data provided by our analysis. Figure~\ref{LMg} plots the oxygen abundance vs absolute 
magnitude in $B$ and \NIR\ filters and including some relationships found by previous studies. In this figure we distinguish between galaxies (blue 
points) and \TDG s candidates (red triangles) found in HCG~31 and Mkn~1087 groups. We also distinguish the dwarf object surrounding POX~4 (labeled 
POX~4~comp in Figures and Tables) because it may be another \TDG, and the galaxy SBS~1054+365 because, as we will see later, its position in the 
metallicity-luminosity diagrams is quite unusual. We estimated the \NIR\ absolute magnitudes for SBS~0948+532 and SBS~1211+540 assuming 
$V-J\sim0.8$, $J-H\sim0.3$ and $H-K_s\sim0.15$, that are the average values found in objects with properties similar to these two galaxies (see 
Paper~I).

The $M_B$--O/H diagram includes the relation given by \citet{RM95} for dwarf and irregular galaxies ($M_B$ $\geq$ $-$18) extrapolated to high 
luminosities and the \citet{vZH06} relation found for isolated dwarf irregular galaxies. Both relations have a similar slope ($-0.147$ and $-0.149$, 
respectively) and intercept (5.67 and 5.65, respectively). 
Our observational data have a rather high dispersion, but the tendency of increasing oxygen abundance with increasing absolute $B$-magnitude is 
clear. 
Most of the objects fainter than $M_B$=$-$18 are located above those relations, but many of them are \TDG\ candidates. On the other hand, a 
substantial fraction of the brighter objects tend to be clearly below the metallicity-luminosity relations obtained by previous authors. The best 
linear fit to our data excluding the \TDG\ candidates --it is well-known that they should not follow the metallicity-luminosity relation \citep{DM98}-- 
provides a slope of $-$0.099$\pm$0.019 and an intercept of 6.37$\pm$0.37. 
The slope we derive for our galaxy sample is shallower than those provided by the \citet{RM95} and \citet{vZH06} relations. However, the 
\citet{Tremonti04} $M_B$--O/H relation for all kinds of galaxies using SDSS data (plotted with a blue dotted line in Fig.~\ref{LMg}) show the steepest 
slope of all relations, which has a value of $-0.185$. That disagree with the conclusions reached by \citet{Tremonti04}, who explained the 
flattening of the $M-Z$ relation at higher masses because efficient galactic winds are able to remove metals from low-mass galaxies. 

Some authors \citep{CA93,PA93,McGaugh94} have already questioned the validity of the luminosity-metallicity relation in starbursting galaxies. As we 
explained in previous papers \citep{LSER04a,LSER04b}, we consider that the emission of the dominant young stellar population in these galaxies 
is increasing their $B$-luminosity, and hence the use of the standard metallicity-luminosity relation is not appropriate for starburst-dominated 
galaxies. Indeed, the increment of the $B$-luminosity is moving all star-forming objects away --towards more negative magnitudes--  from the usual 
relations valid for non-starbursting galaxies, and even producing --incidentally-- the \TDG\ candidates to agree with the relations.   

As proposed by \citet{HGO98} and reviewed by \citet{Salzer05}, perhaps \NIR\ magnitudes are more suitable than the optical $B$-magnitude to built 
metallicity-luminosity diagrams. Indeed, \NIR\ magnitudes are less affected by extinction and more directly related to the stellar mass of the 
galaxy than the optical luminosities. Furthermore, the effect of variations in star-formation histories and stellar populations is less pronounced in 
the \NIR\ than in the optical.
We analysed the behaviour of the oxygen abundance with the $J$, $H$ and $K_S$ absolute magnitudes (see Fig.~\ref{LMg}). As we expected, the oxygen 
abundance increases with the luminosity.
The slopes, $-0.102\pm$0.017, $-0.099\pm$0.016, and $-0.098\pm$0.015 for $M_J$, $M_H$, and $M_{K_S}$, respectively, and intercepts, 6.21$\pm$0.34, 
6.23$\pm$0.33, and 6.22$\pm$0.31, of the lineal fits 
are remarkably similar to the fit parameters of our $M_B$--O/H relation. However, these fits
have a better correlation coefficient (0.773, 0.779 and 0.790) and dispersion (0.18, 0.17 and 0.17) than those derived for the $M_B$--O/H relation 
($r=$0.719 and $\sigma$=0.20). \citet{Salzer05} found a notable decreasing of the rms scatter of the metallicity-luminosity relation between the blue 
and the \NIR.

The $M_{K_S}$--O/H diagram shown in Fig.~\ref{LMg} includes the relations obtained by \citet{Salzer05} --which consider two different empirical 
calibrations of their oxygen abundance, one following the \citet{EP84} method and another considering the \citet*{Kennicutt03} relationships-- and 
\citet{MO06}. These relationships have a very different slope ($-$0.212, $-0.195$, and $-0.14$, respectively) and intercept (3.919, 4.029, and 5.55, 
respectively) than that derived in our analysis of the $M_{K_S}$--O/H relation. However, for objects with $M_{K_S}<-19$, the relations provided by 
\citet{Salzer05} agree fairly well with the position of our data points. 
These authors found that the slopes of the metallicity-luminosity relations change systematically from the shortest to the longest wavelengths, in 
that the relation is steepest in the blue and more shallow in the \NIR. In fact, \citet{Salzer05} also noted that their derived $M_B$--O/H relation 
has a much steeper slope than those found by previously relations (i.e., Skillman et al. 1989, Richer \& McCall 1995).  

The disagreement between our data and previous metallicity-luminosity relations may arise because our sample was chosen considering WR galaxies, which 
host strong starbursts, while other samples consider any star-forming galaxies. It may also occur because dwarf galaxies and \emph{normal} 
galaxies have different relations (i.e. Tremonti et al. 2004).  
Another fact that can contribute to the aforementioned disagreement is
that the oxygen abundances derived in previous analyses have been usually computed via photoionization-based empirical calibrations. As we discussed 
before, they seem to overestimate the actual O/H ratio by about 0.2 dex. 

In Fig~\ref{LMg} we can see that the separation between the positions of \TDG\ candidates and dwarf galaxies is somewhat more evident in the 
metallicity-luminosity diagrams involving \NIR\ magnitudes. POX~4~comp agrees with the relations, and hence further analysis of this object is needed 
to understand its real nature. On the other hand, SBS~1054+365 has a higher oxygen abundance for its optical and \NIR\ absolute luminosities. We do 
not consider the O/H to be overestimated because of the high quality of our spectra (Paper~II) and because previous studies derived the same 
metallicity (i.e., Izotov \& Thuan 1998). We are also quite confident of the correct determination of its optical and \NIR\ magnitudes because of the 
detailed analysis we already presented in Paper~I. Hence we consider that this nearby (8~Mpc) \BCDG\ may have experienced a recent a strong 
pollution of heavy elements which probably has not had enough time to disperse and mix with the surrounding ISM. In Paper~V we will see that this 
galaxy also possesses a very high $M_{dyn}/L_B$ ratio, suggesting that the neutral gas is quite disturbed. Further analysis of this intriguing 
object, including \HI\ radio data and optical spectroscopy, would be necessary to ascertain its true nature. 

Finally, another aspect that should also be addressed is the possibility that some of the sample objects may be a recent merger of two galaxies.  In 
this case, the integrated luminosity would be higher than that expected for a single galaxy, and hence its position would be also far from the 
metallicity-luminosity relations. That may be happening in UM~420 --see Sect.~3.9 in Paper~I, Sect.~3.9.3 in Paper~II and \citet{James09}-- and 
\mbox{Tol~1457-262} (see Sect.~3.18.3 in Paper~II), which show a considerable high luminosity given their low oxygen abundances.
Hence it is more appropriate to compare the stellar mass (and not the $B$ or the \NIR\ luminosities) with the metallicity to reach any conclusions 
about this topic, as we will explain in Paper~V. 

\begin{figure*}[t]
\centering
\includegraphics[angle=270,width=\linewidth]{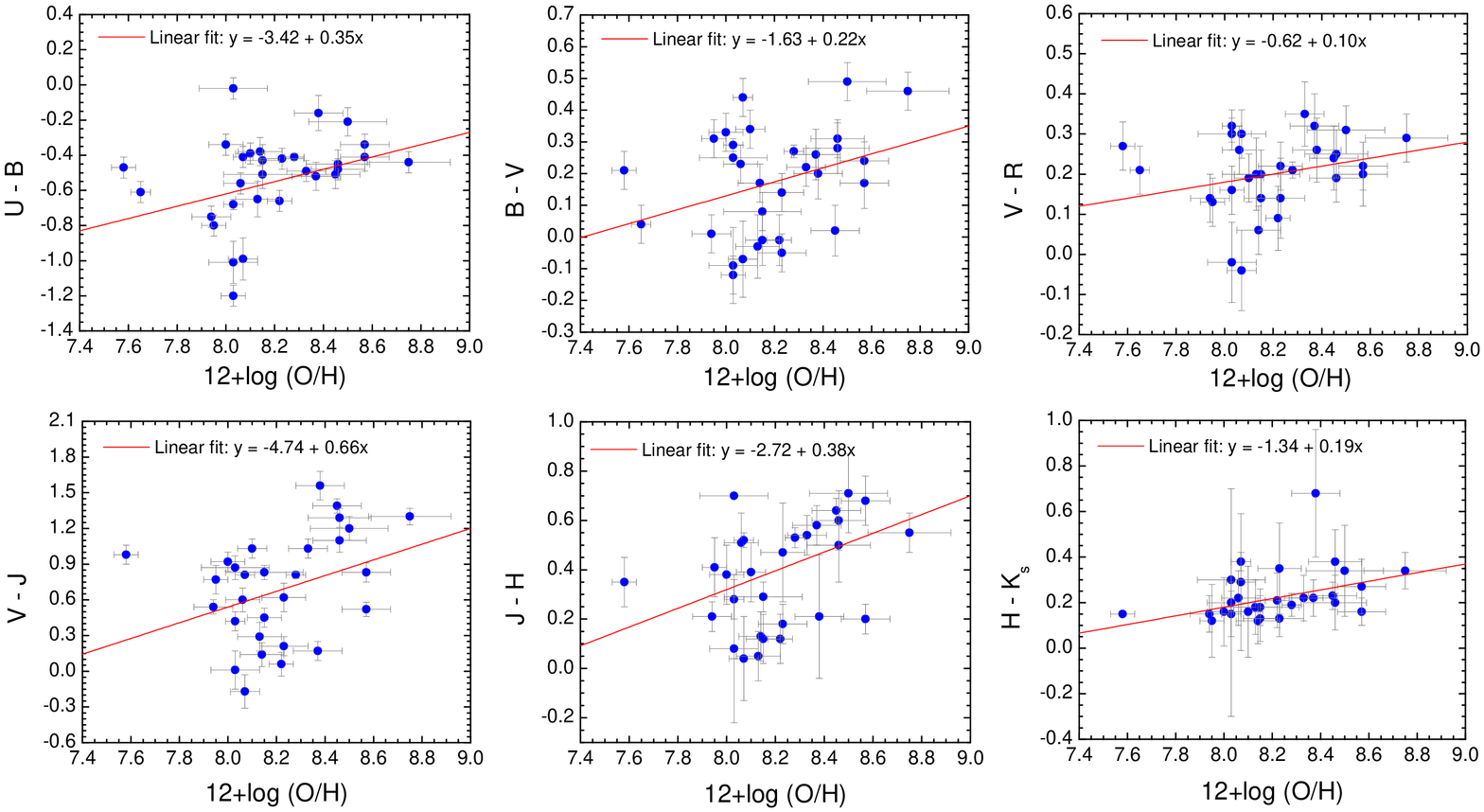}  
\protect\caption[ ]{\footnotesize{Metallicity-colour diagrams for the galaxies analysed in this work. A linear fit is shown with a continuous red 
line.}}
\label{colourabun}
\end{figure*}

The nitrogen abundance also has a strong correlation with the luminosity. Figure~\ref{luminh} plots the nitrogen abundance, expressed in units of 
12+log(N/H), as a function of the absolute $B$ and $K_S$ magnitudes. As in Fig.~\ref{LMg}, we distinguish the \TDG\ candidates and the 
intriguing objects POX~4~comp and SBS~1054+365. From Fig.~\ref{luminh} it is also evident that the nitrogen abundance increases with increasing 
luminosity. We performed a linear fit to our data (without considering the problematic objects already discussed), which is plotted with a continuous 
red line. In the $M_B$--N/H diagram this fit yields a slope of $-0.181\pm0.036$ and an intercept of 3.65$\pm$0.68. This result is virtually identical 
to that found by \citet{vZH06}, their relation is shown with a green dashed line in the left diagram of Fig.~\ref{luminh}. The scatter of the 
$M_B$--N/H relation is higher than that found in the $M_B$--O/H relation, actually, we may distinguish between two kinds of objects in this figure, 
because the more evolved galaxies (such as Mkn~1199, Mkn~1087, Tol~9, ESO~566-8) and the non-starbursting systems (Mkn~1199~NE, IRAS~08339+6517~c, 
ESO~566-7) have a higher N/H value at a given absolute $B$-luminosity than the very starbursting (i.e. IRAS~08339+6517, HCG~31~AC) and blue compact dwarf 
(i.e. UM~420, POX~4, SBS~0926+606~A) galaxies. \TDG\ candidates do not lie far from this relation. These differences are probably a consequence of 
the different star-formation histories of the galaxies, as we also explained the scatter in the observed N/O ratio at \abox$>$7.9 (see 
Fig.~\ref{compabun}).

As it happened in the previous diagrams involving the oxygen abundance, the luminosity-metallicity diagram using \NIR\ data has a smaller scatter 
(Fig.~\ref{luminh}, right). The fit to our data in the $M_{K_S}$--N/H diagram, neglecting \TDG\ candidates and problematic objects, gives a slope 
of $-0.205\pm0.030$ and an intercept of 2.86$\pm$0.61. The correlation coefficient is $r$=0.813 and the dispersion is $\sigma$=0.33. In comparison, 
the $M_B$--N/H diagram has $r$=0.707 and $\sigma$=0.40. We note that again \TDG\ candidates and the galaxy SBS~1054+365 lie far from this fit, which 
is plotted in Fig.~\ref{luminh} (right) with a continuous red line. POX~4~comp agrees quite nicely with the relation. In both diagrams, the 
position of Tol~1457-262 is far from the relation, having a considerable high luminosity for its N/H ratio. As we already discussed, we consider that 
this object may correspond to a recent merger of two independent galaxies. Hence we conclude that luminosity-metallicity relations using \NIR\ band 
magnitudes are very suitable to understand galaxy properties and evolution, providing tighter correlations than those relations involving optical 
magnitudes.

\section{Metallicity-colour relations}

\begin{figure*}[t]
\begin{tabular}{cc}
\includegraphics[angle=270,width=0.45\linewidth]{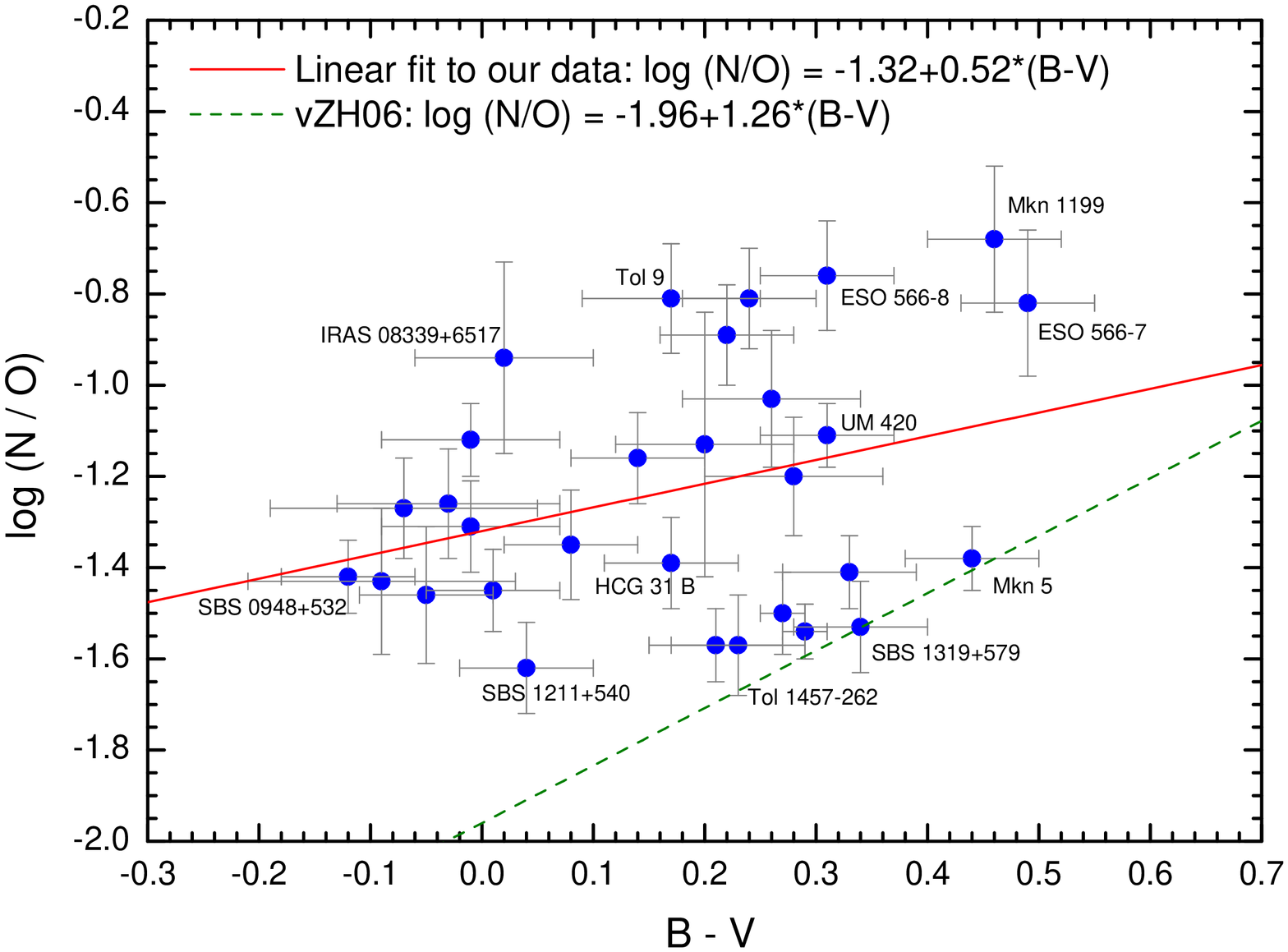}  &  
\includegraphics[angle=270,width=0.45\linewidth]{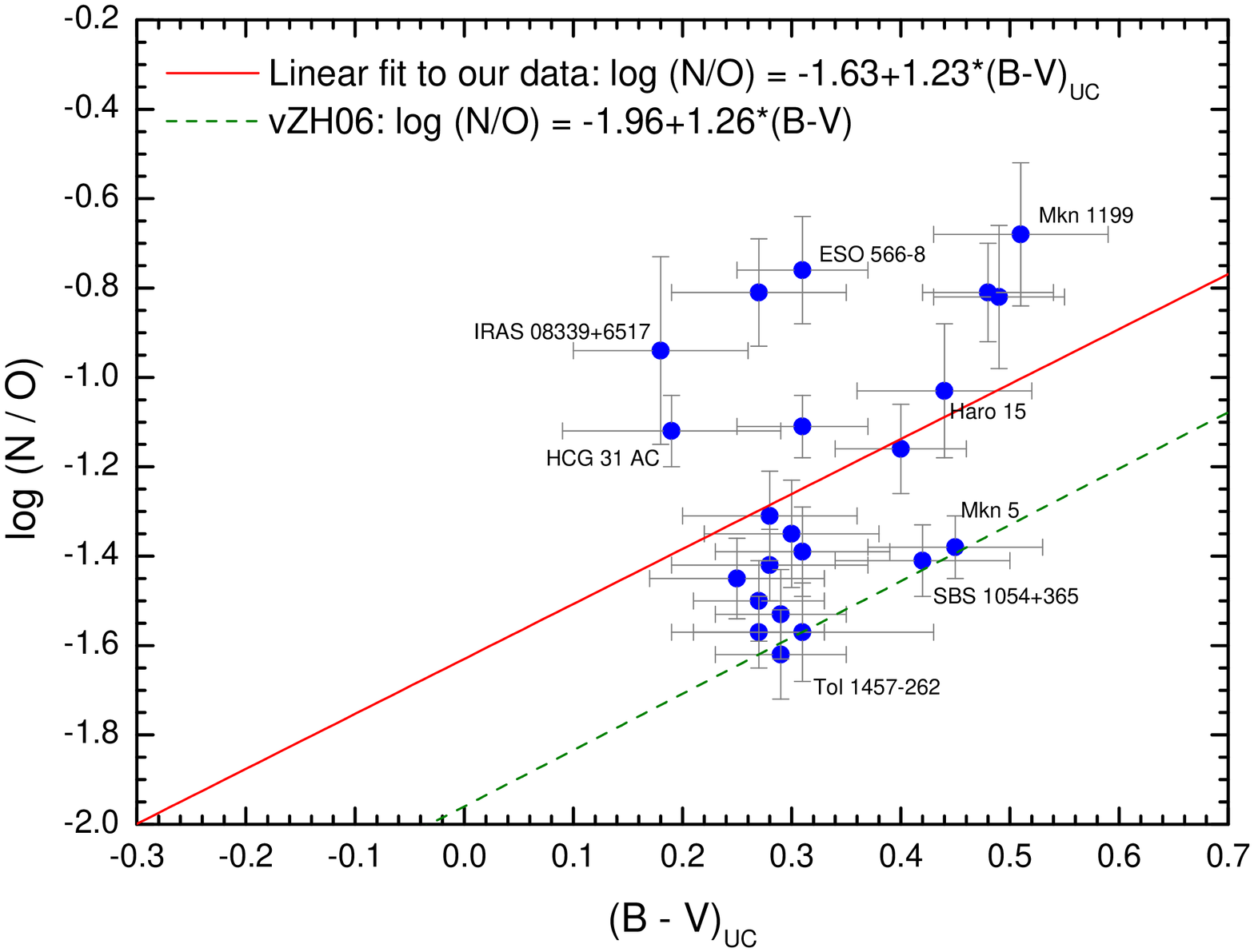}  \\  
\end{tabular}
\protect\caption[ ]{\footnotesize{N/O ratio vs. the reddening-corrected $B-V$ colour considering all the galaxy (left) or just the underlying stellar 
population (right). A linear fit to our data is shown with a continuous red line. The relationship found by \citet{vZH06} in their analysis of dwarf 
irregular galaxies is plotted with a dashed green line.}}
\label{colourno}
\end{figure*}

\begin{figure*}[t]
\begin{tabular}{cc}
\includegraphics[angle=270,width=0.45\linewidth]{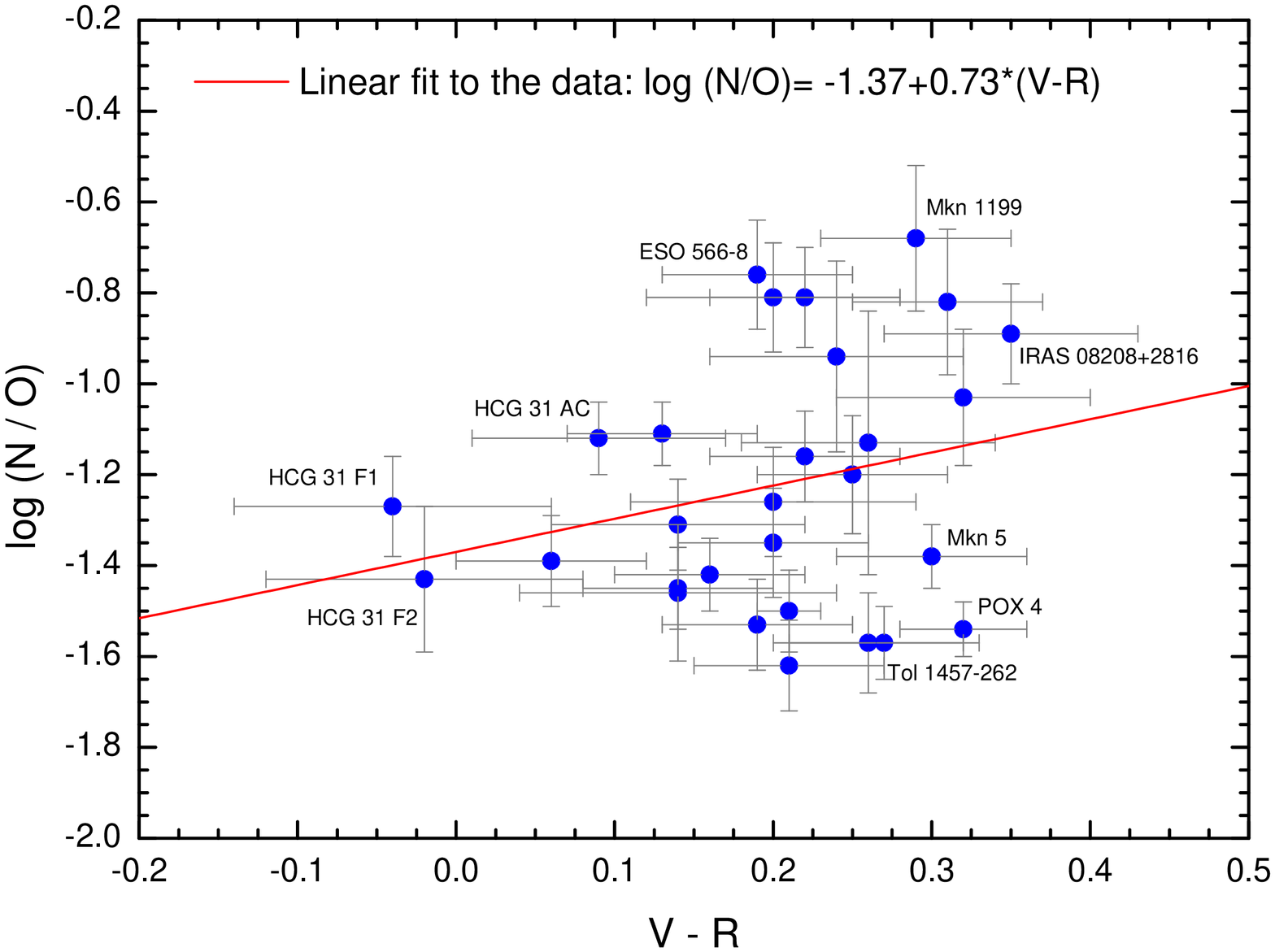}  &  
\includegraphics[angle=270,width=0.45\linewidth]{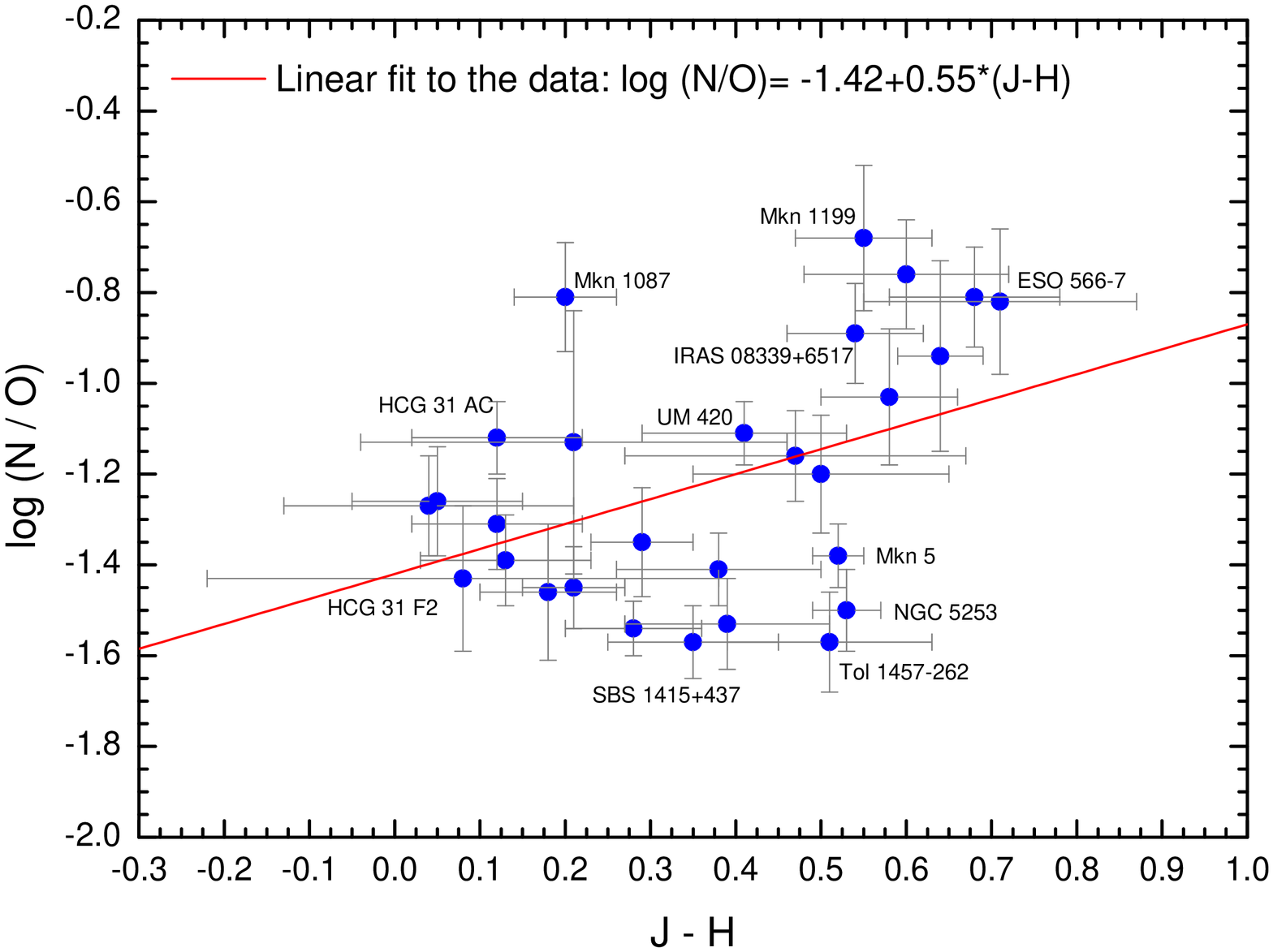}  \\  
\end{tabular}
\protect\caption[ ]{\footnotesize{N/O ratio vs. the reddening-corrected $V-R$ (left) and $J-H$ (right) colours. A linear fit to our data is shown with 
a continuous red line.}}
\label{colourdosno}
\end{figure*}

We analysed the relationship between the optical and \NIR\ colours observed in each galaxy and its metallicity. Figure~\ref{colourabun} plots the 
results for all colours (from Table~\ref{colores}) vs. the oxygen abundance and a fit to the data. Although the dispersion is large, the general 
trend is that galaxies with redder colours have a higher oxygen abundance in their ionized gas. This suggests again the relative importance of 
the evolved stellar populations existing in the galaxies, and agrees both with models of galaxy evolution (i.e., Leitherer et al. 1999; Bruzual \& 
Charlotte 2003) and previous observational analysis (i.e. Jansen et al. 2000; Lilli, Carollo \& Stockton 2003). Neglecting other effects (such as
extinction), the birth of new generations of stars within a galaxy will continuously increase the number of intermediate and low-mass stars, which are 
the typical stellar population that constitutes the low-luminosity component in starburst galaxies. Hence it should be expected that as the system 
evolves --increasing its metallicity-- its optical and \NIR\ colours are more dominated by the evolved population, and therefore they are redder. 
Another interesting comment regarding these correlations is to remark how important it is to have an estimate of the metallicity of the galaxies when 
comparing the optical/\NIR\ colours with the predictions of theoretical evolutionary models (see Paper~I).  

Besides the N/O ratio has a large dispersion in our sample objects, we checked whether a correlation exists between this ratio and the $B-V$ 
colour. Figure~\ref{colourno} plots both quantities considering the $B-V$ colour of the complete galaxy (left panel) and that derived for the old underlying 
population (right panel). Both diagrams include a fit to our data (continuous red line).  
Figure~\ref{colourno} shows the large dispersion of the data points,
but there seems to be a suggestion of a tendency of increasing N/O with increasing $B-V$ colour. 
This tendency does not become clearer when considering the data of the underlying stellar component.
However, the relation found using our data disagrees with that obtained by \citet{vZH06} analysing a sample of dwarf irregular galaxies. We note that 
they referred to the \emph{$B-V$ colour of the underlying population}, but actually it is just the integrated colour of the galaxy. The significant 
difference found between both samples may be a consequence of two reasons: (i) we are comparing different objects --our sample is composed by 
starbursting galaxies, many of them \BCDG s, but \citet{vZH06} analysed quiescent dwarf irregular galaxies with star-formation histories that are 
correctly modelled by simple changes in the SFR-- and (ii) the N/O ratios tabulated by these authors, who used \citet{McGaugh91} photoionization 
models in many cases, are not well estimated --as we saw before, these models overestimate the oxygen abundance, and hence underestimate the N/O 
ratio, as it is clearly seen in the O/H--N/O diagram (Fig.~\ref{compabun}, right) when comparing the \citet*{vZee98} sample with the results 
provided by other authors--.
We consider that the first reason may be the most reasonable explanation: objects with different star-formation histories have different 
relationships between the N/O ratio and other global characteristics such as optical colours. 

However, as we commented before, using the $B$-magnitude may be not ideal to compare the colours and the chemical properties of these galaxies. 
Hence we also compared the N/O ratio with other optical and \NIR\ colours, two of them ($V-R$ and $J-H$) are shown in Fig.~\ref{colourdosno}. 
As we already explained when analysing the N/H--$L$ relationship (Fig.~\ref{luminh}), there are two kinds of objects with similar optical 
luminosities but with different nitrogen abundances. This is also reflected in the N/O ratio (Fig.~\ref{colourno}, left), where normal and dwarf 
non-starbursting galaxies show a higher N/O ratio than starbursts (which have a lower N/O ratio in Fig.~\ref{colourno}, left). This segregation is 
not so evident when comparing the N/O ratio with the \NIR\ luminosities (Fig.~\ref{colourdosno}, right), perhaps because the underlying component 
makes a higher contribution to the \NIR\ magnitudes, but more data are needed to quantify this effect.   

Finally, we also note that we do not see any clear relationship between the underlying population colours and the N/O ratio: for example, the $V-R$ 
colour of the low luminosity component is $0.31\pm0.05$ for all ranges of N/O. It was not possible to analyse the \NIR\ colours of the low-luminosity 
component in the majority of the cases (see Table~5 in Paper~I), but for the few available data it also seems that they are metallicity-independent, 
showing $J-H\sim0.5\pm0.1$ and $H-K_S\sim0.2$.

\section{Conclusions}

We compiled and analysed globally the optical/\NIR\ colours and the physical and chemical properties of the ionized gas for a sample of 20 
Wolf-Rayet galaxies. The individual analysis of the photometry of each galaxy was presented in Paper~I, while the individual spectroscopic analysis 
was discussed in Paper~II.  The metallicity of these galaxies lies between 7.58 and 8.75 --in units of \abox--. The most important conclusions found 
in this study are
\begin{enumerate}
\item The colours estimated for our galaxy sample, which were corrected for both extinction and nebular emission using our spectroscopic data, agree 
quite well with the predictions given by evolutionary synthesis models, 
especially in compact and dwarf objects. Small discrepancies are explained because of the existence of several stellar populations within each galaxy 
and differences in their star-formation history. 
All galaxies show evidence of an old stellar population underlying the starburst, with ages older than 500 Myr. 
\item 
We checked that all objects can be classified as pure star-forming galaxies.  
In total, we compiled 41 independent star-forming regions, of which 31 have a direct estimate of the electron temperature of the ionized gas, and 
hence their element abundances were derived using the direct method. We found that younger bursts have larger ionization budgets and are therefore 
capable to heat the ionized gas to higher electron temperatures. Both \CHb\ and \Wabs\ increase with increasing metallicity, as predicted by galaxy 
evolution models. 
\item We compiled the oxygen abundance and N/O, S/O, Ne/O, Ar/O, and Fe/O ratios in our WR galaxy sample. They generally agree well with previous observations. 
The N/O ratio is found to be rather constant for objects with \abox$\leq$7.6, has an important dispersion in galaxies with 
7.6$\leq$\abox$\leq$8.3, and increases with the metallicity in objects with \abox$\geq$8.3. This behaviour is explained assuming the very different 
star-formation histories that each individual system has experienced.
\item We detected a high N/O ratio in objects showing clear WR features (HCG~31~AC, UM~420, IRAS~0828+2816, III~Zw~107, ESO~566-8 and NGC~5253). The 
ejecta of the WR stars may be the origin of the N enrichment in these galaxies, but further detailed data comparing the chemical properties of a 
larger sample of both WR and non-WR galaxies, as well as careful analyses of galaxies showing a localized high N/O ratio, are needed. 
\item The relative abundance ratios of the $\alpha$-elements to oxygen are approximately constant, which is expected because all four elements are mainly 
produced by massive stars. We found indications of a moderate depletion of oxygen and iron onto grains in the most metal-rich galaxies. 
\item We compared the abundances provided by the direct method with those obtained using the most common empirical calibrations:
   \begin{itemize}
   \item The Pilyugin-method of \citet{P01a,P01b}, which considers the $R_{23}$ and the $P$ parameters and is updated by \citet{PT05}, is nowadays the best 
suitable empirical calibration to derive the oxygen abundance of star-forming galaxies. The cubic fit to $R_{23}$ provided by \citet*{Nagao06} is not 
valid for analysing these star-forming galaxies. 
   \item The relations between the oxygen abundance and the $N_2$ or the $O_3N_2$ parameters provided by \citet{PP04} give acceptable results for 
objects with \abox$>$8.0. 
   \item The results provided by empirical calibrations based on photoionization models \citep{McGaugh91,KD02,KK04} are systematically 0.2 -- 0.3 dex 
higher than the values derived from the direct method. These differences are of the same order as the abundance discrepancy 
found between abundances determined from recombination and collisionally excited lines of heavy-element ions.
This may suggest temperature fluctuations in the ionized gas, as they exist in Galactic and other extragalactic \HII\ regions.
   \end{itemize}
\item We studied the optical/\NIR\ luminosity-metallicity relations for our sample galaxies. We found that our data generally disagree with previous 
relations, perhaps because the objects analysed here host strong starbursts, but maybe also because we used the direct method and not the empirical 
calibrations to derive the oxygen abundances. The $L$--$Z$ relations tend to be tighter when using \NIR\ luminosities, \TDG\ candidates are indeed 
easier detected using the $M_{Ks}$--$Z$ relation. 
\item The nitrogen abundance also has a strong correlation with the luminosity, but 
normal and dwarf non-starbursting galaxies show a higher N/O ratio than strong starbursting galaxies.
\item We found that galaxies with redder colours tend to have a higher oxygen abundance. The N/O ratio also increases with redder colours. Both 
results agree with galaxy evolution models. The colours of the underlying component seem to be metallicity-independent, 
but more data are still needed to confirm this trend.
\end{enumerate}
We finally conclude that it is fundamental to perform a detailed analysis of both the photometric (optical and \NIR\ magnitudes and colours of both 
burst and underlying component) and the chemical (oxygen abundances, N/O ratios) properties of these star-forming galaxies to understand the 
evolutionary stage of every system. A larger galaxy sample following the ideas compiled in this work 
will complement the results derived from huge databases (which do not distinguish the evolutionary state of the galaxies or the relative 
contribution of the burst/underlying populations, and for which the majority of the properties have been derived automatically and, in many cases, 
have considered empirical calibrations to determine the chemical abundances) when analysing the global properties and evolution of star-forming 
galaxies.


\begin{acknowledgements}

\'A.R. L-S thanks C.E. (his formal PhD supervisor) for the help and very valuable explanations, talks and discussions during these years. 
He also acknow\-ledges the help and support given by the Instituto de Astrof\'{\i}sica de Canarias (Spain) while doing his PhD.
\'A.R. L-S. \emph{deeply} thanks the Universidad de La Laguna (Tenerife, Spain) for force him to translate his PhD thesis from English to Spanish; he 
had to translate it from Spanish to English to complete this publication. 
This was the main reason of the delay of the publication of this research, because the main results shown here were already included in the PhD 
dissertation (in Spanish) which the first author finished in 2006 \citep{LS06}. \'A.R. L-S. also thanks the people at the CSIRO/Australia 
Telescope National Facility, especially B\"arbel Koribalski, for their support and friendship while translating his PhD. We are grateful to Janine 
van Eymeren for her comments about this manuscript, and to Mercedes Moll\'a for very fruitful discussions about the chemical evolution of the galaxies.
We are also grateful for the comments provided by an anonymous referee, which improved the quality of this research.
This work has been partially funded by the Spanish Ministerio de Ciencia y Tecnolog\'{\i}a (MCyT) under project AYA2004-07466. 
This research has made use of the NASA/IPAC Extragalactic Database (NED) which is operated by the Jet Propulsion Laboratory, California Institute of 
Technology, under contract with the National Aeronautics and Space Administration.
This research has made extensive use of the SAO/NASA Astrophysics Data System Bibliographic Services (ADS). 

\end{acknowledgements}

\listofobjects

\clearpage
\appendix
\normalsize

\onecolumn
\section{Empirical calibrations of the oxygen abundance\label{empiricalcalibrations}}

When the spectrum of an extragalactic \HII region does not show the [\ion{O}{iii}] $\lambda$4363 emission line or other auroral lines that can be 
used to derive \Te, the so-called \emph{empirical calibrations} are applied to get a rough estimation of its metallicity.
Empirical calibrations are inspired partly by photo-ionization models and partly by observational trends of line strengths with galactocentric 
distance in gas-rich spirals, which are believed to be due to a radial abundance gradient with abundances decreasing outwards. In extragalactic 
objects, the usefulness of the empirical methods goes beyond the derivation of abundance gradients in spirals \citep{P04}, as these methods find 
application in chemical abundance studies of a variety of objects, including low-surface brightness galaxies \citep{deNaray04} and star-forming 
galaxies at intermediate and high redshift, where the advent of 8--10 m class telescopes has made it possible to extend observations (e.g., Teplitz et 
al. 2000, Pettini et al. 2001; Kobulnicky et al 2003; Lilly, Carollo \& Stockton 2003; Steidel et al. 2004; Kobulnicky \& Kewley 2004; Erb et al. 
2006).   
  
\begin{figure}[t!]
\centering
\includegraphics[angle=270,width=0.7\linewidth]{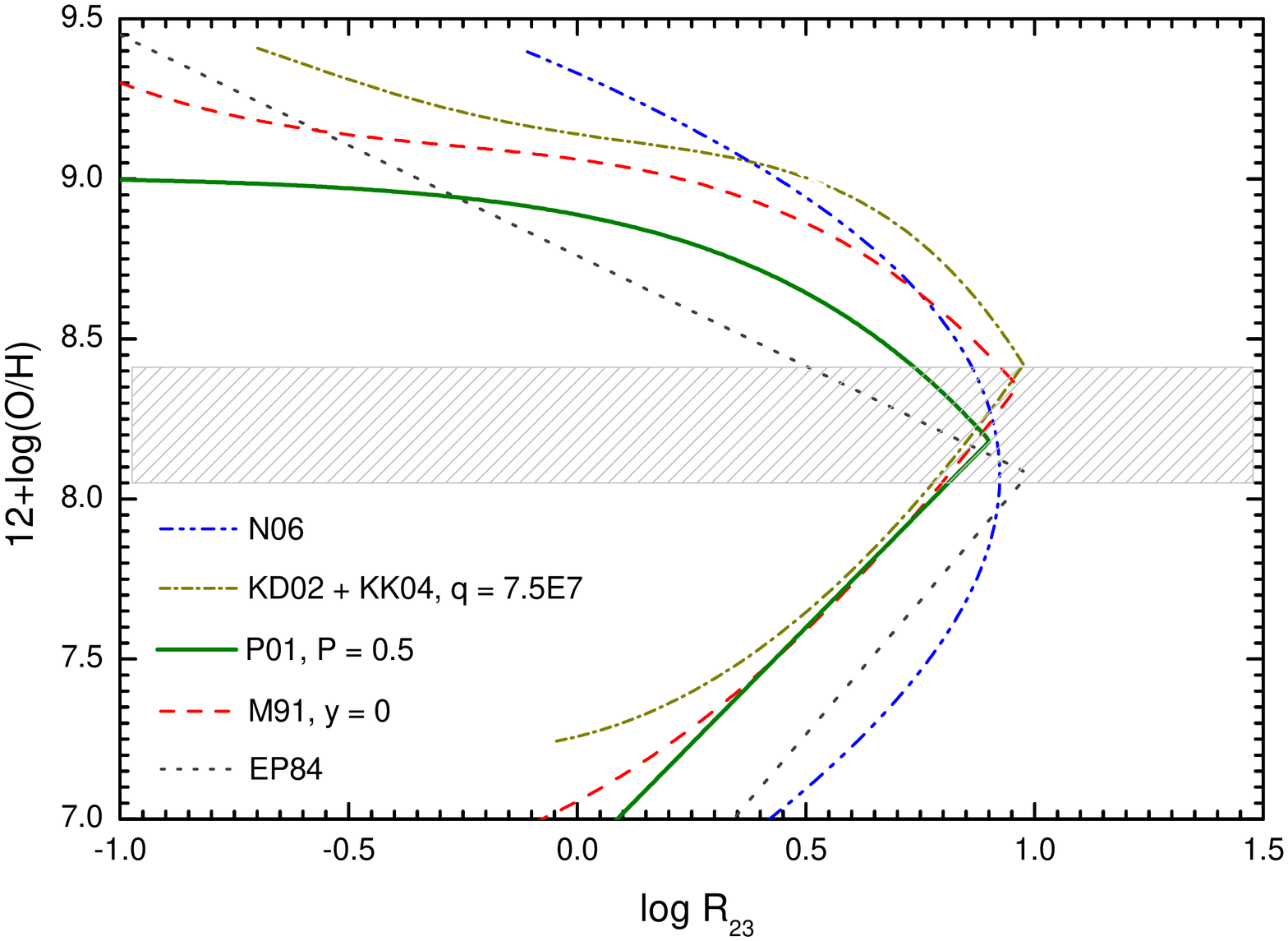}   
\caption{\footnotesize{Empirical calibrations of oxygen abundance using the $R_{23}$ parameter. Note that they are bi-valuated. The dashed zone 
indicates the region with higher uncertainties in O/H. The empirical calibrations plotted in the figure are: EP94: Edmund \& Pagel (1984); M91: 
McGaugh (1991) using $y$=0 ($R_2=R_3$); 
P01: Pilyugin (2001) using $P=0.5$ ($R_2=R_3$); (KD02+KK04): Kewley \& Dopita (2002) using the formulation of Kobulnicky \& Kewley (2004) assuming 
$q=7.5\times 10^7$ cm s$^{-1}$; N06: Nagao et al. (2006) using their cubic fit to $R_{23}$.}}
\label{cempiricas}
\end{figure}

As the brightest metallic lines observed in spectra of \HII regions are those involving oxygen, this element has been extensively used to get a 
suitable empirical calibration. Oxygen abundance is important as one of the fundamental characteristics of a galaxy: its radial 
distribution is combined with radial distributions of gas and star surface mass densities to constrain models of chemical evolution. Parameters 
defined in empirical calibrations evolving bright oxygen lines are
\begin{eqnarray}
R_3=  \frac{I([{\rm O\, III}]) \lambda 4959+I([{\rm O\, III}]) \lambda 5007}{\rm H\beta}, \\
R_2=  \frac{I([{\rm O\, II}]) \lambda 3727}{\rm H\beta}, \\
R_{23} =  R_3 + R_2, \\
P =  \frac{R_3}{R_{23}}, \\
y =  \log \frac{R_3}{R_2} = \log \frac{1}{P^{-1}-1}.
\end{eqnarray}   
\citet{Jensen76} presented the first exploration in this method considering the $R_3$ index, which considers the [\ion{O}{iii}] 
$\lambda\lambda$4959,5007 emission lines. However, were \citet{Pag79} who introduced the most widely used abundance indicator, the $R_{23}$ index, 
which also included the bright [\ion{O}{ii}] $\lambda$3727 emission line. Since then, many studies have been performed to refine the 
calibration of $R_{23}$ \citep{EP84,MRS85,DE86,Torres-Peimbert89,McGaugh91,ZKH94,P00,P01a,P01b,KD02,KK04,PT05,Nagao06}.
The most successful are the calibrations of \citet{McGaugh91} and \citet{KD02}, which are based on photoionization models, and the empirical 
relations provided by \citet{P01a,P01b} and \citet{PT05}. Both kinds of calibrations improve the accuracy by making use of the 
[\ion{O}{iii}]/[\ion{O}{ii}] ratio as ionization parameter, which accounts for the large scatter found in the $R_{23}$ versus oxygen abundance 
calibration, which is larger than observational errors \citep{KKP99}. Figure~\ref{cempiricas} shows the main empirical calibrations that use the 
$R_{23}$ parameter.

The main problem associated with the use of $R_{23}$ parameter is that it is {\bf bivaluated}, i.e., a single value of $R_{23}$ can be caused by two 
very different oxygen abundances. The reason of this behaviour is that the intensity of oxygen lines \emph{does not indefinitely increase} with 
metallicity. Thus, there are two \emph{branches} for each empirical calibration (see Fig.~\ref{cempiricas}): the \emph{low-metallicity} regime, 
with 12+log(O/H)$\leq$8.1, and the \emph{high-metallicity} regime, with 12+log(O/H)$\geq$8.4. That means that a very large fraction of the 
star-forming regions lie in the ill-defined turning zone around 12+log(O/H)$\simeq$8.20, where regions with the same $R_{23}$ value have oxygen 
abundances that differ by almost an order of magnitude. Hence, additional information, such as the [\ion{N}{ii}]/\Ha\ or the 
[\ion{O}{ii}]/[\ion{O}{iii}] ratios, is needed to break the degeneracy between the high and low branches (i.e., Kewley \& Dopita, 2002). 
Besides, the $R_{23}$ method requires that spectrophotometric data are corrected by reddening, which effect is crucial because [\ion{O}{ii}] and 
[\ion{O}{iii}] lines have a considerably separation in wavelength.

Here we list all empirical calibrations that were considered in this work, compiling the equations needed to derive the oxygen abundance from 
bright emission line ratios following every method.

{\bf Edmund \& Pagel (1984)}: Although the $R_{23}$ parameter was firstly proposed by \citet{Pag79}, the first empirical calibration was given by 
\citet{EP84},
\begin{eqnarray}
\nonumber 12+\log ({\rm O/H})_{up}= 8.76 -0.69 \log R_{23}, \\
12+\log ({\rm O/H})_{low} = 6.43 +1.67 \log R_{23},
\end{eqnarray}
with the limit between the lower and the upper branches at \abox$\sim$8.0.

{\bf McCall, Rybski \& Shields (1985)} 
presented an empirical calibration for oxygen abundance using the $R_{23}$ parameter, only valid for 12+log(O/H)$>$8.15. However, they did not give an 
analytic formulae but only listed it numerically (see their Table~15). The four-order polynomical fit for their values gives the following relation:
\begin{eqnarray}
12+\log ({\rm O/H})_{up} = & 9.32546  -0.360465x +0.203494x^2 
                                      &       +0.278702x^3   -1.36351x^4,
\end{eqnarray}
with $x=\log R_{23}$.

{\bf Zaritzky, Kennicutt \& Huchra (1994)} provided a simple analytic relation between oxygen abundance and $R_{23}$: 
\begin{eqnarray}
12+\log ({\rm O/H})_{up} =  9.265-0.33x  -0.202x^2  -0.207x^3 -0.333x^4.
\end{eqnarray}
Their formula is an average of three previous calibrations: \citet{EP84}, McCall et al. (1985) and Dopita \& Evans (1986). Following the authors, 
this calibration is only suitable for 12+log(O/H)$>$8.20, but perhaps a more realistic lower limit is 8.35.

{\bf McGaugh (1991)} calibrated the relationship between the $R_{23}$ ratio and gas-phase oxygen abundance using \HII region models derived from the 
photoionization code \CLOUDY\ \citep{Ferland98}. McGaugh's models include the effects of dust and variations in ionization parameter, $y$. 
\citet*{KKP99} give analytical expressions for the \citet{McGaugh91} calibration based on fits to photoionization models; the middle point between both 
branches is  \abox$\sim$8.4:
\begin{eqnarray}
12+\log ({\rm O/H})_{up} =  & 
7.056  +0.767x +0.602x^2  -y (0.29+0.332x-0.331x^2), \\
\nonumber 
12+\log ({\rm O/H})_{low} = & 9.061-0.2x-0.237x^2-0.305x^3-0.0283x^4 \\
& -y(0.0047-0.0221x-0.102x^2-0.0817x^3-0.00717x^4).
\end{eqnarray}

{\bf Pilyugin (2000)} found that the previous calibrations using the $R_{23}$ para\-meter had a systematic error depending on the hardness of the 
ionizing radiation, suggesting that the excitation parameter, $P$, is a good indicator of it. In several papers, Pilyugin performed a detailed 
analysis of the observational data combined with photoionization models to obtain empirical calibrations for the oxygen abundance. \citet{P00} 
confirmed the idea of \citet{McGaugh91} that the strong lines of [\ion{O}{ii}] and [\ion{O}{iii}] contain the necessary information for the determination 
of accurate abundances in low-metallicity (and may be also in high-metallicity) \HII regions. He used new observational data to propose a linear fit 
involving only the $R_{23}$ parameter,
\begin{eqnarray}
12+\log ({\rm O/H})_{up} = 9.50 - 1.40 \log R_{23},\\
12+\log ({\rm O/H})_{low} = 6.53 + 1.40 \log R_{23},
\end{eqnarray}
assuming a limit of \abox$\sim$8.0 between the two branches. This calibration is close to that given by \citet{EP84}; it has the same slope, but 
\citet{P00} is shifted towards lower abundances by around 0.07 dex. However, this new relation is not sufficient to explain the wide spread of 
observational data. Thus, {\bf Pilyugin (2001a)} give the following, more real and complex, calibration involving also the excitation parameter $P$:
\begin{eqnarray}
12+\log ({\rm O/H})_{up} =\frac{R_{23}+54.2+59.45P+7.31P^2}{6.01+6.71P+0.371P^2+0.243R_{23}}.
\end{eqnarray}
This is the so-called \emph{P-method}, which can be used in moderately high-metallicity \ion{H}{ii} regions (\abox$\geq$8.3). Pilyugin used two-zone 
models of \HII regions and assumed the $T_e$(\ion{O}{ii}) -- $T_e$(\ion{O}{iii}) relation from \citet{G92}. 
For the low metallicity branch, {\bf Pilyugin (2001b)} found that
\begin{eqnarray}
12+\log ({\rm O/H})_{low} = 6.35 + 1.45 \log R_{23} -1.74 \log P.
\end{eqnarray}
Pilyugin estimates that the precision of oxygen abundance determination 
with this method is around 0.1 dex.

{\bf Pilyugin \& Thuan (2005)} revisited these calibrations including more spectroscopic measurements of \HII regions in spiral and irregular 
galaxies with a measured intensity of the [\ion{O}{iii}] $\lambda$4363 line and recalibrate the relation between the oxygen abundance and the 
$R_{23}$ and $P$ parameters, yielding to:
\begin{eqnarray}
12+\log ({\rm O/H})_{low} = \frac{R_{23}+106.4+106.8P-3.40P^2}{17.72+6.60P+6.95P^2-0.302R_{23}},\\
12+\log ({\rm O/H})_{up} = \frac{R_{23}+726.1+842.2P+337.5P^2}{85.96+82.76P+43.98P^2+1.793R_{23}}.
\end{eqnarray}

{\bf Kewley \& Dopita (2002)} used a combination of stellar population synthesis and photoionization models to develop a set of ionization parameters 
and abundance diagnostic based only on the strong optical emission lines. Their \emph{optimal} method uses ratios of [\ion{N}{ii}], [\ion{O}{ii}], 
[\ion{O}{iii}], [\ion{S}{ii}], [\ion{S}{iii}] and Balmer lines, which is the full complement of strong nebular lines accessible from the ground. They 
also recommend procedures for the derivation of abundances in cases where only a subset of these lines is available.  \citet{KD02} models start with 
the assumption that $R_{23}$, and many of the other emission-line abundance diagnostics, also depends on the {\bf ionization parameter} $q \equiv 
c\times U$, that has units of \mbox{cm~s$^{-1}$.} They used the ste\-llar population synthesis codes \STARBURST\ \citep{L99,VL05} and \PEGASE\ 
\citep{PEGASE97} to generate the ionizing radiation field, assuming burst models at zero age with a Salpeter IMF and lower and upper mass limits of 
0.1 and 120 \Mo, respectively, with metallicities between 0.05 and 3 times solar. The ionizing radiation fields were input into the photoionization 
and shock code, \MAPPINGS\ \citep{SDopita93}, which includes self-consistent treatment of nebular and dust physics. \citet{KD02} previously used these 
models to simulate the emission-line spectra of \HII regions and starburst galaxies \citep{Do00}, and are completely described in their study.

{\bf Kobulnicky \& Kewley (2004)} gave a parameterization of the \citet{KD02} $R_{23}$ method with a form similar to that given by \citet{McGaugh91} 
calibration. 
\citet{KK04} presented an iterative scheme to resolve for both the ionization parameter $q$ and the oxygen abundance using only [\ion{O}{iii}], 
[\ion{O}{ii}] and \Hb\ lines. The parameterization they give for $q$ is
\begin{eqnarray}
\log(q)= \frac{32.81 - 1.153y^2 + \big[ 12+\log({\rm O/H}) \big] \big[ -3.396-0.025y+0.1444y^2 \big]}{4.603-0.3119y-0.163y^2 + \big[ 12+\log({\rm 
O/H}) \big] \big[ -0.48+0.0271y+0.02037y^2 \big]},
\end{eqnarray}
where $y=\log$([\ion{O}{iii}]/[\ion{O}{ii}]). This equation is only valid for ionization parameters between 5$\times$10$^6$ and 1.5$\times$10$^8$ cm 
s$^{-1}$. The oxygen abundance is parameterized by
\begin{eqnarray}
\nonumber 12+\log ({\rm O/H})_{up} =  9.72-0.777x-0.951x^2-0.072x^3-0.811x^4 -\log(q) \\
   \times (0.0737-0.0713x-0.141x^2+0.0373x^3-0.058x^4),\\
12+\log ({\rm O/H})_{low} = 9.40+4.65x-3.17x^2-\log(q)(0.272+0.547x-0.513x^2), 
\end{eqnarray}
being $x=\log R_{23}$. The first equation is valid for 12+log(O/H)$\geq$8.4, while the second for 12+log(O/H)$<$8.4. Typically, between two and three 
iterations are required to reach convergence. Following the authors, this parameterization should be regarded as an improved, implementation-friendly 
approach to be preferred over the tabulated $R_{23}$ coefficients given by \citet{KD02}.  

{\bf Nagao, Maiolino \& Marcani (2006)} did not consider any ionization parameter. They merely used data of a large sample of galaxies from the SDSS to 
derive a cubic fit to the relation between $R_{23}$ and the oxygen abundance,
\begin{eqnarray}
\log R_{23} = 1.2299 -4.1926y+1.0246y^2-0.063169y^3,
\end{eqnarray}
with $y$=\abox.

Besides $R_{23}$, additional parameters have been used to derive metallicities in star-forming galaxies.  Without other emission lines, the 
{\bf $N_2$ parameter}, which is defined by
\begin{eqnarray}
N_2 \equiv \log \frac{I([{\rm N\, II}]) \lambda 6583}{\rm H\alpha},
\end{eqnarray}
can be used  as a crude estimator of metallicity. However, we note that the [\ion{N}{ii}]/\Ha\ ratio is particularly sensitive to shock excitation or 
a hard radiation field from an AGN. The $N_2$ parameter was firstly suggested by \citet{SBCK94}, who gave a tentative calibration of 
the oxygen abundance using this parameter. This calibration has been revisited by \citet{vZee98,D02,PP04} and \citet*{Nagao06}.
The {\bf Denicol\'o et al. (2002)} calibration is
\begin{eqnarray}
12+\log ({\rm O/H}) = 9.12 + 0.73 N_2,
\end{eqnarray}
which considerably improves the previous relations because of the inclusion of an extensible sample of nearby extragalactic \HII regions. The 
uncertainty of this method is $\sim$0.2 dex because $N_2$ is sensitive to ionization and O/N variations, so strictly speaking it should be used mainly as an 
indicator of galaxy-wide abundances. \citet*{D02} also compared their method with photoionization models, concluding that the observed $N_2$ is 
consistent with nitrogen being a combination of both primary and secondary origin. 

\citet{PP04} revisited the relation between the $N_2$ parameter and the oxygen abundance including new data for the high- and low-metallicity regimen. 
They only considered those extragalactic \HII regions where the oxygen values are determined either via the $T_e$ method or with detailed photoionization 
modelling. Their linear fit to their data is
\begin{eqnarray}
12+\log ({\rm O/H}) = 8.90 + 0.57 N_2,
\end{eqnarray}
which has both a lower slope and zero-point that the fit given by \citet*{D02}. A somewhat better relation is provided by a third-order polynomical fit 
of the form
\begin{eqnarray}
12+\log ({\rm O/H}) = 9.37 + 2.032 N_2 + 1.26 (N_2)^2 + 0.32 (N_2)^3,
\end{eqnarray}
valid in the range $-2.5 < N_2 < -0.3$. \citet*{Nagao06} also provided a relation between $N_2$ and the oxygen abundance, their cubic fit to their 
SDSS data yields
\begin{eqnarray}
\log N_2 = 96.641 -39.941y + 5.2227y^2 -0.22040y^3,
\end{eqnarray}  
with $y$=\abox.

{\bf Pettini \& Pagel (2004)} revived the O$_3$N$_2$ parameter, previously introduced by \citet{Alloin79} and defined by
\begin{eqnarray}
O_3N_2 \equiv \log\frac{[{\rm O\, III}]\ \lambda 5007/{\rm H\beta}}{[{\rm N\, II}]\ \lambda 6583/{\rm H\alpha}}.
\end{eqnarray} 
\citet{PP04} derived the following least-square linear fit to their data:
\begin{eqnarray}
12+\log({\rm O/H}) = 8.73-0.32 O_3N_2.
\end{eqnarray}
\citet*{Nagao06} also revisited this calibration and derived a cubic fit between the O$_3$N$_2$ parameter and the oxygen abundance,
\begin{eqnarray}
\log O_3N_2 = -232.18 + 84.423y -9.9330y^2 +0.37941y^3,
\end{eqnarray}  
with $y$=\abox.

Other important empirical calibrations that were not used in this study involve the $S_{23}$ parameter, introduced by \citet{Vil96} and 
revisited by \citet{Dia00,OS00} and \citet{PerezMontero05}. In the last years, bright emission line ratios such as  [\ion{Ar}{iii}]/[\ion{O}{iii}] 
and [\ion{S}{iii}]/[\ion{O}{iii}] \citep{Sta06} or [\ion{Ne}{iii}]/[\ion{O}{iii}] and [\ion{O}{iii}]/[\ion{O}{ii}]  \citep*{Nagao06} have been 
explored as indicators of the oxygen abundance in \HII regions and starburst galaxies. \citet{Peimbert07} suggested to use the oxygen recombination 
lines to get a more precise estimation of the oxygen abundance. Nowadays, there is still a lot of observational and theoretical work to do involving 
empirical calibrations (see recent review by Kewley \& Ellison 2008), but these methods should be used only for objects whose \HII regions have the 
same structural properties as those of the calibrating samples \citep{Stasinska09}.

\newpage

\begin{table}[t!]
\centering
  \caption{\footnotesize{List of the parameters used to compute the oxygen abundance in all regions with a direct estimation of \Te\ using empirical calibrations.}}
  
  \label{abempirica1}
  \tiny
  \begin{tabular}{l@{\hspace{5pt}} l@{\hspace{5pt}}   c@{\hspace{5pt}}c@{\hspace{5pt}}c@{\hspace{5pt}}       
c@{\hspace{5pt}}c@{\hspace{5pt}}c@{\hspace{5pt}} }
  \tableline
   \noalign{\smallskip}
Region & $R_{23}$ & $P =R_3/R_{23} $ & $y$ =log($R_3/R_2$) & $N_2$ & $O_3N_2$ & $q_{KD02o}$$^a$ \\ 
\tableline
\noalign{\smallskip}   
       HCG~31~AC  &  5.42  &  0.571  &  0.125  &  0.104  &   1.349  &  3.76E+07  \\  
       HCG~31~B	  &  7.93  &  0.408  & -0.162  &  0.101  &   1.381  &  4.91E+07  \\  
       HCG~31~E   &  7.12  &  0.511  &  0.020  &  0.090  &   1.486  &  7.40E+07  \\  
      HCG~31~F1   &  8.91  &  0.819  &  0.656  &  0.034  &   2.201  &  5.78E+07  \\  
      HCG~31~F2   &  7.60  &  0.724  &  0.418  &  0.036  &   2.064  &  6.28E+07  \\  
       HCG~31~G   &  8.20  &  0.499  & -0.002  &  0.106  &   1.462  &  6.96E+07  \\  
	   Mkn~1199~C &  7.69  &  0.809  &  0.627  &  0.131  &   1.555  &  1.55E+08 \\
    Haro~15~A     &  9.73  &  0.884  &  0.881  &  0.027  &   2.378  &  8.55E+07  \\   
       Mkn~5~A1   &  7.58  &  0.748  &  0.473  &  0.051  &   1.915  &  6.96E+07  \\   
       Mkn~5~A2   &  8.19  &  0.702  &  0.372  &  0.049  &   1.944  &  1.72E+08  \\             
  IRAS~08208+2816~C&  7.77  &  0.793  &  0.583  &  0.129  &   1.558  &  8.55E+07  \\  
  POX~4   & 10.68  &  0.906  &  0.986  &  0.015  &   2.697  &  1.05E+08  \\   
        UM~420   &  6.45  &  0.649  &  0.268  &  0.099  &   1.497  &  4.81E+07  \\ 
 SBS~0926+606A   &  7.40  &  0.811  &  0.632  &  0.026  &   2.227  &  6.68E+07  \\   
SBS~0948+532   &  8.85  &  0.874  &  0.843  &  0.022  &   2.430  &  2.54E+08  \\   
SBS~1054+365   &  9.33  &  0.893  &  0.920  &  0.020  &   2.503  &  9.10E+07  \\  
SBS~1211+540   &  7.22  &  0.892  &  0.918  &  0.008  &   2.788  &  1.16E+08  \\  
SBS~1319+579A   &  9.92  &  0.908  &  0.996  &  0.014  &   2.671  &  1.05E+08  \\  
SBS~1319+579B   &  7.13  &  0.722  &  0.415  &  0.046  &   1.922  &  6.15E+07  \\  
SBS~1319+579C   &  7.11  &  0.710  &  0.389  &  0.052  &   1.860  &  5.91E+07  \\  
SBS~1415+437C   &  5.22  &  0.783  &  0.558  &  0.015  &   2.301  &  5.91E+07  \\  
SBS~1415+437A   &  4.86  &  0.810  &  0.629  &  0.012  &   2.370  &  5.44E+07  \\ 
   III~Zw~107~A    &  7.13  &  0.701  &  0.369  &  0.100  &   1.573  &  5.78E+07  \\ 
Tol~9~INT  &  4.58  &  0.689  &  0.345  &  0.252  &   0.973  &  4.16E+07  \\ 
      Tol~9~NOT   &  4.78  &  0.629  &  0.230  &  0.287  &   0.894  &  3.39E+07  \\ 
Tol~1457-262A   &  9.89  &  0.773  &  0.532  &  0.033  &   2.236  &  9.91E+07  \\ 
Tol~1457-262B   &  9.00  &  0.792  &  0.582  &  0.020  &   2.417  &  1.41E+08  \\ 
Tol~1457-262C   &  8.88  &  0.669  &  0.359  &  0.036  &   2.099  &  7.16E+07  \\ 
      ESO~566-8   &  5.17  &  0.505  &  0.008  &  0.414  &   0.693  &  3.19E+07  \\
     NGC~5253~A   &  9.20  &  0.851  &  0.756  &  0.102  &   1.754  &  6.82E+07  \\ 
     NGC~5253~B   &  9.38  &  0.856  &  0.775  &  0.086  &   1.841  &  7.11E+07  \\ 
     NGC~5253~C   &  8.03  &  0.773  &  0.532  &  0.041  &   2.046  &  2.60E+08  \\ 
     NGC~5253~D   &  7.67  &  0.527  &  0.048  &  0.079  &   1.582  &  7.72E+07  \\
\tableline
  \end{tabular}
      \begin{flushleft}
$^a$ Value derived for the $q$ parameter (in units of cm s$^{-1}$) obtained using the optimal calibration given by \citet{KD02}.
    \end{flushleft}

  \end{table}

\begin{table*}[t]
  \caption{\footnotesize{Results of the oxygen abundance, in the form \abox, for objects with a direct estimation of the metallicity, considering 
several empirical calibrations.}}
  \label{abempirica2}
    \tiny
  \begin{tabular}{l@{\hspace{2pt}} c@{\hspace{2pt}} c@{\hspace{2pt}} c@{\hspace{2pt}} c@{\hspace{2pt}} c@{\hspace{2pt}}  c@{\hspace{2pt}} 
c@{\hspace{2pt}} c@{\hspace{2pt}}  c@{\hspace{2pt}} c@{\hspace{2pt}} c@{\hspace{2pt}} c@{\hspace{2pt}} c@{\hspace{2pt}} c@{\hspace{2pt}} 
c@{\hspace{2pt}} c@{\hspace{2pt}} c@{\hspace{2pt}} c } 
  
  \tableline
   \noalign{\smallskip}
Region & Branch &  \Te & EP84 &  MRS85 & M91 & ZKH94 & P00 & P01$^a$&PT05$^b$& KD02 & KK04 & D02 & PP04a & PP04b & PP04c & N06a$^c$& N06b & N06c \\ 

& & &  R$_{23}$ & R$_{23}$ & R$_{23}$, $y$ & R$_{23}$ & R$_{23}$ & R$_{23}$, P & R$_{23}$, P & R$_{23}$, $q$ & R$_{23}$, $q$ & 
N$_2$ & N$_2$ & N$_2$ & O$_3$N$_2$ & R$_{23}$ & N$_2$ & O$_3$N$_2$  \\

\tableline
\noalign{\smallskip}  
HCG 31 AC &H&8.22$\pm$0.05&8.25& 8.89  & 8.67 & 8.74  & 8.47 & 8.15 & 8.09 & 7.99 & 8.12 & 8.40 & 8.34 & 8.29 & 8.30 & 8.05  &8.16& 8.22\\ 
HCG 31 B  &H&8.14$\pm$0.08&8.14& 8.48  & 8.29 & 8.44  & 8.24 & 8.22 & 8.12 & 8.41 & 8.44 & 8.39 & 8.33 & 8.28 & 8.29 & 8.07  &8.16& 8.20\\ 
HCG 31 E  &H&8.13$\pm$0.09&8.17& 8.62  & 8.14 & 8.53  & 8.31 & 8.18 & 8.13 & 8.19 & 8.32 & 8.35 & 8.30 & 8.26 & 8.25 & 8.07  &8.11& 8.15\\ 
HCG 31 F1 &L&8.07$\pm$0.06&8.02& 8.30  & 8.13 &\nodata& 7.86 & 8.12 & 7.99 & 8.46 & 8.33 & 8.05 & 8.07 & 8.09 & 8.03 &\nodata&7.81& 7.67\\ 
HCG 31 F2 &L&8.03$\pm$0.10&7.90& 8.54  & 8.06 & 8.48  & 7.76 & 8.13 & 7.95 & 8.19 & 8.27 & 8.06 & 8.07 & 8.10 & 8.07 & 8.07  &7.83& 7.79\\ 
HCG 31 G  &H&8.15$\pm$0.08&8.13& 8.43  & 8.26 & 8.40  & 8.22 & 8.11 & 8.17 & 8.31 & 8.42 & 8.41 & 8.34 & 8.29 & 8.26 & 8.07  &8.17& 8.16\\ 
Mkn 1199 C&H&8.75$\pm$0.12&9.37& 9.26  & 9.00 & 9.18  & 9.19 & 8.71 & 8.54 & 9.14 & 9.14 & 8.92 & 8.74 & 8.90 & 8.81 & 9.18  &8.78& 8.94\\ 
Haro 15 A &H&8.10$\pm$0.06&8.08&\nodata& 8.14 &\nodata& 7.91 & 8.12 & 8.12 & 8.48 & 8.34 & 7.98 & 8.01 & 8.05 & 7.97 &\nodata&7.74& 7.38\\ 
Mkn 5 A1  &L&8.07$\pm$0.07&7.90& 8.54  & 8.04 & 8.48  & 7.76 & 8.13 & 8.13 & 8.19 & 8.26 & 8.18 & 8.17 & 8.16 & 8.12 & 8.07  &7.94& 7.89\\ 
Mkn 5 A2  &L&8.08$\pm$0.07&7.95& 8.43  & 8.14 & 8.41  & 7.81 & 7.92 & 8.17 & 8.18 & 8.33 & 8.16 & 8.15 & 8.15 & 8.11 & 8.07  &7.92& 7.87\\ 
IRAS 08208+2816&H&8.33$\pm$0.08&8.15&8.50&8.55& 8.46  & 8.25 & 8.42 & 8.35 & 8.35 & 8.25 & 8.47 & 8.39 & 8.34 & 8.23 & 8.35  &8.23& 8.11\\ 
POX 4    &L&8.03$\pm$0.04&8.15&\nodata& 8.20 & \nodata&7.97 & 7.92 & 8.06 & 8.48 & 8.40 & 7.78 & 7.86 & 7.91 & 7.87 &\nodata&7.53&\nodata\\ 
UM 420    &L&7.95$\pm$0.05&7.78& 8.73  & 7.98 & 8.61  & 7.66 & 7.85 & 7.86 & 8.02 & 8.16 & 8.20 & 8.39 & 8.33 & 8.28 & 7.57  &8.15& 8.14\\ 
SBS 0926+606A&L&7.94$\pm$0.08&7.88&8.57& 7.97 & 8.50  & 7.75 & 7.77 & 7.80 & 8.17 & 8.20 & 7.97 & 8.00 & 8.05 & 8.02 & 7.71  &7.73& 7.64\\ 
SBS 0948+532&L&8.03$\pm$0.05&8.01& 8.31& 8.06 &\nodata& 7.86 & 7.82 & 8.10 & 8.34 & 8.28 & 7.91 & 7.95 & 8.01 & 7.95 &\nodata&8.01&\nodata\\ 
SBS 1054+365&L&8.00$\pm$0.07&8.05& 8.21& 8.09 &\nodata& 7.89 & 7.84 & 7.91 & 8.48 & 8.30 & 7.87 & 7.93 & 7.98 & 7.93 &\nodata&7.63&\nodata\\ 
SBS 1211+540&L&7.65$\pm$0.04&7.86& 8.60& 7.85 & 8.52  & 7.73 & 7.68 & 7.65 & 8.02 & 8.10 & 7.58 & 7.70 & 7.69 & 7.84 & 7.68  &7.31&\nodata\\ 
SBS 1319+579A&L&8.05$\pm$0.06&8.09&\nodata&8.13&\nodata&7.93 & 8.11 & 8.11 & 8.48 & 8.33 & 7.77 & 7.85 & 7.90 & 7.88 &\nodata&7.52&\nodata\\ 
SBS 1319+579B&L&8.12$\pm$0.10&7.85&8.62& 8.01 & 8.53  & 7.72 & 8.13 & 8.12 & 8.13 & 8.23 & 8.14 & 8.14 & 8.14 & 8.11 & 8.07  &7.90& 7.89\\ 
SBS 1319+579 C&L&8.15$\pm$0.07&7.85&8.62&8.02 & 8.53  & 7.72 & 8.13 & 8.13 & 8.12 & 8.23 & 8.18 & 8.17 & 8.16 & 8.13 & 8.06  &7.94& 7.93\\ 
SBS 1415+437 C&L&7.58$\pm$0.05&7.63&8.91&7.72 & 8.76  & 7.53 & 7.57 & 7.55 & 7.86 & 7.99 & 7.79 & 7.86 & 7.92 & 7.99 & 7.39  &7.55& 7.55\\ 
SBS 1415+437 A&L&7.61$\pm$0.06&7.58&8.96&7.64 & 8.80  & 7.49 & 7.50 & 7.48 & 7.82 & 7.92 & 7.72 & 7.81 & 7.86 & 7.97 & 7.34  &7.48& 7.41\\ 
III Zw 107 &H&8.23$\pm$0.09&8.17  &8.62&8.57 & 8.53   & 8.31 & 8.40 & 8.35 & 8.13 & 8.24 & 8.39 & 8.33 & 8.28 & 8.23 & 8.46  &8.15& 8.10\\ 
Tol 9 INT  &H&8.58$\pm$0.15&8.30& 9.00 & 8.76 & 8.84  & 8.58 & 8.61 & 8.55 & 8.95 & 8.90 & 8.68 & 8.56 & 8.54 & 8.42 & 8.77  &8.46& 8.40\\ 
Tol 9 NOT  &H&8.55$\pm$0.16&8.29& 8.97 & 8.73 & 8.81  & 8.55 & 8.56 & 8.50 & 8.94 & 8.88 & 8.72 & 8.59 & 8.59 & 8.44 & 8.75  &8.51& 8.44\\ 
Tol 1457-262A&L&8.05$\pm$0.07&8.09&\nodata&8.26&\nodata&7.92 & 8.11 & 8.20 & 8.58 & 8.42 & 8.04 & 8.06 & 8.09 & 8.02 &\nodata&7.80& 7.63\\ 
Tol 1457-262B&L&7.88$\pm$0.07&8.55&\nodata&8.21&\nodata&7.87 & 7.91 & 8.21 & 8.38 & 8.29 & 7.88 & 7.93 & 7.99 & 7.96 &\nodata&7.60&\nodata\\ 
Tol 1457-262C&L&8.06$\pm$0.11&8.48&\nodata&8.21&\nodata&7.88 & 8.00 & 8.24 & 8.48 & 8.37 & 8.07 & 8.08 & 8.10 & 8.06 &\nodata&7.83& 7.77\\ 
ESO 566-8 &H&8.46$\pm$0.11&8.27& 8.92 & 8.68 &  8.77  & 8.50 & 8.44 & 8.38 & 8.92 & 8.84 & 8.84 & 8.68 & 8.76 & 8.51 & 8.70  &8.66& 8.53\\ 
NGC 5253 A&H&8.18$\pm$0.04&8.09& 8.24 & 8.13 &\nodata & 8.15 & 8.11 & 8.13 & 8.53 & 8.33 & 8.40 & 8.34 & 8.28 & 8.17 &\nodata&8.16& 8.00\\ 
NGC 5253 B&H&8.19$\pm$0.04&8.09& 8.21 & 8.14 &\nodata & 8.14 & 8.11 & 8.13 & 8.48 & 8.34 & 8.34 & 8.29 & 8.25 & 8.14 &\nodata&8.10& 7.94\\ 
NGC 5253 C&L&8.28$\pm$0.04&8.14& 8.46  & 8.53 & 8.42  & 8.23 & 8.38 & 8.32 & 8.67 & 8.63 & 8.11 & 8.11 & 8.13 & 8.08 & 8.30  &7.87& 7.80\\ 
NGC 5253 D&L&8.31$\pm$0.07&8.15& 8.52  & 8.19 & 8.47  & 8.26 & 8.23 & 8.17 & 8.32 & 8.37 & 8.31 & 8.27 & 8.23 & 8.22 & 8.37  &8.07& 8.10\\ 
\tableline
  \end{tabular}
    \begin{flushleft}
 NOTE: The empirical calibrations and the parameters used for each of them are:   
 EP84: Edmunds \& Pagel (1984) that involves the $R_{23}$ parameter; MRS85: McCall, Rybski \& Shields (1985) using
$R_{23}$; M91: McGaugh (1991) using $R_{23}$ and $y$; ZKH94: Zaritzky, Kennicutt \& Huchra (1994) using $R_{23}$; P00: Pilyugin (2000) using 
$R_{23}$; P01: Pilyugin (2001a,b) using $R_{23}$ and $P$; KD02: Kewley \& Dopita (2002) using $R_{23}$ and $q$; KK04: Kobulnicky \& Kewley (2004) 
using $R_{23}$ \& $q$; D02: Denicol\'o, Terlevich \& Terlevich (2002) using the $N_2$ parameter; PP04: Pettini \& Pagel (2004), using (a) $N_2$ with 
a linear fit, (b) $N_2$ with a cubic fit, (c) the $O_3N_2$ parameter; N06: Nagao et al. (2006) using the cubic relations involving the $R_{23}$ (a), 
$N_2$ (b) and $O_3N_2$ (c) parameters. The value compiled in the column labeled \Te\ is the oxygen abundance derived by the direct method. \\

  $^a$ The value listed for P01 is the average value between the high- and the low-metallicity branches for objects with 7.90$<$\abox$<$8.20.\\
  $^b$ The value listed for PT05 is the average value between the high- and the low-metallicity branches for objects with 8.05$<$\abox$<$8.20.\\
  $^c$ The value listed for N06 is the average value between the high- and the low-metallicity branches for objects with 8.00$<$\abox$<$8.15.\\
  \end{flushleft}
  \end{table*}


\end{document}